\documentclass{iopjournal}[centering]
\usepackage{amssymb}
\usepackage{physics}
\usepackage{siunitx}
\usepackage{mathtools}
\usepackage{upgreek}
\usepackage{booktabs}
\usepackage{multirow}
\usepackage{makecell}
\AtBeginDocument{
	\heavyrulewidth=.08em
	\lightrulewidth=.05em
	\cmidrulewidth=.03em
	\belowrulesep=.65ex
	\belowbottomsep=0pt
	\aboverulesep=.4ex
	\abovetopsep=0pt
	\cmidrulesep=\doublerulesep
	\cmidrulekern=.5em
	\defaultaddspace=.5em
}
\usepackage{enumitem}   
\usepackage[font=small,labelfont=bf, justification=Justified]{subcaption}
\usepackage[font=small,labelfont=bf, justification=Justified]{caption}

\usepackage{txfonts}
\usepackage{bm}
\usepackage{placeins}
\usepackage{setspace}
\usepackage[colorlinks=true]{hyperref}
\usepackage{cleveref}
\crefrangelabelformat{equation}{(#3#1#4--#5\crefstripprefix{#1}{#2}#6)}
\usepackage[nolist]{acronym}

\usepackage{tikz}
\usetikzlibrary{calc,positioning,shapes.misc}
\usetikzlibrary{arrows.meta,trees,shadings}
\usetikzlibrary{shapes.geometric}
\usetikzlibrary{intersections}
\newcommand{\pasqrthz}{\ensuremath{\text{pA}/\sqrt{\text{Hz }}}}
\newcommand{\hzsqrthz}{\ensuremath{\text{Hz}/\sqrt{\text{Hz }}}}
\newcommand{\pmsqrthz}{\ensuremath{\text{pm}/\sqrt{\text{Hz }}}}
\newcommand{\nradsqrthz}{\ensuremath{\text{nrad}/\sqrt{\text{Hz }}}}
\newcommand{\weff}{\ensuremath{w_\text{eff}}}
\newcommand{\rqpd}{\ensuremath{r_\text{QPD}}}
\newcommand{\rsepd}{\ensuremath{r_\text{SEPD}}}
\newcommand{\eqpd}{\ensuremath{\eta_\text{QPD}}}
\newcommand{\he}{\ensuremath{\eta_\text{het}}}
\newcommand{\she}{\ensuremath{\sqrt{\he}}}
\newcommand{\phe}{\ensuremath{\psi_\text{het}}}
\newcommand{\sheunc}{\ensuremath{k_\text{unc}}}
\newcommand{\shecorS}{\ensuremath{k_{\text{cor, }\Sigma}}}
\newcommand{\shecorD}{\ensuremath{k_{\text{cor, }\Delta}}}
\newcommand{\radsqrthz}{\ensuremath{\text{rad}/\sqrt{\text{Hz }}}}

\newcommand{\erfi}{\ensuremath{\text{ erfi}} }
\newcommand{\sn}[2]{\ensuremath{{#1}\times 10^{#2}}}
\newcommand{\revw}[1]{\textcolor{blue}{#1}}

\begin{document}

\articletype{Paper} 

\title{Mathematical derivation and verification of the amplitude of LISA's interferometric signals on an ultra-stable interferometer testbed}

%

\author{Alvise Pizzella$^{1,2, *}$\orcid{0000-0003-0542-7807}, Lennart Wissel$^{1,2}$\orcid{0000-0002-0688-3233}, Miguel Dovale-\'Alvarez$^{1,2}$\orcid{0000-0002-6300-5226}, Pablo Mart\'inez Cano$^{1,3}$\orcid{0009-0001-9463-0614}, Rodrigo Garc\'ia \'Alvarez$^{1,2}$\orcid{0009-0008-3480-2710}, Christoph Bode$^{1,2}$\orcid{0000-0002-9203-3955}, Juan Jos\'e Esteban Delgado$^{1,2}$\orcid{0000-0002-7613-3681} and Gerhard Heinzel$^{1,2}$\orcid{0000-0003-1661-7868}}

\affil{$^1$Max Planck Institute for Gravitational Physics (Albert Einstein Institute), D-30167 Hannover, Germany}
\affil{$^2$Leibniz Universit\"at Hannover, D-30167 Hannover, Germany}
\affil{$^3$University of Granada, 18012-Granada, Spain}
\affil{$^*$Author to whom any correspondence should be addressed.}

\email{alvise.pizzella@aei.mpg.de}
\email{gerhard.heinzel@aei.mpg.de}

\begin{abstract}
The \acf{lisa} mission aims to detect gravitational waves by interferometrically measuring the change of separation between free-falling \acfp{tm}. \ac{lisa}'s interferometers must deliver $\pmsqrthz$ sensitivity while accommodating beam tilts up to 1\,mrad at the \aclp{pd}, which degrade the interferometric amplitude and increase the induced readout noise coupling. This paper uses an analytical framework developed by the authors in a previous work, based on minimal and justified approximations, that relates beam tilt to the resulting heterodyne signal amplitude in a generic two-beam interferometer with circular-area \acp{pd}. A set of interferometric topologies is analyzed, all of high relevance for \ac{lisa}. We derive the exact amplitude response for an infinite detector and a closed-form approximation for finite detectors, and we validate both against numerical simulations and experimental measurements on an ultra-stable \ac{lisa}-representative testbed. We then use this model to quantify the phase-noise amplification arising from reduced \ac{snr} under tilt, showing that curvature mismatches between the interfering beams substantially enhance this effect. Finally, we introduce a compact function that captures the angular dependence of correlated and uncorrelated phase noises in \ac{qpd}-based readouts. Here, a new noise feature, caused by wavefront curvature mismatch, is predicted and measured for the first time. These results indicate that controlling wavefront curvature mismatch in the \ac{tmi} is essential to limit excess phase noise. The models and results derived in this paper, although originating in the context of \ac{lisa}, are general and can be applied to any interferometric topology undergoing tilts with pivot on the detector plane.
\end{abstract}

\section{Introduction} \label{section:introduction}

The quest for the detection of gravitational waves started with Joseph Weber at the University of Maryland in the early nineteen sixties, using resonant bar detectors at room temperature~\cite{pizzella_historyGW}. This first generation of detectors was followed by a second generation of cryogenic resonant bar antennas, which operated between 1985 and 2010~\cite{pizzella_historyGW}. As these failed to obtain sufficient evidence to claim a detection, an alternative technology for gravitational-wave detection, laser interferometry, was proposed by Weber~\cite{Moss:71}. The foundations for its use in the \ac{ligo} were established by Rainer Weiss in~\cite{Weiss1972}.

After several years of effort, finally, on 14~September 2015, the Advanced \ac{ligo} detectors, a second-generation laser interferometer sensitive in the Hz-kHz band, first detected a gravitational wave signal originating from a binary black hole merger, designated event GW150914~\cite{first_detection}. This landmark observation firmly established laser interferometry between free-falling \acfp{tm} as a mature and effective technology for gravitational-wave detection, thereby advancing its further development. Currently, a third-generation ground-based detectors, such as the Einstein Telescope~\cite{Branchesi_2023} and Cosmic Explorer~\cite{evans2023cosmicexplorersubmissionnsf}, is under development; their low-frequency sensitivity, however, is ultimately constrained by ground motion and by the residual noise due to the length and angular control systems needed to maintain the interferometers in a suitable operating point~\cite{Cahillane2022}. Accessing the rich sub-Hz gravitational-wave spectrum requires moving away from such noise sources, and hence moving to space; this is the motivation of the \ac{lisa} mission \cite{LISA_red_book}.

\Ac{lisa} will enable the exploration of a rich region of the gravitational-wave spectrum, in the frequency range between 0.1\,mHz and 1\,Hz. The mission was officially adopted by \ac{esa} in 2024~\cite{lisa_adoption}. Core enabling technologies — residual \ac{tm} acceleration at the $10^{-15}\,{\rm m/s^2}$ level in interplanetary orbit, drag-free spacecraft control, and picometer-precision interferometric displacement sensing — were successfully demonstrated by the \ac{lpf} mission, launched on 3~December~2015~\cite{pathfinder1, pathfinder2, pathfinder3}. 

Unlike \ac{lpf}, however, \ac{lisa} is designed to have the longest measurement baseline ever implemented in a scientific experiment: $2.5$\,million\,km between spacecraft. This design requires a novel heterodyne interferometry scheme capable of operating over interplanetary distances. Demonstrating its performance on ground is indeed profoundly challenging.

\ac{lisa} probes the separation between free-falling \acp{tm} that are shielded from external disturbances by their host spacecraft. The \ac{tm}-to-\ac{tm} measurement is split into three sub-measurements: 
\begin{enumerate}[label=(\roman*)]
    \item a \ac{tm} to spacecraft displacement measurement in the transmitting spacecraft, performed in the \ac{tmi};
    \item a spacecraft-to-spacecraft displacement measurement, performed in the \ac{isi};
    \item a second \ac{tm} to spacecraft \ac{tmi} measurement in the receiving spacecraft.
\end{enumerate}
An \ac{rfi} is added as a necessary phase reference within each spacecraft, linking the two onboard lasers~\cite{LISA_yellow_book}. 

The \ac{lisa} spacecraft exchange laser beams that propagate for $\sim2.5\times 10^9$\,m, or $\sim8.3\,$ light-seconds. During this journey, the TEM$_{00}$ Gaussian beam diverges into a several-kilometer-wide profile, of which only a 30\,cm portion is captured by a receiving telescope. This results in a ''flat-top'', or ''\ac{th}'', beam profile~\cite{Chwalla_2020}, which further propagates through the interferometer. This received beam (Rx) is interfered with a local TEM$_{00}$ beam called Tx beam -- as it is a fraction of the beam being transmitted -- producing a heterodyne beat note whose differential phase variations encode path length variations in the optical link. The combination and detection of these beams occurs on the \ac{lob}~\cite{LOB_design}. The \ac{lob} consists of an ultra-low-expansion glass baseplate, to which optical elements are bonded via hydroxide-catalysis bonding~\cite{hcb}. Each spacecraft hosts two \acp{lob}, establishing the triangular links of the constellation.

The ambitious sensitivity of \unit{\pico \meter \per \hertz^{1/2}} is challenged by several noise sources. According to recent estimates, \Ac{ttl} coupling noise -- which was a major noise source in \ac{lpf}~\cite{pathfinder4} -- is one of the main contributors to the overall noise floor in \ac{lisa} \cite{IDS_performance_model_error_budget}. \ac{ttl} coupling is caused by the unavoidable angular jitter of the spacecraft, and its optical components, with respect to the \acp{tm} and the line of sight, and the consequent spurious path length signals that the jitter produces. 

In \ac{lisa}, beam tilts mainly originate from a well-defined pivot, i.e. the \ac{tm}'s surface for the \ac{tmi} and the entrance pupil of the telescope for the \ac{isi}, which allows the use of custom-designed imaging systems to suppress \ac{ttl}. Tilts of the spacecraft result in rotations of the Rx beam wrt. the receiving telescope entrance pupil, which is in turn imaged into a fixed aperture located on the \ac{lob}, called Rx-clip. The Rx-clip is then imaged onto the \ac{isi} \aclp{pd}, ensuring that the impact of spacecraft tilts on the pathlength signals is minimized. 

\Ac{ttl} primarily affects the \ac{isi}, and its influence on the optically measured path length variations $\Delta s$ is mitigated through three mechanisms. First, the \ac{dfacs} actively stabilizes the attitudes and alignment of both spacecraft and \acp{tm}, reducing the overall angular jitter. Second, custom imaging systems passively suppress \ac{ttl} by mapping the relevant beam pivot points onto the \acp{pd}~\cite{Troebs_2018, Chwalla_2020}. Third, the residual path length variations $\Delta s(\varphi,\eta)$ are removed in post-processing using the independently measured tilt angles~\cite{PhysRevD.111.043048}. Accurate subtraction and control require an angular readout with precision near $1\,\mathrm{nrad}/\sqrt{\mathrm{Hz}}$ in the \ac{lisa} band, provided by the \ac{dws} technique. 

The performance of both the longitudinal and angular readout is affected by the alignment of the interfering beams. The presence of large tilts degrades the overlap, or \ac{he}, of the two beams, ultimately leading to a smaller signal and a consequently reduced \ac{snr}. The induced asymmetry in the case of non-zero tilts may also affect the otherwise very high common-mode rejection, a key factor in the very high sensitivity of \ac{dws}.

A complete description of the \ac{he} generally requires numerical evaluation of detector-plane integrals. In this article, we instead develop an analytical approach based on physically justified approximations. This yields compact expressions for the first few orders Maclaurin coefficients of the \ac{he} response on \acp{sepd} and \acp{qpd}, expressed in terms of fundamental geometrical parameters: wavelength, beam profile and sizes, wavefront curvatures, and \ac{pd} diameter. We then validate these expressions through comparison with a high-stability optical testbed representative of the \ac{lisa} \ac{lob}.

We then use these results to quantify the impact of beam tilts on the noise in the \ac{lps} and \ac{dws} signals in \ac{lisa}'s readout architecture. We investigate, in particular, the impact of wavefront curvature mismatches, as these lead to diverse \acp{he} across the segments of a \ac{qpd}. Our considerations suggest introducing a new function, derived from the \ac{he} and encompassing the overall \ac{qpd}'s response to tilts, to enable a straightforward derivation of the \ac{qpd} signal's noise. We derive this in a general context using two \acp{gb}, which can be applied to any interferometric topology subject to tilts. We further derive this for a \ac{gb}-\ac{th} interferometric topology, which is relevant mostly only for spaceborne interferometers as \ac{lisa}, Taiji \cite{Wu2021} and Tianqin \cite{Luo_2016}, or Grace-Follow On \cite{Sheard2012}.

Finally, we apply these results to \ac{lisa}'s \ac{tmi}, to quantify the noise benefits of intentionally tilting the \acp{tm} to compensate for misalignments in the \ac{lob}'s build. 

\Cref{section:setup} introduces the testbed; \Cref{section:heff-model} and \Cref{section:heff-imaging-systems} develop the analytical model; \Cref{section:results} presents the experimental verification; \Cref{section:qpd-noise} analyzes the consequences for phase noise; and \Cref{section:tm-tilt} considers the noise trade-offs associated with operating the \ac{tm} at a small static tilt.

\section{Setup} \label{section:setup}

\subsection{Overview of the testbed}

\begin{figure*}[!htpb]
\centering
\includegraphics[width=\textwidth]{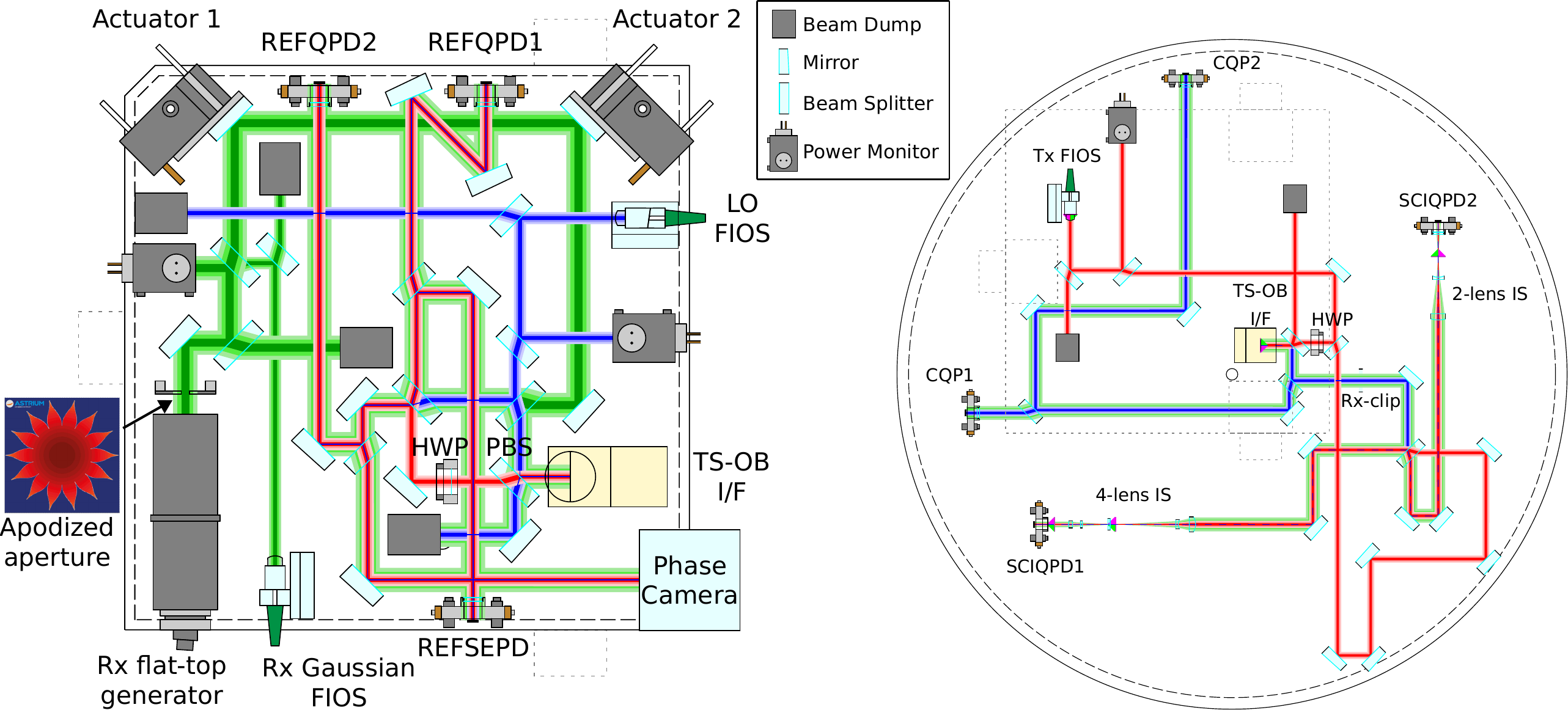}
\caption{Optical layout of the testbed. \textit{Left:} The \acf{ts} provides the \acf{ob} (right) with laser beams (\ac{rxgb} and \ac{rxft}, both in green) to simulate \ac{lisa}'s \ac{tmi} and \ac{isi}. The first beam is a TEM\textsubscript{00} Gaussian beam provided by a \ac{fios} with a width of $\sim$2 mm. The second is a flat-top beam generated by clipping a 9 mm radius beam Gaussian beam on an apodized aperture; this generates a beam flat in both intensity and phase with a diameter of $\sim$4\,mm \cite{Chwalla_2016}. The \ac{ts} sits on top of the \ac{ob} and is held by three Zerodur$^\circledR$ feet, which are indicated by the dashed lines. An optical link is established via a vertical interface (\ac{ts}/\ac{ob} I/F). A pair of actuated steering mirrors, named ''act1'' and ''act2'' on the left, on the \ac{ts} tilt the Rx beams around the Rx-clip (this is an aperture, which is physically located on the \ac{lob}) to simulate either the spacecraft jitter in the \ac{isi} of the \ac{tm} jitter on the \ac{tmi}. An additional \ac{fios} generated TEM\textsubscript{00} Gaussian beam, the \acs{lo} (blue), is used as an alignment aid for the \ac{ts} with respect to the \ac{ob} and as a phase reference. The \ac{ts} hosts four optical copies of the Rx-clip (located on the \ac{ob}). In these optical copies, one \ac{refsepd}, two \acp{refqpd} and a phase camera are placed.
\textit{Right:} The \ac{ob} is a simplified version of the \acp{ob} in \ac{lisa}. It features a local \ac{fios} generated TEM\textsubscript{00} Gaussian beam, the Tx beam, and only one interferometer, where all three the Tx, Rx, and the \acs{lo} beams interfere. At the interferometer's two output ports, the two \acp{sciqpd} are placed. These output ports feature imaging systems to image the point of rotation of the Rx beam onto the center of the \acsp{qpd} and reduce \ac{ttl} coupling. In this Figure we show the two types of imaging systems that were developed for \ac{tdobs}, one featuring 2 lenses, and one featuring 4 lenses. During the work reported in this article, two 4-lense imaging systems were used. The \ac{ts}'s nominal position on top of the \ac{ob} is indicated by the dashed lines. Such position is defined with the aid of a \ac{cqp}; this is a pair of large silicon \acsp{qpd} positioned on the \ac{lob}. When the \acs{lo} beam impinges on the center of both \acsp{qpd} of the \ac{cqp}, the Tx beam \ac{ts} also propagates to the center of the \acp{pd} on the \ac{ts}. Figure and caption from \cite{alvise_2024}.}
\label{fig::tdobs_layout}
\end{figure*}

A testbed for simulating the above-described effects on the \ac{lob} was developed by the authors of~\cite{Chwalla_2016}. The \ac{tdobs} testbed was primarily designed to demonstrate, in a \ac{lisa}-representative manner, that imaging systems can effectively suppress \ac{ttl} coupling~\cite{Chwalla_2016}. In addition to its initial scope, \ac{tdobs} can serve as a versatile platform for testing several other aspects of \ac{lisa}'s metrology, such as imaging systems, photoreceivers, readout systems, and mechanisms for compensating linear \ac{ttl}~\cite{Chwalla_2016}. The optical layout is shown in Fig.~\ref{fig::tdobs_layout}.

The testbed consists of two Zerodur$^\circledR$ glass-ceramic baseplates: the \ac{ob} and the \ac{ts}. The \ac{ob} is a simplified realization of the \ac{lob}, featuring a single interferometer that can emulate any of \ac{lisa}'s three interferometers. The \ac{ts}, an optical ground-support equipment mounted above the \ac{ob}, provides beams representative of either the \ac{tmi} (a tilting \ac{gb}) or the \ac{isi} (a tilting \ac{th} beam), referred to as \ac{rxgb} and \ac{rxft}, respectively. The \ac{th} beam allows for testing the \ac{isi} in a relatively compact space.

A pair of steering mirrors on the \ac{ts} synthesizes beam tilts representative of \ac{tm} motion or spacecraft attitude fluctuations. The testbed was recently upgraded to operate with beat notes in the 5–28\,MHz range (i.e., the \ac{lisa} heterodyne band) and with low-noise \acp{pr} comparable to those foreseen for \ac{lisa}~\cite{me_thesis, alvise_2024}.

\subsection{Laser modulation bench}

The laser modulation bench generates and delivers three beams (Rx, \ac{lo}, and Tx) to \ac{tdobs}. All beams originate from a single NPRO-type Nd:YAG laser at 1064\,\unit{\nano\meter} stabilized to a molecular iodine reference, achieving a frequency stability of approximately 1000\,\hzsqrthz at 1\,\unit{\milli\hertz}~\cite{Huarcaya2023}.

The testbed uses three tunable optical frequencies to simulate the three \ac{lisa} beams. To generate these, the main beam is split into three paths and sent through a double-pass \ac{aom} stage~\cite{AOM_DP}, providing MHz-level frequency shifts with high tuning bandwidth and efficient fiber coupling. The Rx beam is further split using a half-wave plate and a \ac{pbs}. The resulting four beams are coupled into single-mode, polarization-maintaining optical fibers and routed to \ac{tdobs}. Of the two Rx beams, one is sent to the \ac{rxgb} \acf{fios}, while the other is directed to the flat-top beam generator (Rx-FT)~\cite{Chwalla_2020}.

\subsection{Photoreceivers} \label{subsection:pr}

\begin{figure}
    \centering
    \includegraphics[width=0.25\columnwidth]{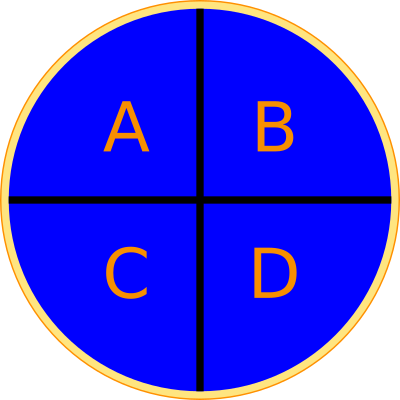}
    \includegraphics[width=0.25\columnwidth]{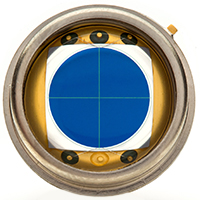}
    \caption{\textit{Right:} Picture of a large commercial \ac{qpd}. \textit{Left:} definition of the used segment-nomenclature throughout the paper. The area between the segments is called \textit{slit}. Depending on the specific \ac{qpd} model, the slit can be either insensitive, partially sensitive, or even more sensitive than the active area itself. Note that the \acp{qpd} in this experiment use a diameter of 1\,mm.}
    \label{fig:qpd-pic}
\end{figure}

In this paper, we refer to the \ac{pr} as the ensemble formed by the \ac{pd} and its associated \ac{tia}. \ac{lisa} employs round \acfp{qpd}, i.e.\ \acp{pd} whose active area is divided into four identical, albeit rotated, segments (see Fig.~\ref{fig:qpd-pic}). The used material for the absorption layer is \ac{ingaas}, chosen due to its high quantum efficiency at 1064\,nm despite the higher capacitance. Each \ac{qpd} segment is read out by a dedicated \ac{tia}.

Each \ac{qpr} segment must satisfy an input-equivalent current noise density requirement  $\tilde i_\text{en}(f) < 2$ \pasqrthz within the \ac{lisa} heterodyne band. An expression valid for beat note frequencies in this band is given in \cref{eq::tia-eq-in-current-noise} from \cite{german_thesis}, noting its explicit frequency dependence:
\begin{equation}
\tilde i_\text{en}(f) \approx \sqrt{ \frac{4 k_B T}{R_\text{F}} + \tilde i_\text{ACE}^2 + \Big(2 \pi f (C_\text{PD} + C_\text{ACE} ) \, \tilde e_\text{ACE}\Big)^2 },
\label{eq::tia-eq-in-current-noise}
\end{equation}
where the subscript ACE denotes the active circuit element used as the amplifier; $\tilde i_\text{ACE}$, $\tilde e_\text{ACE}$ and $C_\text{ACE}$ represent its current noise density, voltage noise density, and input capacitance, respectively. $C_\text{PD}$ is the \ac{pd} (or \ac{qpd} segment)'s capacitance, $R_\text{F}$ is the feedback resistance of the \ac{tia}, $k_B$ is the Boltzmann constant and $T$ is the temperature of the \ac{tia}, which both for \ac{lisa} and \ac{tdobs} is room temperature $\sim$\,300\,K.

Minimizing this electronic noise requires careful optimization of both the \ac{tia} and the \acp{qpd}. In particular, the \ac{qpd}-segment's capacitance must be kept as small as possible to avoid noise increases at higher heterodyne frequencies. Since $C_\text{PD}\propto r_\text{QPD}^2$, where $r_\text{QPD}$ is the \ac{qpd} radius, this constrains the detector's size, and further introduces a trade-off with the \ac{dws} gain~\cite[Section~III]{alvise_2024}.

\begin{figure}[!htpb]
\centering
\includegraphics{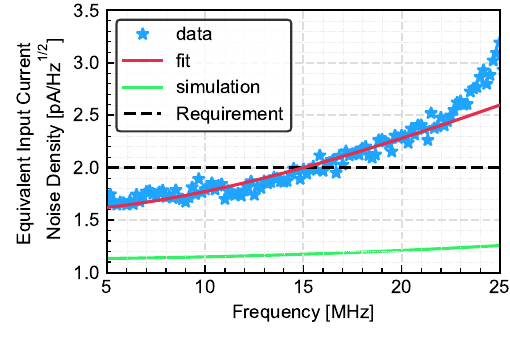}
\caption{Measurement with three different light intensities, SPICE simulation, and fit of the electronic noise from the \ac{tia} used for the \aclp{pd} in \ac{tdobs}. The fit function is $f(x) = \sqrt{p_0^2 + (p_1x)^2}$, reproducing \cref{eq::tia-eq-in-current-noise}. Figure from \cite{alvise_2024}.}
\label{fig::tia_noise}
\end{figure}

The \acp{qpr} used in \ac{tdobs} were developed close to the current \ac{lisa} baseline~\cite{german_discrete}. As the flight-model \acp{qpd} are still under developement, we had to chose commercial counterparts. For the \acp{qpd}, the GAP1000Q model from GPD Optoelectronics was selected. This features a 1\,mm diameter active area with 20\,\unit{\micro\meter} slits. The associated \acp{tia} follow the present \ac{lisa} baseline and use a transistor first-stage amplifier~\cite{german_discrete}. Characterized using the white-light method~\cite{german_thesis}, these \acp{tia} exhibit input-equivalent current noise below 2\,\pasqrthz up to $\sim15$\,\unit{\mega\hertz} (see Fig.~\ref{fig::tia_noise}).

A total of four such \acp{qpr} are used in \ac{tdobs}: two serve as \acp{refqpd} on the \ac{ts} in optical copies of the Rx-clip, while the remaining two are \acp{sciqpd} placed at the interferometer output ports on the \ac{ob}.

\subsection{Imaging Systems} \label{subsection:imaging-systems}

Pupil-to-pupil imaging systems are used to mitigate \ac{ttl} coupling, as well as beam walk, defined as the lateral displacement of the beam spot on the \ac{pd} surface along one or both transverse axes~\cite{Chwalla_2020}. Mathematically, they can be described using the ray-transfer matrix formalism~\cite[chapter 15]{siegman1986lasers} by a $2\times2$ $\big(\begin{smallmatrix} A & B\\ C & D \end{smallmatrix}\big)$ matrix with null $B$ element. As ray-transfer matrices have a determinant of one when there is no change of refraction index, this implies that $D = A^{-1}$:
\begin{align}
\textbf{r'} &= M_\text{IS} \textbf{r},  & M_\text{IS} &= \begin{pmatrix}
m^{-1} & 0\\
C & m
\end{pmatrix}, \label{eq::generic-is}
\end{align}
where $\textbf{r} = (x, \, \theta)$ is the vector describing the ray's position and angle, and $m$ is the \textit{angular magnification}. Moreover, such systems can be designed to produce a collimated output for a collimated input by imposing $C=0$ in \cref{eq::generic-is}. The imaging systems used in \ac{tdobs} and in \ac{lisa}, including the telescope, are of this collimating type.

When a \ac{gb} is propagated through an imaging system, its mode changes. For a collimating system, a \ac{gb} characterized by spot radius $w_\text{in}$ and \ac{roc} $R_\text{in}$ at the input pupil plane emerges with
\begin{align}
    w_\text{out} &= m^{-1}\, w_\text{in}, \label{eq:IS-beam-waist}\\
    R_\text{out} &= m^{-2}\, R_\text{in}. \label{eq:IS-beam-roc}
\end{align}
In \ac{lisa}, imaging systems also serve to compress the beam size, allowing for smaller \ac{qpr} dimensions. As a consequence, the beam’s incidence angle is magnified by a factor $m$ (see \cref{eq::generic-is}).

\begin{figure*}[!htpb]
\centering
\includegraphics[width=0.47\textwidth]{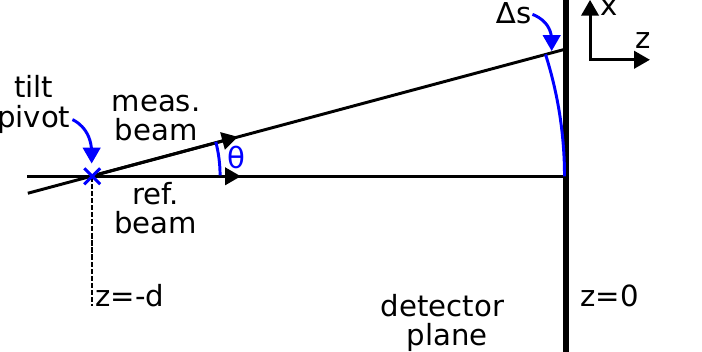}
\includegraphics[width=0.47\textwidth]{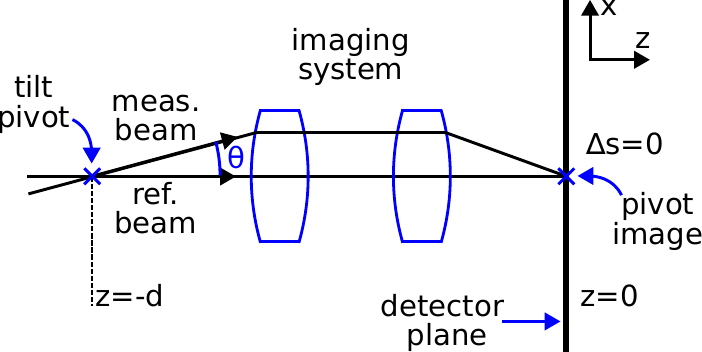}
\caption{Description of the lever arm effect between two beams, which dominates the geometrical \ac{ttl}. Two beams, the reference beam and the measurement beam, interfere. The resulting intensity is measured by a \ac{pr} located on the $z=0$ plane. The measurement beam can rotate about the pivot point $(0, 0, -d)^T$, representing either the Rx-clip or the \ac{tm}'s surface in \ac{lisa}. Such a tilt causes the measurement beam to propagate over a longer path; we denote the additional path-length as $\Delta s$. The extra path length is interferometrically measured at the \ac{pr}; if $\Delta s$ is not constant in time, it generates a noise term, and such noise is called \ac{ttl}. Imaging systems reduce the lever arm effect between two beams and beam walk by imaging the point of rotation of the beam into the center of the \ac{pr}. This means the optical path length is unchanged by the tilt, as light always takes the shortest path. As a consequence of this design, beam tilts in \ac{lisa} have the center of the \ac{qpd} as a rotation pivot and reduced undesired beam walk on the \ac{qpd}. Figure from \cite{alvise_2024}.}
\label{fig::imaging_system}
\end{figure*}

\ac{lisa} implements the imaging systems such that the tilt pivot of the beam arriving at the \ac{pr} coincides with the detector center. Furthermore, the telescope functions as a four-mirror collimating imaging system that images the Rx beam onto the Rx-clip, and additional collimating imaging systems further image the Rx-clip onto each \ac{qpr} surface. This peculiarity of \ac{lisa}'s design allows for a description of tilts as only rotations about the centres of the \acp{qpd}. This principle is depicted in \Cref{fig::imaging_system}.

\subsection{Phasemeters}

The phasemeter is responsible for reading out the phases of the electrical beat notes from the \ac{qpr}; in addition to the phases, also the amplitude of each beat note is returned. \ac{tdobs} employs two phasemeters, both sampling the input signals at $f_s = 80$\,MHz. The first is based on I/Q demodulation and requires precise knowledge of the beat note frequencies to be tracked. It is an older 16-channel model, not designed to meet the stringent \ac{lisa} phase noise requirement, but remains essential to \ac{tdobs} because its control software drives the steering mirrors on the \ac{ts}. 

The second phasemeter is an engineering-model precursor of the \ac{lisa} phasemeter. This 8-channel version features \ac{dpll}-based beat-note tracking and is designed to meet the required \ac{lisa} performance up to its Nyquist frequency. Its dominant noise sources are thermally induced phase fluctuations and \ac{adc}-clock jitter. The latter is mitigated by injecting a common electrical pilot tone that is tracked across all channels, enabling post-processing corrections~\cite{oliver_pm, gerberding_thesis, bode_thesis}.

\subsection{Longitudinal Path Length Signal} \label{ssec:lps}

The \ac{lps} signal is the main \ac{qpd} signal, and encodes the gravitational wave's signal. This signal combines the readout of the four segments to synthesize a \ac{sepd}-like response. We define the \ac{lps} as the \ac{ap} across the four segments normalized by the wave number $k = 2\pi/\lambda$,
\begin{equation}
\text{LPS}_\text{AP}
= \frac{1}{k}\,\phi_\text{AP}
= \frac{1}{k}\,\frac{\phi_A + \phi_B + \phi_C + \phi_D}{4},
\label{eq::lps-ap-definition}
\end{equation}
This average-phase \ac{lps} is used throughout the paper as the effective longitudinal readout of the \ac{qpd}. Another possible combination of the four segments' readout to recover a longitudinal signal, the \ac{lpf}-\ac{lps}, is discussed in \cite{Hartig_2023}.

\subsection{Differential Wave Front Sensing} \label{subsection:dws}

\begin{figure*}[!htpb]
\centering
\includegraphics[width=0.47\textwidth]{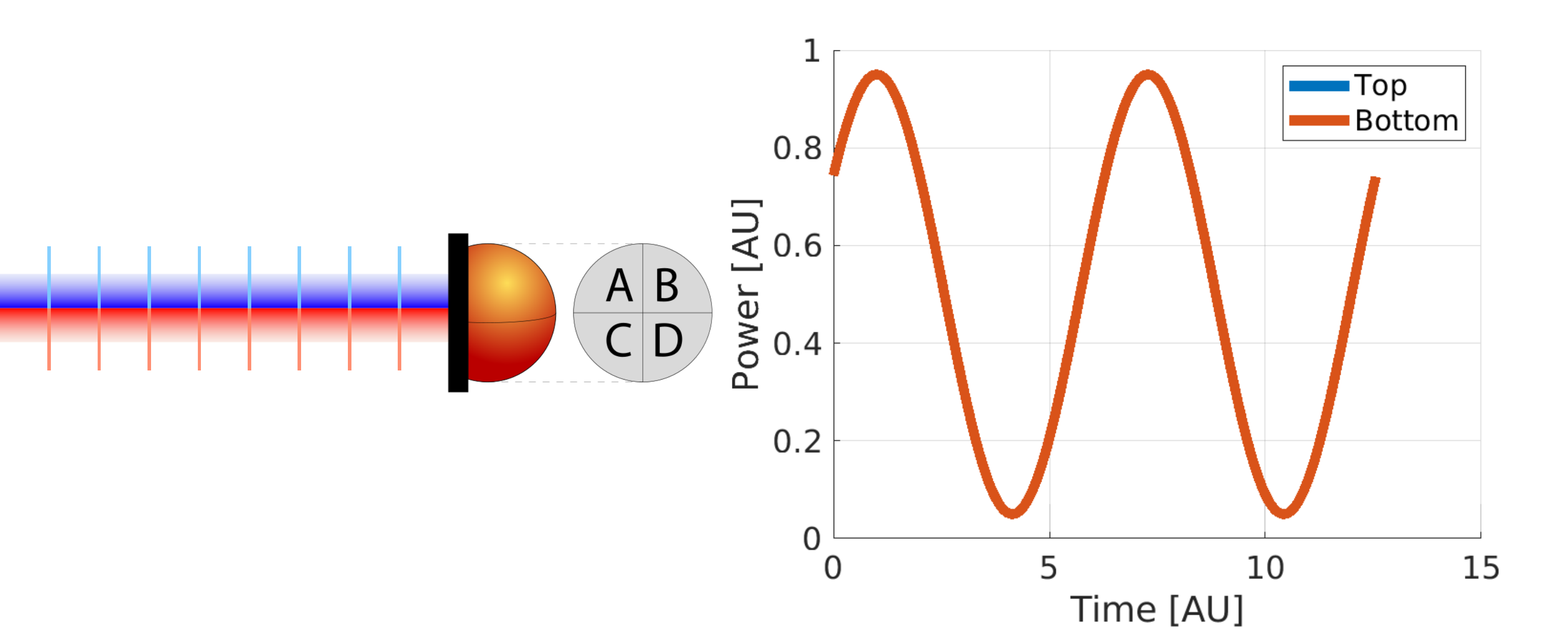}
\includegraphics[width=0.47\textwidth]{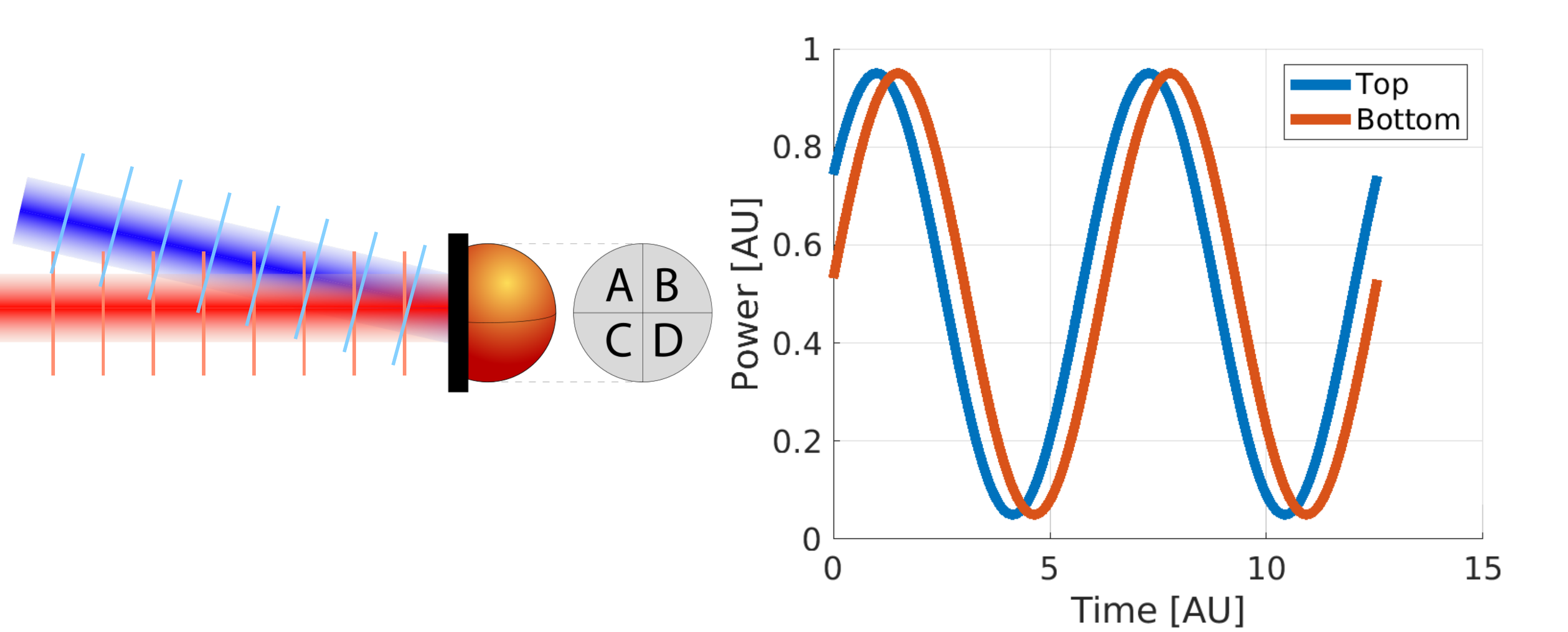}
\caption{Principle of \ac{dws}. A reference beam (red) and measurement beam (blue) with a MHz-scale frequency difference produce a heterodyne beat tone on each \ac{qpd} segment. When aligned (left), all beat notes share the same phase. When the measurement beam is tilted (right), the wave fronts reach the top segments earlier than the bottom ones, producing a phase difference proportional to the tilt angle. Figure from \cite{alvise_2024}.}
\label{fig::dws-description}
\end{figure*}

The \ac{dws}~\cite{Morrison:94, Morrison:94_exp} signal is the \ac{qpd} signal used for the angular readout. Its principle is to measure the relative wave-front angle between two interfering beams using the phase differences of their heterodyne beat notes across the \ac{qpd}. As sketched in Fig.~\ref{fig::dws-description}, a reference beam is assumed to be centered and normally incident, while the measurement beam rotates around the detector center. The segment phases $\phi_i$ are combined to form the vertical and horizontal \ac{dws} combinations:
\begin{align}
    \text{DWS}_\text{v} &= \phi_\text{top} - \phi_\text{bottom}
    = \frac{\phi_A + \phi_B - \phi_C - \phi_D}{2}, \label{eq:dwsv-def}\\
    \text{DWS}_\text{h} &= \phi_\text{right} - \phi_\text{left}
    = \frac{\phi_A - \phi_B + \phi_C - \phi_D}{2}, \label{eq:dwsh-def}
\end{align}
which depend on the vertical and horizontal tilt angles $\eta$ and $\varphi$, respectively. For small tilts, the response is approximately linear \cite{Hechenblaikner:10, alvise_2024},
\begin{align}
\text{DWS} \approx \kappa_1\,\theta, \qquad
\kappa_1 \sim r_\text{QPD}/\lambda \sim 10^3, 
\label{eq::dws-approx-linear}
\end{align}
with $\theta$ being the tilt angle representative of either $\eta$ or $\varphi$, and $\kappa_1$ the zero-angle \ac{dws} gain. We refer to $\theta$ as the \ac{wfa} when indicating it is the angle recovered from \ac{dws}. A more accurate treatment that accounts for the \ac{gb} structure and larger tilt angles is reported in~\cite{alvise_2024}. A practical phasemeter implementation is described in~\cite{PhysRevApplied.14.054013}.

\section{Tilted beam intensity} \label{section:heff-model}
This section derives the relation between the \ac{he} and the tilt angle $\theta$ in the case of the interference of two fundamental-mode \acp{gb}, or one fundamental-mode \ac{gb} and a \ac{th}, beam being detected by a \ac{pd}. Two \ac{pd} shapes are considered: a \ac{sepd} and a \ac{qpd}. The former, although of no interest for \ac{lisa}, is included in the analysis for completeness, as it is easier to model and also furnishes a useful comparison. This estimation can be done analytically under the assumption that the \ac{pd}’s radius is much larger than the beam’s spot radius; this is calculated in subsections~\ref{ssec:infinite-sepd} and~\ref{ssec:infinite-qpd} in the case of two \acp{gb} impinging on an infinite \ac{sepd} and infinite \ac{qpd}, respectively. Subsection~\ref{ssec:th-gb-pd-inf} extends these calculations to the case where the measurement beam is a \ac{th}. An analytical extension of this method to a finite and circular \ac{pd} is also possible; this is calculated in subsections~\ref{ssec:finite-sepd} and~\ref{ssec:finite-qpd} in the case of two \acp{gb} impinging on a finite \ac{sepd} and finite \ac{qpd}, respectively. This last case also requires the wavefront curvature mismatch between the beams to be small. Subsection~\ref{ssec:th-gb-pd-fin} extends these models to the case where the measurement beam is a \ac{th}. A numerical model is also developed, to describe the cases that don't respect these conditions; we report one such model in subsection~\ref{subsection:heff-numeric-model}, which we compare against the analytical expressions.

\subsection{Framework}
This subsection defines the mathematical framework of the model and the overlap and \ac{he} of two beams. The framework presented in this paper is a further development of that presented in~\cite[III.A]{alvise_2024}; here, as the model was used to calculate the \ac{dws} signal, only the interference phase was considered, whereas in this work, for the \ac{he}, also the beam's amplitude must be considered. Readers who are already familiar with \cite{alvise_2024} can skip the following lines until \cref{eq::itf_tilt_simp}. For the sake of self-consistency, the initial assumptions are repeated in the following lines.

To start, two fundamental-mode \acp{gb} are assumed. The presence of imaging systems, as mentioned in \cite[II.D]{alvise_2024} and in \Cref{fig::imaging_system}, justifies the restriction of the analyzed rotations to those with the pivot located at the \ac{pd}'s center. Let $\vec x = (x, y, z)^T$ be the laboratory's \ac{rf}. In this \ac{rf}, the \ac{pd} is a surface defined by $z=0$, and the reference beam propagates along the $z$-axis. The electric field of the reference beam in the lab's \ac{rf} is given by
\begin{align}
\vec E(\vec x)_r &= \Re \left[E_r \hat \varepsilon_{r} \mathcal{E}_r(\vec x) e^{i\omega_r t}\right], \label{eq:GB-Vfield}\\
\begin{split}
\mathcal{E}_r(\vec x) &= \sqrt{\frac{2}{\pi}}\frac{w_{0, \, r}}{w_r(z_r)} \exp \left( - \frac{x^2 + y^2}{w_r(z_r)^2} + \right.\\
& \left. +i \left( k \frac{x^2 + y^2}{2 R_r(z_r)} +kz_r + \eta(z_r)\right)\right),
\label{eq::GB-amplitude}
\end{split}
\end{align}
where $E_r \in \mathbb{C}$ [V/m] is a complex scalar determining the beam's amplitude and initial phase, $\hat \varepsilon_{r} = (\epsilon_{x, \, r}, \epsilon_{y, \, r}, 0)^T$, $\epsilon_{x, \, r}$, $\epsilon_{y, \, r} \in \mathbb{C}$, $\hat \varepsilon_r \perp \hat \varepsilon_z$ is a generic polarization unitary vector of an electromagnetic wave propagating along the $z$-axis, $\mathcal{E}_r(\vec x)$ is the complex scalar wave amplitude, $z_r = z - z_{0, \, r}$ is the waist position adjusted $z$-coordinate and $w_{0, \, r}$ is the reference beam's waist, $w_r(z_r)$, $R_r(z_r)$ and $\eta(z_r)$ are the reference beam's spot size, \ac{roc} and Gouy phase at the position $z_r$, respectively, and $\omega_r = 2\pi \frac{c}{\lambda}$ is the angular frequency. From here onward, we will mostly omit the argument $z_r$. \Cref{eq::GB-amplitude} is normalized such that
\begin{equation}
    \int_{\mathbb{R}^2} |\mathcal{E}_r(\vec x)|^2 \dd{x} \dd{y} = w_{0, \, r}^2 \, \forall z.
\end{equation}
The total power of the beam can be recovered as
\begin{equation}
    P_{0, \, r} =  \frac{1}{2Z} \int_{\mathbb{R}^2} |\vec E_r(\vec x)|^2 \dd{x} \dd{y} = \frac{1}{2Z} |E_r|^2 w_{0, \, r}^2 = I_{0, \, r} w_{0, \, r}^2, \label{eq:beam-total-power}
\end{equation}
where $Z[\Omega]$ is the wave impedance of the medium in which the beam is propagating and $I_{0, \, r}$ is the reference beam's intensity [\unit{\watt \per \meter^2}]. A dual statement holds for $P_{0, \, m}$, \textit{mutatis mutandis}.

\begin{figure}
    \centering
    \includegraphics[width=0.6\columnwidth]{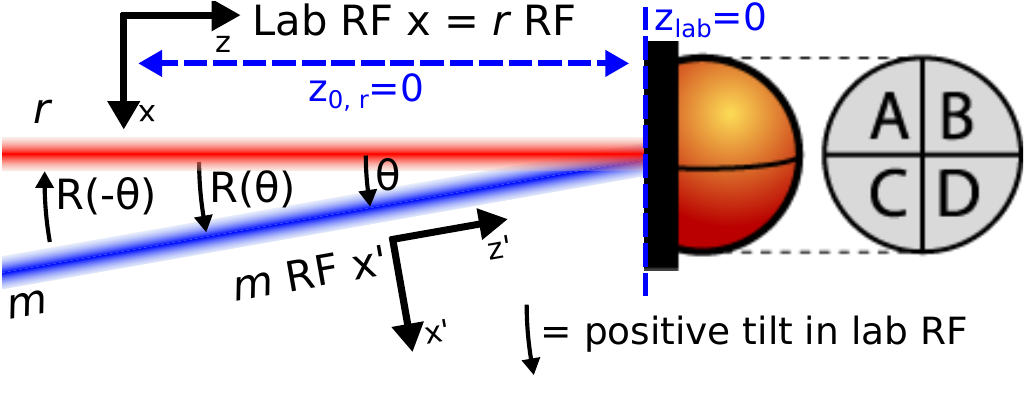}
    \caption{Framework of the used \aclp{rf}. The \ac{qpd} defines the lab's \ac{rf}, with the \ac{qpd}'s center being located at $\vec x = \vec 0$, and the slits are aligned with the $x$ and $y$ axes. The reference beam's \ac{rf} shares the same coordinates with the lab's \ac{rf}, but is shifted along the $z$-axis. The measurement beam's \ac{rf} is rotated and shifted along the $z$ axis with respect to the lab's \ac{rf}. A positive tilt of the beam is defined as an anticlockwise tilt. The measurement beam's \ac{rf} can be obtained by rotating the lab's \ac{rf} by an angle $\theta$. Dually, the lab's \ac{rf} can be obtained by rotating the measurement beam's \ac{rf} by an angle $-\theta$. The tilt shown in the figure causes the measurement beam's path length to the top segments A and B to be larger than that to the bottom segments C and D. Therefore, in such a configuration, the \ac{dws}\textsubscript{v} signal is positive. Figure from \cite{alvise_2024}.}
    \label{fig:analytic-beam-rotation}
\end{figure}

The measurement beam propagates on a generic axis, which is rotated by a small angle $\theta\ll1$ around the \ac{qpd}'s center with respect to the $z$-axis called $z'$. This axis is used to define the measurement beam's \ac{rf} in the primed coordinates $\vec x' = (x', y', z')^T$. Choosing this rotation to be either a vertical (pitch) rotation or a horizontal (yaw) rotation allows for analytical solutions. In the proposed linearized framework, any generic rotation can be decomposed in a combination of these two rotations, hence allowing a description of any rotation. As justified in \Cref{fig::imaging_system}, the rotation pivot can be assumed to be the \ac{qpd}'s center $(x, y, z)^T = (0, 0, 0)^T$. Therefore, the lab's \ac{rf} and measurement beam's \ac{rf} are related by a rotation matrix
\begin{align}
    &\begin{pmatrix}
    x\\
    y\\
    z
    \end{pmatrix}
    =
    \begin{pmatrix}
    1 & 0 &0\\
    0 & \cos(\theta)  & \sin(\theta)\\
    0 & -\sin(\theta) & \, \cos(\theta)
    \end{pmatrix}
    \begin{pmatrix}
    x'\\
    y'\\
    z'
    \end{pmatrix} &\text{ or } & \vec x = R(-\theta) \vec x'.
    \label{eq::meas-to-lab-RF}
\end{align}

The electric field of the measurement beam is expressed by the same \cref{eq:GB-Vfield,eq::GB-amplitude} in the measurement beam's \ac{rf}. We name this quantity
\begin{equation}
\vec E'_m(\vec x') =  \Re \left[E_m\hat \varepsilon_{m}' \mathcal{E}'_m(\vec x') e^{-i \omega_m t}\right],
\label{eq::GB-vector}
\end{equation}
Expressing $\vec E'_m(\vec x')$ in the lab's \ac{rf} requires a change of coordinates. It can be expressed in the same form as
\begin{equation}
\vec E_m(\vec x, \theta) =  \Re \left[E_m \hat \varepsilon_{m}(\theta) \mathcal{E}_m(\vec x, \theta) e^{-i \omega_m t} \right],
\label{eq::GB-vector-rotated}
\end{equation}
We now evaluate the relation between all primed and unprimed quantities. The beam's amplitude $E_m'$ is a scalar and does not vary in the change of coordinates: $E_m = E_m'$. The electric field is a vector field; therefore a generic electric field transforms as
\begin{equation}
\vec E(\vec x) = R(-\theta) \vec E'( \vec x') = R(-\theta) \vec E'\Big( R(\theta)\vec x \Big), \label{eq:vector-field-rotation}
\end{equation}
where in the second step the inverse of \cref{eq::meas-to-lab-RF} has been used. The rotation of the complex scalar wave amplitude of the measurement beam results in \cref{eq::GB-full-tilt} as shown in \cite{alvise_2024}. The polarization unitary vector is also rotated, resulting in \cref{eq::versor-rotation}.
\begin{equation}
\hat \varepsilon_m' = \begin{pmatrix}
\epsilon_{x, \, m}\\
\epsilon_{y, \, m}\\
0
\end{pmatrix} \rightarrow \hat \varepsilon_m(\theta) = R(-\theta) \hat \varepsilon_m' = \begin{pmatrix}
\epsilon_{x, \, m}\\
\cos(\theta)\epsilon_{y, \, m}\\
\sin(\theta) \epsilon_{y, \, m}
\end{pmatrix}
\label{eq::versor-rotation}
\end{equation}
The resulting full expression for the complex scalar wave amplitude is shown in \cref{eq::GB-full-tilt} in generic form (note that in our coordinate choice $z=0$). This rather complicated expression can be simplified by neglecting the dependence of $w_m$, $R_m$ and $\eta$ on the transverse coordinate $y$ \cite{alvise_2024}. This leads to \cref{eq::GB-tilt} as shown in \cite[section III]{alvise_2024}, making this a linear framework.
\begin{equation}
\begin{split}
\mathcal{E}_{m \text{, full}}(\vec x, \theta) &= \mathcal{E}_m'( R(\theta) \vec x)\\
&= \sqrt{\frac{2}{\pi}}\frac{w_{m, \, 0}}{w_m(\sin(\theta)y + z_m)} \exp \left( - \frac{x^2 + (\cos(\theta)y + \sin(\theta)z)^2}{w_m^2(\sin(\theta)y + z_m)} \right.+\\
& \quad \left.+ i \left( k \frac{x^2 + (\cos(\theta)y + \sin(\theta)z)^2}{2 R_m(\sin(\theta)y + z_m)} + k(\sin(\theta)y + z_m) + \eta(\sin(\theta)y + z_m)\right)\right), \label{eq::GB-full-tilt}
\end{split}
\end{equation}
\begin{equation}
\mathcal{E}_{m \text{ , approx}}(\vec x, \theta) = \sqrt{\frac{2}{\pi}}\frac{w_{m, \, 0}}{w_m} \exp \left( - \frac{x^2 + y^2}{w_m^2} + i \left( k \frac{x^2 + y^2}{2 R_m} + \underbrace{k y \theta}_\text{tilt} + kz_m + \eta(z_m) \right)\right).
\label{eq::GB-tilt}
\end{equation}

The remaining tilt coupling is given by the extra pathlength term $\exp(i k \theta y)$. The intensity of the superimposed reference and measurement beams $I_\text{ifm}$ is recovered by taking the square modulus of the sum of the two electric fields, and dividing it by twice the wave impedance of the medium in which the beam is propagating $Z[\Omega]$:
\begin{small}
\begin{equation}
\begin{split}
\text{GB-GB: }\\ 
I_\text{ifm}(x, y, \theta) &= \frac{1}{2 Z} \left[ \underbrace{\left(E_r \sqrt{\frac{2}{\pi}}\frac{w_{r, \, 0}}{w_r(z_r)}\right)^2 \exp \left( -2 \frac{x^2 + y^2}{w_r(z_r)^2} \right)}_{|\vec E_r|^2}+ \underbrace{\left(E_m \sqrt{\frac{2}{\pi}} \frac{w_{m, \, 0}}{w_m(z_m)}\right)^2 \exp \left( -2 \frac{x^2 + y^2}{w_m(z_m)^2} \right)}_{|\vec E_m|^2} + \right.\\
&\quad +\left. \underbrace{\hat \varepsilon_{r} \cdot \hat \varepsilon_{m}\left( E_r E_m \frac{2}{\pi} \frac{w_{r, \, 0}}{w_r(z_r)} \frac{w_{m, \, 0}}{w_m(z_m)} \exp \left( -2\frac{x^2 + y^2}{w_\text{eff}^2}(1 - i\rho) +i\Big(kz_r + \eta(z_r) -k y \theta - kz_m -\eta(z_m) + (\omega_m - \omega_r)t \Big)\right) + \, c.c. \,\right)}_{\vec E_r \cdot \vec E_m^* + \vec E_r^* \cdot \vec E_m} \right],
\end{split}
\label{eq::itf_tilt_full}
\end{equation}
\end{small}
where $\cdot^*$ indicates complex conjugation and the quantities
\begin{align}
\frac{2}{w_\text{eff}^2} &= \frac{1}{w_r^2(z)}+\frac{1}{w_m^2(z)}, \label{eq::weff}\\
\frac{1}{R_\text{rel}} &= \frac{1}{R_r(z)}-\frac{1}{R_m(z)}, \label{eq::rrel}\\
\rho &= \frac{w_\text{eff}^2 k}{4 R_\text{rel}}, \label{eq::rho}
\end{align}
have been introduced \cite{alvise_2024}. The quantity $w_\text{eff}$ is the effective spot radius of the two overlapped beams, and $R_\text{rel}$ is their absolute wavefront mismatch \cite{alvise_2024}. The quantity $\rho$ is the effective-spot-radius-normalized wavefront curvature mismatch; an expression for $\rho$ depending only on the fundamental parameters of the interfering beams is derived in \cite[eq. (33)]{alvise_2024} in the case of two fundamental mode \acp{gb}. Both $R_\text{rel}$ and $\rho$ depend on the beam's labelling, and are hence odd under the swapping of the two beams. In this linearized framework, the same holds also for the tilt angle $\theta$. Consequently, any physical effect of a combination of these parameters must respect this symmetry \cite{alvise_2024}.

Our goal is deriving the variation of the amplitude of the beam interference overlap as a function of the tilt and with respect to the individual beam's powers; we hence split the beam's intensity into the three components mentioned in \cref{eq::itf_tilt_full}. From here on, we make the assumption that both beams have linear and horizontal polarization, given by $\varepsilon_m = \varepsilon_r = (1, 0, 0)^\text{T}$. Since the tilt is vertical, this results in $\hat \varepsilon_{r} \cdot \hat \varepsilon_{m} = 1$ independently of $\theta$. Note that if both polarization and tilt were vertical $\varepsilon_m = \varepsilon_r = (0, 1, 0)^\text{T}$, then the scalar product $\hat \varepsilon_{r} \cdot \hat \varepsilon_{m} = \cos(\theta)$ would depend on the tilt angle. This effect, although negligible, must be taken into account due to the vector nature of the electric field. A simplified expression for the overlap intensity term in \cref{eq::itf_tilt_full}, which still contains the relevant phenomena, is given by
\begin{equation}
I_{\text{smp}}(x, y, \theta) =  \frac{2}{\pi}\exp \left( -2\frac{x^2 + y^2}{w_\text{eff}^2}(1 - i\rho) - ik y \theta\right) , \label{eq::itf_tilt_simp}
\end{equation}
to which the complex conjugate is to be added to make it a real quantity. The full equation of the intensity \cref{eq::itf_tilt_full} can be recovered using \cref{eq::itf_tilt_simp-to-full}
\begin{equation}
\begin{split}
\text{GB-GB: } \\
I_{\text{ifm}}(x, y, \theta) &=  I_r(\vec x) + I_m(\vec x) + C \left(I_{\text{smp}}(x, y, \theta) e^{i \Psi} + c.c. \right), \label{eq::itf_tilt_simp-to-full}
\end{split}
\end{equation}
where the quantities $I_r$, $I_m$, $C$ and $\Psi$ are
\begin{align}
    I_r(\vec x) &= \frac{2}{\pi} I_{r, \, 0} \left(\frac{w_{r, \, 0}}{w_r(z_r)}\right)^2 \exp \left( -2 \frac{x^2 + y^2}{w_r(z_r)^2} \right), \label{eq:ref-beam-power}\\
    I_m(\vec x) &= \frac{2}{\pi} I_{m, \, 0}  \left(\frac{w_{m, \, 0}}{w_m(z_m)}\right)^2 \exp \left( -2 \frac{x^2 + y^2}{w_m(z_m)^2} \right),\\
    C     &= \hat \varepsilon_{r} \cdot \hat \varepsilon_{m} \sqrt{I_{r, \, 0} I_{m, \, 0} } \left( \frac{w_{r, \, 0}}{w_r(z_r)}\frac{w_{m, \, 0}}{w_m(z_m)} \right),\\
    \Psi &= kz_r + \eta(z_r) - kz_m -\eta(z_m) + (\omega_m - \omega_r)t.
\end{align}
Note that by taking the square modulus of the measurement beam's electric field $\vec E_m(\vec x, \theta)$, the dependence on $\theta$ vanishes. We use \cref{eq::itf_tilt_simp} to evaluate the detector-plane integrals, and later revert the result back into the form of $I_{\text{ifm}}$ to determine the \ac{he}.

Given a \ac{pd} of any geometry, the power measured by a \ac{pd} is the integral of the intensity over the \ac{pd}'s surface. Throughout the whole paper, the beams are assumed to be centered on the \ac{pd}.
\begin{equation}
    P_\text{PD}(\theta) = \int_{S_\text{PD}} \dd{x} \dd{y} I_\text{ifm}(x, y, \theta) 
\end{equation}
The total power can be split into three terms corresponding to the ones underbraced in \cref{eq::itf_tilt_full}. These are the power of the reference beam $P_\text{PD, r}$, the power of the measurement beam $P_\text{PD, m}$, and the \textit{overlap power term} $P_\text{pd}(\theta)$. $P_{\text{PD, }m}$ depends, in principle, on the tilt angle $\theta$. Such dependency vanishes if the \ac{pd}'s size is much larger than the measurement beam's size, and is negligible otherwise. Hence, the dependence of $P_{\text{PD, }m}$ on $\theta$ is neglected. Hence, only the \textit{overlap power term} depends on the tilt angle $\theta$. 
\begin{align}
    P_{\text{PD, }r}         &= \frac{1}{2Z} \int_{S_\text{PD}} \dd{x} \dd{y} |\vec E_r(\vec x)|^2  &= \int_{S_\text{PD}} \dd{x} \dd{y} I_r(\vec x) \label{eq:r-pd-power}\\
    P_{\text{PD, }m}         &= \frac{1}{2Z} \int_{S_\text{PD}} \dd{x} \dd{y} |\vec E_m(\vec x, \theta)|^2  &=  \int_{S_\text{PD}} \dd{x} \dd{y} I_m(\vec x)\label{eq:m-pd-power}\\
    P_\text{PD, ifm}(\theta) &= \frac{1}{2Z} \int_{S_\text{PD}} \dd{x} \dd{y}  \vec E_r(\vec x) \cdot \vec E_m^*(\vec x, \theta) &=   \int_{S_\text{PD}} \dd{x} \dd{y} \left[ I_\text{smp} + c.c. \right](\vec x, \theta)\label{eq:ifm-pd-power}
\end{align}
The power measured by the \ac{pd} can hence be written as
\begin{equation}
    P_\text{PD}(\theta) = P_{\text{PD, }r} +  P_{\text{PD, }m} + 2 \Re \left[ P_\text{PD, ifm}(\theta)  \right].
    \label{eq:power1}
\end{equation}

The \ac{he} and \textit{heterodyne phase shift} $\she(\theta), \psi_\text{het}(\theta) \in \mathbb{R}$ of two interfering beams are defined in this paper as
\begin{equation}
    \she(\theta) \, e^{i \psi_\text{het}(\theta)} = \frac{P_\text{PD, ifm}(\theta)}{\sqrt{P_{\text{PD, }r}}\sqrt{P_{\text{PD, }m}}}.
    \label{eq:het-eff-definition}
\end{equation}
These are two purely geometrical quantities, which are independent of the beams' powers. In particular, the \ac{he} is limited in the range $0 \leq \she(\theta) \leq 1$, while for the heterodyne phase shift it must hold $0 \leq \psi_\text{het}(\theta) < 2\pi$. The \ac{he} quantifies the overlap of the two beams, and hence the "quality" of the interference. It should not be confused with the \textit{visibility}, or \textit{contrast}, \cite{me_thesis}, which is often used in homodyne interferometry.
\begin{equation}
    V = \frac{P_\text{PD, max}-P_\text{PD, min}}{P_\text{PD, max}+P_\text{PD, min}} , \label{eq:visibility}
\end{equation}
which is instead beam-power-dependent. The power in \cref{eq:power1} can finally be rewritten as
\begin{equation}
    \begin{split}
    P_\text{PD}(\theta) &= P_{\text{PD, }r} +  P_{\text{PD, }m} + 2 \she(\theta) \, \times \\
    &\quad \times \sqrt{ P_{\text{PD, }r} P_{\text{PD, }m} }  \cos \left(2\pi f_\text{het} t + \psi_\text{het}\left(\theta\right) \right).
    \end{split}
\label{eq:power2}
\end{equation}
to show its dependence on the \ac{he}, and where in the argument of the cosine the \ac{hf} term $f_\text{het} t, \, f_\text{het} = \frac{1}{2\pi}(\omega_m - \omega_r)$ was added to describe the heterodyne interference, which is the case in \ac{lisa}, where $f_\text{het} \in [5-25]$\,MHz. The visibility in \cref{eq:visibility} can hence be re-written as
\begin{equation}
    V = \frac{4 \she(\theta) \sqrt{ P_{\text{PD, }r} P_{\text{PD, }m} }  }{P_{\text{PD, }r} +  P_{\text{PD, }m}}. \label{eq:het-to-visibility}
\end{equation}

In \cref{eq:power2}, the $P_{\text{PD, }r}$ and $P_{\text{PD, }m}$ terms are usually called \textit{DC powers}. The remaining term is called the \textit{beat note}, and is the signal encoding the differential path length variation caused by gravitational waves. While the angle-dependent heterodyne phase shift $\psi_\text{het}(\theta)$ is responsible for \ac{ng-ttl} in this specific case of rotations about the \ac{qpd}'s center, the angular dependence of the \ac{he} is responsible for variations of the signal's amplitude. This induces variations of the \ac{snr}, defined as~\cite{RIN_lennart}:
\begin{equation}
    \text{SNR} = \frac{\text{Signal RMS}}{\text{Noise ASD}}, \label{eq:SNR-def}
\end{equation}
where, shortly, "Signal RMS" is the signal's amplitude and "Noise ASD" is the noise level. We refer to subsection~\ref{ssection:qpd-phase-noise} for further detail on \cref{eq:SNR-def}. Since the noise level does not depend on the beam alignment, while the signal's amplitude does, this mechanism leads to a tilt-dependent coupling of additive interferometric readout noise, distinct from \ac{ttl}, where a real path length variation occurs. As beam tilts are going to be present in \ac{lisa}, this article is focused on understanding how these affect the \ac{he} and, therefore, how they also affect \ac{lisa}'s interferometric readout noise, i.e. the readout noise of the \ac{qpd} signals, both \ac{lps} and \ac{dws}. Even though \ac{lisa} only employs \acp{qpd}, for completeness, this phenomenon is also analyzed on \acp{sepd}, as these are an extremely valid example to show additional features that arise by segmenting the detector's active area.

\subsection{Two Gaussian Beams on an Infinite SEPD} \label{ssec:infinite-sepd}
We now derive the \ac{he} as a function of the beam tilt angle for an infinite \ac{sepd}. This result can be obtained exactly. In order to calculate the \ac{he} defined in \cref{eq:het-eff-definition}, one has to first calculate the two individual beam powers $P_{\text{PD, }r/m}$ as defined in \cref{eq:beam-total-power}. As the area of the \ac{sepd} is infinite, the \ac{sepd} detects the totality of each beam's power, hence $P_{\text{PD, }r/m} = P_{0, \, r/m}$. By integrating the simplified beam intensity in \cref{eq::itf_tilt_simp} over the \ac{sepd}'s active area, one gets the infinite \ac{sepd} overlap power:
\begin{equation}
\begin{split}
    P_{\text{SEPD, }\infty}(\theta) &= \int_{-\infty}^{+\infty}\dd{x}  \int_{-\infty}^{+\infty} \dd{y} I_\text{smp}(x, y, \theta) \\
    &= \frac{w_\text{eff}^2}{1 - i\rho}\exp\left( -\frac{k^2 w_\text{eff}^2 \theta^2}{8(1-i\rho)} \right).
    \end{split} \label{eq:sepd-integral-inf}
\end{equation}

\begin{figure}[!htpb]
    \centering
    \includegraphics{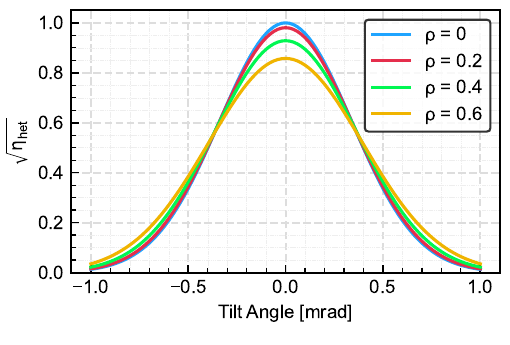}
    \caption{\ac{he} on a \ac{sepd} as a function of the measurement beam's tilt angle for two same spot-radius beams resulting in $w_r = w_m = \weff = 1$\,mm, $\lambda=1064$\,nm, plotted for four different values of the effective-spot-radius-normalized relative wavefront curvature mismatch parameter $\rho$. The plotted curves are derived from \cref{eq:het_eff_SEPD-inf}. The presence of wavefront curvature mismatch on one side strongly reduces the maximum achievable \ac{he}. On the other side, it mildly broadens the angular range of the interferometric signals (\ac{lps} and \ac{dws}). The expected value of $\weff$ in the \ac{lisa} interferometers is $\sim400$\,\unit{\micro \meter}~\cite[Table III]{alvise_2024}, giving an angular range of $~\sim1.5$\,mrad where the \ac{he} stays above the value of 0.5 (not shown in this plot).}
    \label{fig:heff_SEPD_inf}
\end{figure}

Noting that $\frac{\weff^2}{w_m w_r} = \frac{2 w_r w_m}{w_r^2 + w_m^2}$, when taking the ratio in \cref{eq:het-eff-definition}, this leads to the \ac{he} and heterodyne phase for an infinite \ac{sepd} of
\begin{align}
     \she_{\text{, SEPD, }\infty}(\theta) &= \frac{1}{\sqrt{1 + \rho^2}}\frac{2 w_r w_m}{w_r^2 + w_m^2} \exp \left( -\frac{k^2 w_\text{eff}^2 \theta^2}{8(1 +\rho^2)} \right),
    \label{eq:het_eff_SEPD-inf}\\
     \psi_{\text{het, SEPD, }\infty}(\theta) &= \arctan(\rho) - \underbrace{\rho \frac{k^2 w_\text{eff}^2 \theta^2}{8(1 + \rho^2)}}_{\text{NG-TTL}}. \label{eq:het_psi_SEPD-inf}
\end{align}
Note that \cref{eq:het_eff_SEPD-inf,eq:het_psi_SEPD-inf} depend on $\theta$, $w_\text{eff}$ and $k$ only through their product $\Theta = k w_\text{eff} \theta$, meaning that the result for specific values of $k$ and $w_\text{eff}$ can be carried over to different values by an appropriate rescaling of the angle $\theta$. Also note that \cref{eq:het_eff_SEPD-inf} is even in both $\theta$ and $\rho$, respecting the previously mentioned symmetries under beam-swap. \Cref{eq:het_psi_SEPD-inf}, instead, depends on the sign of $\rho$, as to be expected for a differential phase. It is hence maximum at $\theta=0$, where
\begin{align}
    \she_{\text{, SEPD, }\infty}(\theta=0) &= \frac{1}{\sqrt{1 + \rho^2}}\frac{2 w_r w_m}{w_r^2 + w_m^2}, \label{eq:heff-sepd-inf-0}\\
     \psi_{\text{het, SEPD, }\infty}(\theta=0) &= \arctan{\rho}.
\end{align}
Lastly, note that the second term in \cref{eq:het_psi_SEPD-inf} is responsible for the \ac{ng-ttl}, as it generates a tilt varying phase corresponding to no real path length variation; such term vanishes in the case of perfect beam mode-match. The \ac{he} at $\theta=0$ is equal to one only in the case of perfect mode match, defined by $w_r = w_m$ and $\rho=0$, and no tilt. In all other cases, the \ac{he} degrades. \Cref{fig:heff_SEPD_inf} plots the \ac{he} as a function of the beam tilt on an infinite \ac{sepd} in \cref{eq:het_eff_SEPD-inf} for some specific values of $\rho$.

What can be concluded from \Cref{fig:heff_SEPD_inf} is that the phase readout noise from a \ac{sepd} strongly depends on the beam's alignment. The presence of large beam tilts reduces the overlap between the beams and, consequently, the amplitude of the interferometric signal. Such a signal, hence, results in a lower \ac{snr}, as it is compared against noises which are unaffected by the tilt. The additional presence of wavefront curvature mismatch exacerbates the \ac{snr} loss at null tilt angles, while it moderately improves the \ac{he} at larger tilts.

\subsection{Two Gaussian Beams on an Infinite QPD} \label{ssec:infinite-qpd}
We now derive the \ac{he} as a function of the beam tilt angle for the individual segments of an infinite \ac{qpd}, assumed for simplicity to be slit-less. This case is the first step toward a description of the \ac{tmi} in \ac{lisa}; it captures most of the features but misses some implications due to the comparable sizes of \ac{qpd} and the beam.

This result differs from that on a \ac{sepd}, as the integration domain is halved. We further stress that a) adding the individual \ac{qpd} voltages and b) adding the phase readout from the individual \ac{qpd} segments, as \ac{lisa}'s architecture does, leads to structurally different results; The first case yields a \ac{sepd} of equivalent size, hence featuring the same \ac{he}. The second case cannot be --- at least directly --- related to an equivalent size \ac{sepd}. 

This result can also be obtained exactly. Due to the system's symmetry, we can simplify the \ac{qpd} calculation by dividing it into just two sections: a top part and a bottom part of the \ac{qpd}. This is because the two top segments are unchanged by the symmetry $x \rightarrow -x$, and similarly for the two bottom segments. Therefore, we only need to derive one equation for the top segments and one for the bottom segments, streamlining the overall calculation by avoiding the need to analyze all four segments individually. 
As a first step, we calculate the individual beam powers impinging on each \ac{qpd} half. This is simply half of the total power, or $P_{\text{PD, }r/m} = \frac{1}{2} P_{0, \, r/m}$. As a second step, we integrate the simplified beam intensity in \cref{eq::itf_tilt_simp} over the \ac{qpd}'s top and bottom segments' active area, one gets the infinite \ac{qpd}-half overlap powers:
\begin{align}
    P_{\text{QPD, top, }\infty}(\theta) &= \int_{-\infty}^{+\infty} \dd{x} \int_{0}^{+\infty} \dd{y} I_\text{smp}(x, y, \theta) \\
    &= \frac{w_\text{eff}^2}{2(1 - i\rho)}\left(1-i \erfi \left( \frac{k w_\text{eff} \theta}{2^\frac{3}{2} \sqrt{1-i\rho}} \right) \right) \exp\left( -\frac{k^2 w_\text{eff}^2 \theta^2}{8(1-i\rho)} \right)
    \label{eq:int_QPD_inf_t}\\
    P_{\text{QPD, btm, }\infty}(\theta) &= \int_{-\infty}^{+\infty} \dd{x} \int_{-\infty}^{0} \dd{y} I_\text{smp}(x, y, \theta) \\
    &= \frac{w_\text{eff}^2}{2(1 - i\rho)}\left(1+i \erfi \left( \frac{k w_\text{eff} \theta}{2^\frac{3}{2} \sqrt{1-i\rho}} \right) \right) \exp\left( -\frac{k^2 w_\text{eff}^2 \theta^2}{8(1-i\rho)} \right)\\
    &= P_{\text{QPD, top, }\infty}(-\theta)
    \label{eq:int_QPD_inf_b}
\end{align}

Note that, also here, \cref{eq:int_QPD_inf_t,eq:int_QPD_inf_b} only depend on $\theta$, $\weff$ and $k$ only through their product $\Theta = k w_\text{eff} \theta$. These two \cref{eq:int_QPD_inf_t,eq:int_QPD_inf_b} only differ for the sign of the \textit{imaginary error function} erfi(x) term. Since erfi(x) is an odd function, $P_{\text{QPD, top, }\infty}(\theta) = P_{\text{QPD, btm, }\infty}(-\theta)$. 

From \cref{eq:int_QPD_inf_t,eq:int_QPD_inf_b}, it is possible to recover an exact equation for the \ac{he} and heterodyne phase of an infinite \ac{qpd} by taking, respectively, the absolute value and the argument. Using the property of the imaginary error function $\erfi(x^*)\,=\,\erfi^*(x)$ and calculating the ratio in \cref{eq:het-eff-definition}, one gets the \ac{he} and heterodyne phase for the top and bottom segments of a \ac{qpd}, reported in \cref{eq:heff-QPD-inf-top,eq:peff-QPD-inf-top,eq:heff-QPD-inf-bottom,eq:peff-QPD-inf-bottom} respectively.
\begin{align}
\she_{\text{, QPD, top, }\infty}(\theta) &=  \frac{1}{\sqrt{1+\rho^2}} \frac{2 w_r w_m}{w_r^2 + w_m^2} \exp\left( -\frac{k^2 w_\text{eff}^2 \theta^2}{8(1+\rho^2)} \right) \sqrt{1 +2 \Im \left[ \erfi \left( \frac{k \weff \theta}{2^\frac{3}{2}\sqrt{1 -i\rho}} \right) \right] + \left|\erfi\left( \frac{k \weff \theta}{2^\frac{3}{2}\sqrt{1 -i\rho}} \right)\right|^2}
\label{eq:heff-QPD-inf-top}\\
\psi_{\text{het, QPD, top, }\infty}(\theta) &= \arctan(\rho) - \rho \frac{k^2 w_\text{eff}^2 \theta^2}{8(1 + \rho^2)} + \arg \left( 1 - i \erfi \left( \frac{k \weff \theta}{2^\frac{3}{2}\sqrt{1 -i\rho}} \right) \right) \label{eq:peff-QPD-inf-top}
\end{align}
\begin{align}
\she_{\text{, QPD, btm, }\infty}(\theta) &=  \frac{1}{\sqrt{1+\rho^2}} \frac{2 w_r w_m}{w_r^2 + w_m^2} \exp\left( -\frac{k^2 w_\text{eff}^2 \theta^2}{8(1+\rho^2)} \right) \sqrt{1 -2 \Im \left[ \erfi\left( \frac{k \weff \theta}{2^\frac{3}{2}\sqrt{1 -i\rho}} \right) \right] + \left|\erfi\left( \frac{k \weff \theta}{2^\frac{3}{2}\sqrt{1 -i\rho}} \right)\right|^2}
\label{eq:heff-QPD-inf-bottom}\\
\psi_{\text{het, QPD, btm, }\infty}(\theta) &= \arctan(\rho) - \rho \frac{k^2 w_\text{eff}^2 \theta^2}{8(1 + \rho^2)} + \arg \left( 1 + i \erfi \left( \frac{k \weff \theta}{2^\frac{3}{2}\sqrt{1 -i\rho}} \right) \right) \label{eq:peff-QPD-inf-bottom}
\end{align}

Note that the heterodyne phase in \cref{eq:peff-QPD-inf-top,eq:peff-QPD-inf-bottom} contains the same \ac{ng-ttl} term in \cref{eq:het_psi_SEPD-inf}; they also contain a further term, which leads to the \ac{dws} signal described in \cite[eqs. (38, 39)]{alvise_2024}. Also note that, when calculating the \ac{ap}-\ac{lps} defined in \cref{eq::lps-ap-definition}, measuring the interference with a \ac{qpd} instead of a \ac{sepd} gives an extra \ac{qpd}-\ac{ap} term, leading to additional \ac{ng-ttl} with respect to a \ac{sepd}. This can be seen by taking the average of \cref{eq:peff-QPD-inf-top,eq:peff-QPD-inf-bottom}, shown in \cref{eq:lps-qpd-infinite}. This term also vanishes in the case of perfectly matched \acp{gb}, as found also numerically in \cite[Figure 12]{Hartig_2023} in the case of square \acp{qpd}. Also, using the \ac{lpf}-\ac{lps} definition would cancel out this extra term.
\begin{equation}
    \psi_{\text{het, QPD, AP, }\infty}(\theta) = \underbrace{\arctan(\rho) - \rho \frac{k^2 w_\text{eff}^2 \theta^2}{8(1 + \rho^2)}}_\text{SEPD} + \underbrace{\frac{1}{2}\arg \left( 1 + \erfi^2 \left( \frac{k \weff \theta}{2^\frac{3}{2}\sqrt{1 -i\rho}} \right)\right)}_\text{QPD-AP term} \label{eq:lps-qpd-infinite}
\end{equation}

Analogous expressions can be derived in the case of a horizontal tilt, with the correspondences 
\begin{align}
    \she_{\text{, QPD, top, }\infty}(\theta) &\Leftrightarrow \she_{\text{, QPD, left, }\infty}(\theta)\\
    \she_{\text{, QPD, btm, }\infty}(\theta) &\Leftrightarrow \she_{\text{, QPD, right, }\infty}(\theta)
\end{align}

These two equations, as well as \cref{eq:heff-QPD-inf-top,eq:heff-QPD-inf-bottom}, only differ for the sign of the $\Im[\erfi(\cdot)]$ term, due to the fact that $\erfi(x)$ is an odd function. Note that, also including the \ac{dws} signal in \cite{alvise_2024}, this is the first \ac{qpd} function to depend on the sign of $\rho$. As previously argued, the substitution $\rho \rightarrow -\rho$ is in fact equivalent to $\theta \rightarrow -\theta$, as this corresponds to swapping the two beams in the linear framework. This respects the fact that both $\theta$ and $\rho$ change signs under beam-swap. In case of perfect beam mode-match $\rho=0$, \cref{eq:heff-QPD-inf-top,eq:peff-QPD-inf-top,eq:heff-QPD-inf-bottom,eq:peff-QPD-inf-bottom} take the simpler, and same in the case of the \ac{he}, form
\begin{align}
    \she_{\text{, QPD, }\infty}(\theta) &= \frac{2 w_r w_m}{w_r^2 + w_m^2} \sqrt{1 + \erfi^2 \left( \frac{k w_\text{eff} \theta}{2^\frac{3}{2}} \right)} \, \exp\left( -\frac{k^2 w_\text{eff}^2 \theta^2}{8} \right) ,    \label{eq:hef-QPD-inf-rho0}\\
    \psi_{\text{het, QPD, }\infty}(\theta) &= \arg \left( 1 \pm i \erfi \left( \frac{k \weff \theta}{2^\frac{3}{2}} \right) \right)  \label{eq:psi-QPD-inf-rho0},
\end{align}
where in \cref{eq:psi-QPD-inf-rho0} the + is to be taken for the top segments, while the - for the bottom ones, and we remind that $\erfi(x) \in \mathbb{R} \, \forall x \in \mathbb{R}$. 

\begin{figure}[!htpb]
    \centering
    \includegraphics{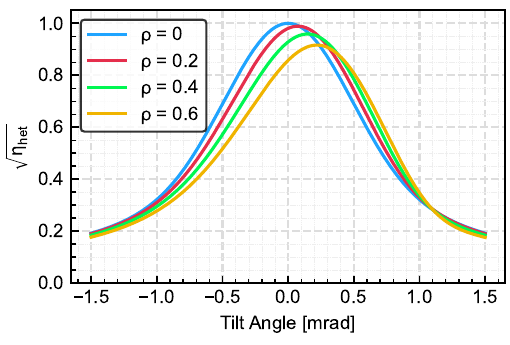}
    \caption{Plot of the \ac{he} of the top segments of an infinite \ac{qpd} as a function of the measurement beam's tilt angle for two same spot-radius beams resulting in $w_r = w_m = \weff = 1$\,mm, $\lambda=1064$\,nm, plotted for four different values of the effective-spot-radius-normalized relative wavefront curvature mismatch parameter $\rho$. The plotted curves are derived analytically from \cref{eq:heff-QPD-inf-top}. If \cref{eq:heff-QPD-inf-bottom} was used for the bottom angle, the same effect would happen, but in the opposite angle. If negative values of $\rho$ were used, the \ac{he}'s peak would shift into the opposite direction. The presence of wavefront curvature mismatch not only strongly reduces the maximum achievable \ac{he} but also shifts the angle at which the \ac{he} peaks. Furthermore, asymmetry of the plot increases with $\rho$. An intuitive explanation of this phenomenon is given in \Cref{fig:wavefront-curvature-interpret}.}
    \label{fig:heff_QPD_inf}
\end{figure}

The resulting \ac{he} for the top segments of an infinite \ac{qpd} in \cref{eq:heff-QPD-inf-top} is plotted in \Cref{fig:heff_QPD_inf} for a few values of $\rho$. What can be concluded from this Figure is, similarly to the \ac{sepd} case in \Cref{fig:heff_SEPD_inf}, that the phase readout noise from a \ac{qpd}-segment strongly depends on the beam's alignment. The consequent decrease in \ac{snr} affects all \ac{qpd} signals, meaning both \ac{lps} and \ac{dws}. The reduction in \ac{dws} sensitivity is not to be confused with what is presented in \cite[Figure 7]{alvise_2024}, where the only cause is the reduction of \ac{dws} response; the result derived in this subsection concerns only the segment's signal amplitude, which is the main driver of the readout phase noise. These two noise contributions, in fact, add up, as later derived in \Cref{section:tm-tilt}.

A key difference in the \ac{he} of an infinite \ac{sepd} segment and that of an infinite \ac{qpd} is that the former always peaks at $\theta=0$, while the latter peaks at $\theta=0$ only if $\rho=0$. In other words, in the case of perfect beam mode-match, the \ac{he} is maximum at zero tilt; otherwise, the peak angle is not zero, and its exact value depends on $\rho$; in particular, it must be an odd function of $\rho$. This is intuitive, but not straightforward, when looking at \cref{eq:heff-QPD-inf-top,eq:heff-QPD-inf-bottom}. It can, however, be made evident, for instance, by taking a Maclaurin expansion. For instance, expanding \cref{eq:heff-QPD-inf-top} to second-order in both $\rho$ and $\theta$ leads to \cref{eq:heff-QPD-infinite-top-2nd}. 
\begin{equation}
    \she_{\text{, QPD, top, }\infty}(\theta) = \frac{2 w_r w_m}{w_r^2 + w_m^2} \left( \sqrt{1-\rho^2} + k w_\text{eff} \frac{\rho}{\sqrt{2 \pi}}\theta - (k w_\text{eff})^2\left( \frac{2\pi -4 - (3\pi-5)\rho^2}{16\pi}\right)\theta^2 \right) + \mathcal{O}(\rho^3, \theta^3).
    \label{eq:heff-QPD-infinite-top-2nd}
\end{equation}
In this series, the odd terms in $\theta$ are not zero. With this second-order series, one can easily find the value of $\theta$ at which the maximum of $\she_{\text{, QPD, top, }\infty}$ is reached:
\begin{equation}
\theta_\text{max, QPD, top} \approx \frac{\sqrt{2 \pi} \rho}{(k w_\text{eff})(2\pi -4 -(3\pi-5)\rho^2))}. \label{eq:heff-QPD-finite-top-max}
\end{equation}
Dually, the tilt angle at which $\she_{\text{, QPD, btm, }\infty}(\theta)$ is maximum is $\theta_\text{max, QPD, btm} = - \theta_\text{max, QPD, top}$. Last, for symmetry reasons, and according to \cref{eq:heff-QPD-inf-top,eq:heff-QPD-inf-bottom}, it holds
\begin{equation}
    \she_{\text{QPD, top, }\infty}(\theta=0) = \she_{\text{QPD, btm, }\infty}(\theta=0).
\end{equation}

As evident from \cref{eq:heff-QPD-finite-top-max}, this phenomenon is caused by the presence of wavefront curvature mismatch, combined with the fact that the \ac{qpd}'s active area is not symmetric around the measurement beam's tilt pivot, as it was in the case of an infinite \ac{sepd}. A geometrical interpretation of it is depicted in \Cref{fig:wavefront-curvature-interpret}. The presence of a wavefront curvature mismatch breaks the \ac{qpd}-segment symmetry of tilts, hence leading to the top and bottom segments of a \ac{qpd} to be illuminated by different intensities. 

\begin{figure}[!htpb]
    \centering
    \includegraphics{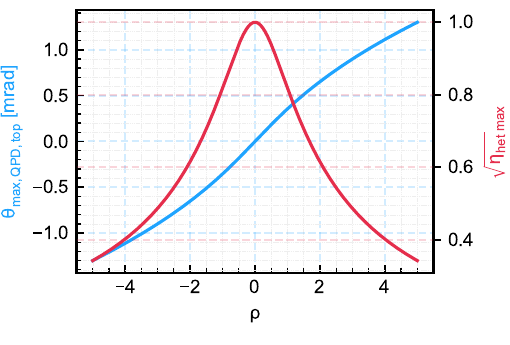}
    \caption{Plot of the peak angle $\theta_\text{max, QPD, top}(\rho)$, defined as the angle at which the \ac{he} for a given value of $\rho$ $\she_{\text{, QPD, top, }\infty}(\theta)$ is maximum, and of the \ac{he} evaluated at $\theta_\text{max, QPD, top}$, for two interfering beams with $w_r = w_m = \weff=1$\,mm. These are two non-linear functions of the effective-spot-radius-normalized relative wavefront curvature mismatch parameter $\rho$, which must be, respectively, even and odd. One can see that the presence of a wavefront curvature mismatch always degrades the maximum reachable \ac{he}.}
    \label{fig:heff-QPD-inf-max}
\end{figure}

\Cref{eq:heff-QPD-finite-top-max} is a rough approximation of a function that can not be calculated analytically, and is a short-lived approximation of the wanted result. A close determination of the peak angle of \cref{eq:heff-QPD-inf-top,eq:heff-QPD-inf-bottom} as a function of $\rho$ is possible only numerically, as \cref{eq:heff-QPD-finite-top-max} is unfortunately a very bad approximation, which, however, suffices in showing that the peak angle is non-zero. \Cref{fig:heff-QPD-inf-max} plots the values of $\theta_\text{max, QPD, top}$ as a function of $\rho$, together with the peak value of the \ac{he} $\she_{\text{, QPD, top, }\infty}(\theta=\theta_\text{max, QPD, top})$. 

\begin{figure}[!htpb]
    \centering
    \includegraphics{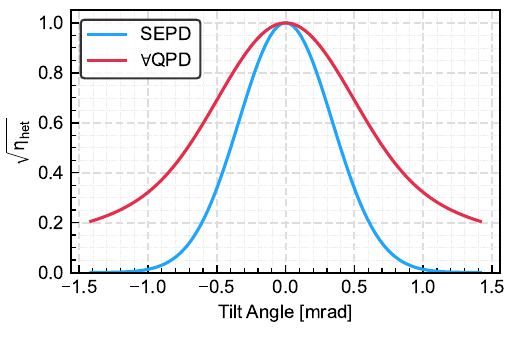}
    \caption{Plot of the \ac{he} for an infinite \ac{sepd} in \cref{eq:het_eff_SEPD-inf} and the whole-\ac{qpd} \ac{he} for an infinite \ac{qpd} (calculated numerically, approximated in \cref{eq:heff-QPD-infinite-tot-2nd} ). The used beam parameters are $\weff= w_r = w_m = 1$\,mm and $\rho = 0$. The \ac{he} for a \ac{sepd} degrades noticeably quicker as a function of the beam tilt than the whole-\ac{qpd} \ac{he}.}
    \label{fig:heff-SEPD-QPD-inf}
\end{figure}

\begin{figure*}[!htpb]
    \centering
    \begin{subfigure}{.49\textwidth}
        \centering
        \includegraphics{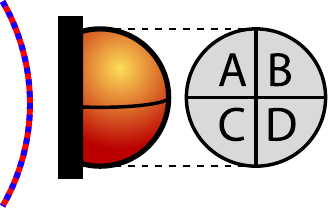}  
        \caption{$\rho=0$, $\theta=0$.}
        \label{sfig:same-null}
    \end{subfigure}
    \begin{subfigure}{.49\textwidth}
        \centering
        \includegraphics{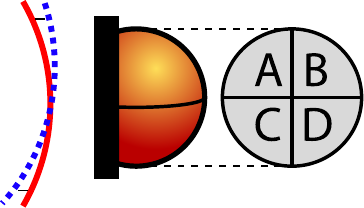}  
        \caption{$\rho=0$, $\theta\neq0$.}
        \label{sfig:same-tilt}
    \end{subfigure}
    \begin{subfigure}{.49\textwidth}
        \centering
        \includegraphics{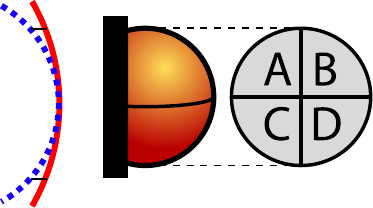}  
        \caption{$\rho\neq0$, $\theta=0$.}
        \label{sfig:diff-null}
    \end{subfigure}
    \begin{subfigure}{.49\textwidth}
        \centering
        \includegraphics{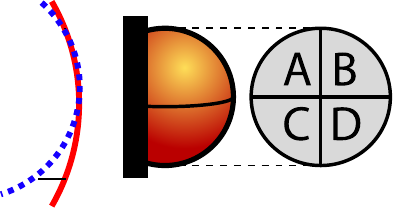}  
        \caption{$\rho\neq0$, $\theta\neq0$.}
        \label{sfig:diff-tilt}
    \end{subfigure}
    \caption{Geometric interpretation of the shift in the \ac{he}'s maximum angle on a \ac{qpd}. The principle followed in this explanation can be found also in \cite[Figure 2]{Hartig_2023}. The four figures represent four situations where the wavefronts of a reference beam (red) and of a measurement beam (blue) impinge on a \ac{qpd}. The black lines indicatively represent the wavefront curvature mismatch. The measurement beam is free to rotate around the center of the \ac{qpd}'s surface, hence so does its wavefront. In subfigure~\ref{sfig:same-null}, there is no wavefront curvature mismatch, and the beams are parallel. In this situation, the two beams have the same local phase difference $\phi_\text{off}(y)$ at every point of the \ac{qpd}'s surface, which we parametrize only with $y$ as there is no dependence on $x$ due to transaltional symmetry; as this argument is independent of the absolute phase difference, we can say $\phi_\text{off}(y) = 0,~\, \forall y$. This results in coherent interference of the two beams in the totality of the \ac{qpd} surface, and hence maximizes the \ac{he}. In subfigure~\ref{sfig:same-tilt}, there is no wavefront curvature mismatch, but a beam tilt is present. Due to the tilt, the wavefronts of the two beams manifest a $y$-coordinate dependent phase offset, given by the tilt term in \cref{eq::GB-tilt}. Setting for simplicity $\phi_\text{off}(y=0)=0$, this is $\phi_\text{off}(y) = ky \propto |y|$. The presence of this phase offset inhibits constructive interference over the whole \ac{qpd}, hence decreasing the \ac{he}. This effect is independent of the sign of $\phi_\text{off}(y)$, hence the measured \ac{he} is the same on both halves of the \ac{qpd}. In subfigure~\ref{sfig:diff-null}, a wavefront curvature mismatch is present, and the beams are parallel. Due to the mismatch, the wavefronts of the two beams are in phase only at $y=0$, while in all other points of the \ac{qpd} a wavefront mismatch proportional to the absolute value of the distance to the \ac{qpd}'s center arises, or $\phi_\text{off}(y)$. Such a phase offset is an even function of $y$; in fact, in this case the resulting \ac{dws} signal is zero. This leads to a similar situation as in subfigure~\ref{sfig:same-tilt}: the \ac{he} degrades symmetrically in the two \ac{qpd} halves. In subfigure~\ref{sfig:diff-tilt}, both a wavefront curvature mismatch and a tilt are present. In comparison to subfigure~\ref{sfig:diff-null}, the tilt has reduced the phase offset on the top segments, while on the other hand, the phase offset is increased in the bottom segments. The local intensity and the measured power are hence increased in the top segments, and decreased in the bottom ones. The presence of a wavefront curvature mismatch, hence, breaks the tilting symmetry, and the top segments manifest a higher \ac{he} than the bottom segments: this is exactly what \Cref{fig:heff_QPD_inf} shows. Looking only at subfigures~\ref{sfig:same-tilt} and~\ref{sfig:diff-tilt}, one can also see why a beam curvature mismatch results in a varying heterodyne phase on a \ac{sepd}. }
    \label{fig:wavefront-curvature-interpret}
\end{figure*}

By adding the powers measured by the top and bottom halves of an infinite \ac{qpd}, one gets the total beam power impinging on the \ac{qpd}. This principle can be used to define the \ac{he} of the \ac{qpd} as a whole, resembling \ac{lisa}'s architecture, where the phase measurements from the individual \ac{qpd} segments are combined in post-processing. We name this quantity whole-\ac{qpd} \ac{he} $\she_{\text{, }\forall\text{QPD, }\infty}(\theta)$. This can be calculated by adding up the two \cref{eq:heff-QPD-inf-top,eq:heff-QPD-inf-bottom} and dividing the result by two, as done in \cref{eq:heff-QPD-inf-tot}. This result, once more, would be of difficult analytic interpretation, while it's more intuitive to take a look at its second-order Maclaurin expansion in both $\theta$ and $\rho$ as reported in \cref{eq:heff-QPD-infinite-tot-2nd}. This can be compared with the analogous second-order Maclaurin expansion of the \ac{he} on an infinite \ac{sepd} in \cref{eq:het_eff_SEPD-inf}, reported in \cref{eq:heff-SEPD-infinite-series}. It is important to stress that this quantity differs from the \ac{he} of a \ac{sepd}. While it must, for symmetry reasons, be an even function and hence peak at $\theta=0$, it degrades significantly slower as a function of the tilt angle. This can mathematically be seen by comparing the quadratic terms in \cref{eq:heff-QPD-infinite-tot-2nd} and \cref{eq:heff-SEPD-infinite-series}. The quadratic term of \cref{eq:heff-QPD-infinite-tot-2nd} is always smaller than that of \cref{eq:heff-SEPD-infinite-series}, indicating that the whole-\ac{qpd} \ac{he} is less affected by beam tilts than the \ac{he} of \ac{sepd}. The reason why the whole-\ac{qpd} \ac{he} decreases slower as a function of the beam tilt angle can be intuitively understood from \Cref{fig:wavefront-curvature-interpret}: in the presence of a beam tilt, a local phase offset builds up between the wavefronts of the interfering beams. Such local phase offset is monotonically related to the distance from the \ac{pd}'s center. Its presence reduces the constructiveness of the interference pattern of the two beams over the detector; such a pattern is integrated by the \ac{pd}, which hence outputs a beat note with smaller amplitude. The full plot of the two heterodyne efficiencies is shown in \Cref{fig:heff-SEPD-QPD-inf}. 

This article is, to our knowledge, the first to acknowledge this wavefront-curvature mismatch-driven feature of the beam overlap in the presence of tilts around the centre of a \ac{qpd}. This has further implications in the \ac{qpd} signals noise which are discussed in \Cref{section:qpd-noise}.

\begin{align}
    \she_{\text{, }\forall\text{QPD, }\infty}(\theta) &= \frac{1}{2}\left( \she_{\text{, QPD, top, }\infty}(\theta) + \she_{\text{, QPD, btm, }\infty}(\theta)\right) \label{eq:heff-QPD-inf-tot}\\
    &\approx \frac{2 w_r w_m}{w_r^2 + w_m^2} \left( \sqrt{1-\rho^2} + \frac{(k w_\text{eff})^2(4-5\rho^2 + \pi(-2+3\rho^2))}{16 \pi}\theta^2 \right) + \mathcal{O}(\rho^3, \theta^3)    \label{eq:heff-QPD-infinite-tot-2nd}\\
    \she_{\text{, SEPD, }\infty}(\theta) &\approx  \frac{2 w_r w_m}{w_r^2 + w_m^2} \left( \frac{1}{\sqrt{1 + \rho^2}} - \frac{(k w_\text{eff})^2}{4(1+\rho^2)^\frac{3}{2}} \theta^2 \right) + \mathcal{O}(\theta^3) \label{eq:heff-SEPD-infinite-series}
\end{align}

We furthermore stress that in \ac{lisa}'s architecture, where the phase of each \ac{qpd}-segment is extracted separately, and the phase combinations are extracted afterwards, that the \ac{sepd} \ac{he} and the whole-\ac{qpd} \ac{he} can be recovered only in post-processing. \revw{Among these two, at first glance, the whole-\ac{qpd} \ac{he} gives the closest description of the amplitude decrease on the \ac{qpd} as a function of tilts. In \Cref{section:qpd-noise} we derive a more appropriate coupling factor for \ac{qpd}-signals noise.}

\subsection{Top-Hat - Gaussian Beam on an Infinite PD}\label{ssec:th-gb-pd-inf}
This subsection addresses the calculation of the heterodyne interference of a \ac{gb} with a tilting \ac{th} beam; this case is a first step towards the description of the \ac{isi} in \ac{lisa}. This result cannot be obtained by approximating the measurement beam's waist value to infinity, as it was performed in \cite{alvise_2024} for the calculation of the \ac{dws} signal of a \ac{th} beam and a \ac{gb}, as this would make the \ac{th} beam's power diverge. Instead, this requires a proper modelling of the \ac{th} beam to obtain its total power. 

In this subsection, the tilting \ac{gb}, i.e. the measurement beam defined in \cref{eq:GB-Vfield,eq::GB-amplitude}, is replaced by a \ac{th} beam, defined as
\begin{equation}
    \vec E (\vec x)_\text{TH} = \begin{cases}
    \Re \left[ \hat \varepsilon_m E_\text{TH} e^{i k z}  \right] & x^2 + y^2 \leq r_\text{TH}^2\\
    0                                         & \text{elsewhere}
    \end{cases} \label{eq:th-ef-boundary}
\end{equation}
in its coordinate frame. The tilts of the \ac{th} beam are obtained by rotating the vector field using \cref{eq:vector-field-rotation}. This gives the same result as of \cref{eq::versor-rotation} for the polarization unitary vector, while the \ac{th} beam gains the tilt term $e^{ik\theta y}$. The electric field in \cref{eq:th-ef-boundary}, hence, becomes
\begin{equation}
    \vec E (\vec x)_\text{TH} = \Re \left[ \hat \varepsilon_m E_\text{TH} e^{i k \theta y + i k z}  \right]
\end{equation}
when different from zero. We neglect variations of the boundary conditions in \cref{eq:th-ef-boundary} due to the tilt, as these are second-order in $\theta$. The intensity of the plane wave is
\begin{equation}
    I (\vec x)_\text{TH} = \begin{cases}
    \frac{1}{2Z}  |E_\text{TH}|^2 = I_\text{TH} & x^2 + y^2 \leq r_\text{TH}^2\\
    0                                       & \text{elsewhere}
    \end{cases} \label{eq:th-int-boundary}
\end{equation}
Note that \cref{eq:th-int-boundary} is independent of the $z$ coordinate. The integration of this intensity on an infinite \ac{sepd} results in the total \ac{th} beam's power
\begin{equation}
    P_\text{TH} = I_\text{TH} \pi r_\text{TH}^2. \label{eq:pw-power}
\end{equation}

The overlap of the \ac{th} beam and of the \ac{gb} behaves as the overlap of two \acp{gb}, of which one has an asymptotically infinite spot size. Assuming $w_{m, 0} = \infty$, this results in $w_\text{eff}=\sqrt{2}w_r(z)$, $R_\text{rel}=R_r(z)$ and $\rho = \frac{k w_r(z)^2 }{2 R_r(z)} = \frac{z}{z_{\text{R, }r}}$. Therefore, the same simplified expression for the overlap intensity term in \cref{eq::itf_tilt_simp} can be used to describe the main geometrical dependencies of this configuration. To recover the full beam intensity, however, \cref{eq::itf_tilt_simp-to-full-TH} has to be used
\begin{equation}
\text{GB-TH: }I_{\text{ifm}}(x, y, \theta) =  I_r + I_\text{TH} + C e^{i \Psi} \left(I_{\text{smp}}(x, y, \theta) + c.c. \right), \label{eq::itf_tilt_simp-to-full-TH}
\end{equation}
which relies on \cref{eq:ref-beam-power,eq:th-int-boundary} as well as on the quantities $C$ and $\Psi$:
\begin{align}
    C     &= \hat \varepsilon_{r} \cdot \hat \varepsilon_{m} \sqrt{I_{r, \, 0} I_\text{TH} } \left( \frac{w_{r, \, 0}}{w_r(z_r)} \right) \sqrt{\frac{\pi}{2}}\\
    \Psi &= kz_r + \eta(z_r) - kz_m.
\end{align}
An important remark is that, within this framework, when evaluating the integral of \cref{eq::itf_tilt_simp-to-full-TH}, the constraint set by the finite radius of the \ac{th} beam is mathematically equivalent to that set by the size of the \ac{pd} --- which is assumed to be infinite in this subsection. To reasonably assume that the \ac{pd}'s size is infinite, and hence to integrate \cref{eq::itf_tilt_simp} on an infinite \ac{pd} surface, it must hold $r_\text{TH} \sim r_\text{PD} \gg w_r(z_r)$. With these assumptions, the resulting overlap powers on a \ac{sepd} and the top segments of a \ac{qpd} are again described by \cref{eq:sepd-integral-inf,eq:int_QPD_inf_t}, respectively. However, the resulting \ac{he} differs in the normalization, as the expression for the \ac{th} beam's power, which appears in the denominator of \cref{eq:het-eff-definition}, has changed. The resulting expression includes the quantity $ w_r(1 + \rho^2)^{-\frac{1}{2}}$. However, due to the simplified expression of the parameter $\rho$ in this specific case, one can check that this results simply in $w_{0, \, r}$. The resulting \ac{he} on a \ac{sepd} and the top segments of a \ac{qpd} are reported as a function of the \ac{gb}'s spot size in \cref{eq:het_eff_SEPD-inf-THGB,eq:het_eff_QPD-top-inf-THGB}. The result for the bottom segments of a \ac{qpd} can be obtained by switching the sign of $\theta$ in \cref{eq:het_eff_QPD-top-inf-THGB}. Note that equation \cref{eq:het_eff_SEPD-inf-THGB} depends only on the waist size of the \ac{gb} $w_{0, \, r}$, and not on its spot size. It also does not depend on the wavefront curvature mismatch, as it is independent of $\rho$. On the other hand, \cref{eq:het_eff_QPD-top-inf-THGB} depends on $w_r$ and $\rho$ only in the square root. Regarding the heterodyne phase, there is no variation for the heterodyne phase with respect to that of a \ac{gb}-\ac{gb} interference, and this case is still described with \cref{eq:het_psi_SEPD-inf,eq:peff-QPD-inf-top}.

\begin{align}
\she_{\text{, SEPD, }\infty}(\theta) &= \frac{ \sqrt{2} w_{0, \, r}}{r_\text{TH}} \exp \left( -\frac{k^2 w_{0, \, r}^2 \theta^2}{4} \right) \label{eq:het_eff_SEPD-inf-THGB}\\
\she_{\text{, QPD, top, }\infty}(\theta) &=  \frac{ \sqrt{2} w_{0, \, r}}{r_\text{TH}} \exp\left( -\frac{k^2 w_{0, \, r}^2 \theta^2}{4} \right) \sqrt{1 +2 \Im \left[ \erfi \left( \frac{k w_r \theta}{2\sqrt{1 -i\rho}} \right) \right] + \left|\erfi\left( \frac{k w_r \theta}{2\sqrt{1 -i\rho}} \right)\right|^2}
\label{eq:het_eff_QPD-top-inf-THGB}
\end{align}

This result is, however, of little use, as it is valid only for $r_\text{TH} \gg w_r(z_r)$, and hence $\she_{\text{, PD, }\infty}(\theta)$ tends to zero. Note also that, for $w_{0, \, r} > r_\text{PD}/\sqrt{2}$, \cref{eq:het_psi_SEPD-inf,eq:peff-QPD-inf-top} give an unphysical result. As the interesting cases are those with $r_\text{th} \sim r_\text{PD}$, the \ac{gb}-\ac{th} interference requires, hence, a finite \ac{qpd} modelling, which is performed in \Cref{ssec:th-gb-pd-fin}.

\subsection{Two Gaussian Beams on an Finite SEPD} \label{ssec:finite-sepd}
We now derive the \ac{he} as a function of the beam tilt angle for a finite \ac{sepd}. This result can be obtained exactly only for $\theta=0$; such a condition is useful to calculate as a reference. As a first step for this derivation, we calculate the beam power impinging on the \ac{sepd} to be derived. This is no longer in the total power of the beam, but a fraction of it, given by \cref{eq:SEPD-fin-power}, and depends on the ratio between the \ac{sepd}'s radius, $\rsepd$, and the beam's spot radius.
\begin{equation}
    P_{\text{SEPD, }\circ, \, r/m} = P_{0, \, r/m} \left(1 - \exp\left( -2\frac{\rsepd^2}{w_{r/m}^2(z_{r/m})}\right) \right) \label{eq:SEPD-fin-power}
\end{equation}
Integrating \cref{eq::itf_tilt_simp} on a finite circle or radius $\rsepd$ gives
\begin{equation}
    \begin{split}
    P_{\text{SEPD, }\circ}(\theta=0) &= \int_0^{\rsepd} r \dd{r} \int_{0}^{2\pi} \dd{\phi} I_{\text{smp}}(x, y, \theta=0)\\
    &= \frac{\weff^2}{1-i\rho}\left(1- e^{-\beta^2(1-i\rho)} \right). \label{eq:p_sepd-fin-theta0}
    \end{split}
\end{equation}
where we defined $\beta=\sqrt{2}r_\text{SEPD}/\weff$. Calculating the \ac{he} and heterodyne phase shift gives
\begin{align}
\she_{\text{SEPD, }\circ, \,\theta=0} &= \frac{1}{\sqrt{1+\rho^2}}\frac{2 w_r w_m}{w_r^2 + w_m^2} \frac{1}{\sqrt{1-\exp(- \beta_r^2)}\sqrt{1-\exp(-\beta_m^2)}} \sqrt{1 - 2 e^{-\beta^2}\cos(\beta^2 \rho) + e^{-2\beta^2}}\\
\psi_{\text{het, SEPD, }\circ, \,\theta=0} &= \arctan{\rho} + \arctan \left( \frac{e^{-\beta^2}\sin(\beta^2 \rho)}{1-e^{-\beta^2}\cos(\beta^2 \rho)}\right)
\end{align}
These two derived quantities tend to \cref{eq:het_eff_SEPD-inf,eq:het_psi_SEPD-inf}, respectively, in the limit $\beta\rightarrow\infty$. As one can see, both expressions include the oscillatory terms $\cos/\sin(\beta^2 \rho)$. Their impact vanishes in the case of a perfect beam match $\rho=0$. Otherwise, when not suppressed by the term $e^{-2\beta^2}$; however, so if $\beta\rightarrow1$, largely different outcomes of $\she_{\text{SEPD, }\circ, \,\theta=0}$ and $\psi_{\text{het, SEPD, }\circ, \,\theta=0}$ may occur for different values of $\rho$. Note that, due to the cylindrical symmetry specific to the case $\theta=0$, \cref{eq:het_eff_SEPD-inf,eq:het_psi_SEPD-inf} also represent the \ac{he} and heterodyne phase shift of a \ac{qpd}-segment.

\begin{figure*}
    \centering
    \begin{subfigure}{\columnwidth}
        \centering
        \includegraphics{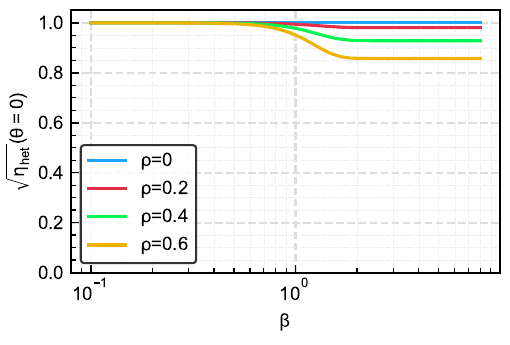}  
        \caption{$\she_{\text{SEPD, }\circ, \,\theta=0}$}
    \end{subfigure}
    \begin{subfigure}{\columnwidth}
        \centering
        \includegraphics{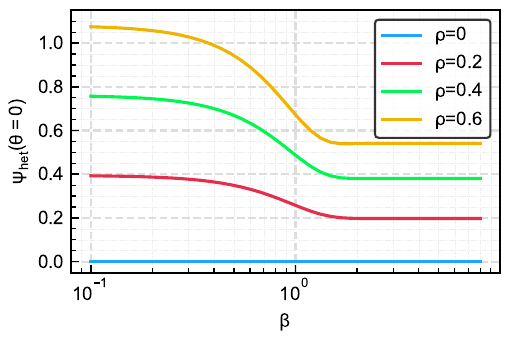}  
        \caption{$\psi_{\text{het, SEPD, }\circ, \,\theta=0}$}
    \end{subfigure}
    \caption{Plot of the \ac{he} and heterodyne phase shift for two \acp{gb} impinging on a finite \ac{sepd} in  \cref{eq:het_eff_SEPD-inf,eq:het_psi_SEPD-inf}. The used beam parameters are $\weff= w_r = w_m = 1$\,mm. Both figures manifest a large variation of the plotted quantity around $\beta \sim 1$.}
    \label{fig:she_pe_SEPD_t0}
\end{figure*}

We proceed now to derive an expression of the \ac{he} as a function of the beam tilt angle. To make the integration possible, we expand in $\theta$ of the simplified beam intensity in \cref{eq::itf_tilt_simp}. We hence integrate the overlap intensity term \cref{eq::itf_tilt_simp} on the finite \ac{sepd}. This calculation cannot be done analytically on a finite circular area centered at $(x, y)=(0, 0)$. A way to circumvent this issue is by Maclaurin expanding in the simplified overlap intensity term $I_\text{smp}(x, y, \theta)$ the term that prevents the integration of the beam intensity, which is the pathlength term $\exp(ik\theta y)$; this comes at the cost of reducing the angular range of validity of the result. The term $\exp(ik\theta y)$ can hence be expanded in $\theta$, giving an expression which is valid only for small tilts. This procedure is justified, as the tilt range that we aim to describe is indeed small, of the order of a few mrad at most. Conversely, as argued in \cite[subsection III B]{alvise_2024}, the finite size of the \ac{pd} can increase the angular range of validity of the resulting expressions if the \ac{sepd}'s size is very small, counteracting the shrinking of such range due to the expansion in $\theta$. Note that, for a \ac{sepd}, this expansion must be at least second-order in $\theta$, as the integration on the \ac{sepd} cancels out the odd powers of $y$, and hence of $\theta$. This also implies that the resulting expression gains one order in precision in the expansion. The expansion of \cref{eq::itf_tilt_simp} to the second-order in $\theta$ gives \cref{eq::itf_tilt_simp-2nd}. The integration can then be performed in polar coordinates for convenience. The finite \ac{sepd} overlap power is reported in \cref{eq:sepd-finite-power}, at second-order.
\begin{align}
I_{\text{smp}}(x, y, \theta) &\approx  I_{\text{smp-}\theta2}(x, y, \theta) = \frac{2}{\pi}\exp \left( -2\frac{x^2 + y^2}{w_\text{eff}^2}(1 - i\rho) \right) \left(1 -iky\theta - \frac{1}{2}(ky\theta)^2 \right)+ \mathcal{O}(ky\theta)^3\,. \label{eq::itf_tilt_simp-2nd}\\
\begin{split}
P_{\text{SEPD, }\circ}(\theta) &= \int_0^{\rsepd} r \dd{r} \int_{0}^{2\pi} \dd{\phi} I_{\text{smp-}\theta2}(r, \phi, \theta) \\
&= \frac{\weff^2}{8(1-i\rho)^2}\left( 8(1-i\rho)- k^2 \weff^2 \theta^2 - e^{-\beta^2(1-i\rho)} \left( 8(1-i\rho) -2k^2\theta^2 \left( r_\text{QPD}^2(1-i\rho) + \weff^2 \right) \right) \right) + \mathcal{O}(k\theta)^4 \label{eq:sepd-finite-power}
\end{split}
\end{align}

Note that this procedure is exact in $\rho$: to calculate the \ac{he} from the power term in \cref{eq:sepd-finite-power}, no expansion in $\rho$ is required, while such expansion was needed for the \ac{dws} signal in \cite{alvise_2024}. This is due to the fact that the integration domain has circular symmetry. From \cref{eq:sepd-finite-power}, the \ac{he} can be derived by taking the absolute value and normalizing by the square root of the beams' powers. The resulting expression is a polynomial, which has the same order as the Maclaurin expansion in \cref{eq::itf_tilt_simp-2nd}. We report the resulting \ac{he} for a finite \ac{sepd} in \cref{eq:heff-fin-sepd}, where we name the resulting Maclaurin coefficients $t_i$ with $i\in[1-7]$, as we went two orders further and derived it to seventh-order in $\theta$. The parameters 
\begin{align}
\beta &= \sqrt{2} \rsepd/w_\text{eff} & \beta_r &= \sqrt{2} \rsepd/w_r & \beta_m &= \sqrt{2} \rsepd/w_m \label{eq:beta-definitions}
\end{align}
have been introduced for simplicity. The non-listed $t_1$, $t_3$, $t_5$ and $t_7$ coefficients are zero.

From \cref{eq:heff-fin-sepd}, new knowledge can be gained regarding the limit of the coefficients $t_0$, $t_2$ and $t_4$ for small \ac{sepd} radii. In this limit, the \ac{he} tends to one independently of the tilt and of the wavefront curvature mismatch $\rho$ and tilt angle $\theta$.
\begin{align}
    \lim_{\beta\rightarrow0} \she_\text{, 0} \, t_0 &= 1 \quad \forall \rho, \, \theta \label{eq:sepd-lim-t0}\\
    \lim_{\beta\rightarrow0} \she_\text{, 0} \, t_2 &= 0 \quad \forall \rho, \, \theta \label{eq:sepd-lim-t2}\\
    \lim_{\beta\rightarrow0} \she_\text{, 0} \, t_4 &= 0 \quad \forall \rho, \, \theta\\
    \lim_{\beta\rightarrow0} \she_\text{, 0} \, t_6 &= 0 \quad \forall \rho, \, \theta
\end{align}

\begin{figure}[!htpb]
    \centering
    \includegraphics{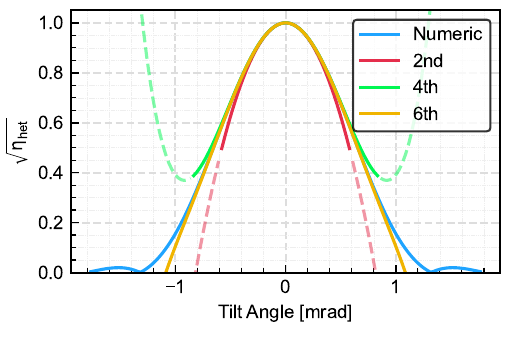}
    \caption{Plot of the \ac{he} as a function of the beam tilt calculated for a finite \ac{sepd} for the parameters $w_r = w_m = \weff=1$\,mm, $r_\text{QPD}=1$\,mm and $\rho=0$. The \ac{he} is calculated both numerically and analytically at various orders. Only one value of $\rho$ is shown, as this method is exact in $\rho$. Dashed lines indicate where the analytic value deviates from the numerical result by more than 0.1. Note that, in the numeric curve, diffraction fringes are visible for $|\theta|>0.75$\,mrad. These are related to the sign reversal in the \ac{dws} signal observed in \cite[Figure 11]{alvise_2024}, which takes place only in finite \acp{qpd}.}
    \label{fig:het-eff-SEPD-fin}
\end{figure}

\begin{figure}
    \centering
    \includegraphics{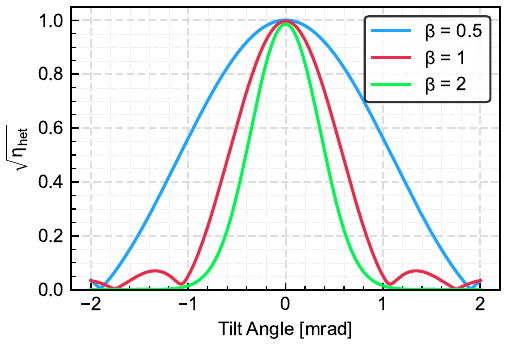}
    \caption{Plot of the \ac{he} as a function of the beam tilt numerically calculated for a \ac{gb}-\ac{gb} beam pair impinging on a finite \ac{sepd}. The beam parameters are $w_r = w_m = \weff=1$\,mm, $ r_\text{SEPD}=[0.5,\,1,\,2]/\sqrt{2}$\,mm and $\rho=0.2$. Note that, in the $\beta=0.5$ and $\beta=1$ plots, diffraction fringes are visible outside the main peak. These are the causes of the \ac{dws} sign reversal discussed in \cite[Figure 11]{alvise_2024}.}
    \label{fig:heff_GB_3betas}
\end{figure}

\begin{figure}
    \centering
    \begin{subfigure}{0.45\textwidth}
        \centering
        \includegraphics[width=\columnwidth]{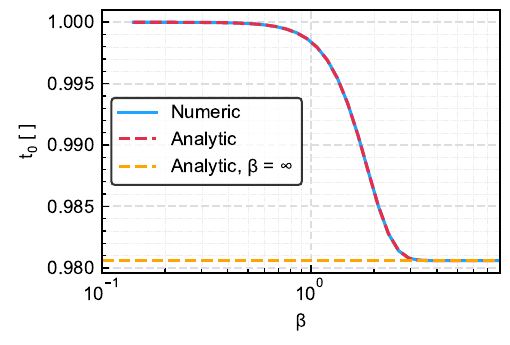}  
        \caption{$t_0$}
    \end{subfigure}
    \begin{subfigure}{0.45\textwidth}
        \centering
        \includegraphics[width=\columnwidth]{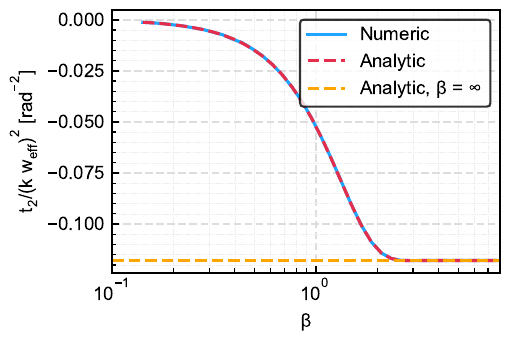}  
        \caption{$t_2$}
    \end{subfigure}
    \begin{subfigure}{0.45\textwidth}
        \centering
        \includegraphics[width=\columnwidth]{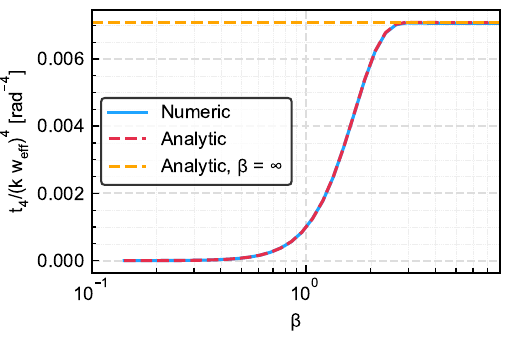}  
        \caption{$t_4$.}
    \end{subfigure}
    \begin{subfigure}{0.45\textwidth}
        \centering
        \includegraphics[width=\columnwidth]{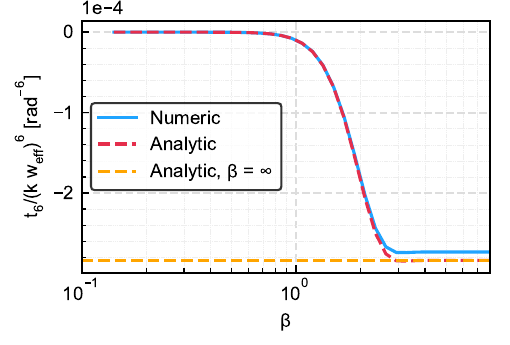}  
        \caption{$t_6$.}
    \end{subfigure}
    \caption{Plots of the $t_0$, $t_2$, $t_4$ and $t_6$ coefficients as a function of the ratio between \ac{sepd} radius and effective beam spot radius $\beta = \sqrt{2} r_\text{SEPD}/\weff$ calculated both analytically using \cref{eq:heff-fin-sepd}, and numerically (see subsection~\ref{subsection:heff-numeric-model}). The used beam parameters are $w_r = w_m = \weff=1$\,mm and $\rho=0.2$. The numeric parameters are obtained by numerically integrating \cref{eq::itf_tilt_simp} and linearly fitting the resulting \ac{he}, as explained in \cref{subsection:heff-numeric-model}. The deviation of the numerically calculated $t_0$ for low $\beta$ values is attributed to fitting error. In the $\beta \rightarrow \infty$ limit, the expansion coefficients of \cref{eq:het_eff_SEPD-inf} are recovered. In the $\beta \rightarrow 0$ limit, $t_0$ tends to 1 while $t_2$, $t_4$ and $t_6$ tend to zero, confirming \cref{eq:sepd-lim-t0,eq:sepd-lim-t2}. The plot shows the same large change in \ac{he} around $\beta=1$ found in \Cref{fig:she_pe_SEPD_t0}.}
    \label{fig:heff-SEPD-fin-t0t2}
\end{figure}

Numerical simulations (see subsection~\ref{subsection:heff-numeric-model} for more details) are used to assess the validity of the approximations for finite \ac{sepd} sizes. \Cref{fig:het-eff-SEPD-fin} compares the \ac{he} on a finite \ac{sepd} of two beams calculated either numerically using the method described in subsection~\ref{subsection:heff-numeric-model}, or analytically to second-, fourth- and sixth-order using \cref{eq:heff-fin-sepd}. As one can see, the angular range of this method is strongly limited by the order of the used expansion. \Cref{fig:heff_GB_3betas} compares the resulting \ac{he} for three values of the $\beta$ parameter around the value of one. One can see that a small value of $\beta$ corresponds to a broader main peak in the interference pattern. \Cref{fig:heff-SEPD-fin-t0t2} compares the analytically calculated $t_0$, $t_2$, $t_4$ and $t_6$ coefficients in \cref{eq:heff-fin-sepd} to the numerically calculated ones for a specific value of $\rho$. The numerical model confirms the analytically estimated coefficients. 

\subsection{Two Gaussian Beams on an Finite QPD} \label{ssec:finite-qpd}
We now derive the \ac{he} as a function of the beam tilt angle for a finite \ac{qpd}. This calculation follows that in Subsection~\ref{ssec:infinite-qpd}, where the \ac{qpd} is assumed slit-less and split into two parts according to the symmetry of the system. This result can, albeit approximated, deliver a full description of the \ac{tmi} in \ac{lisa}.

As in the case of the finite \ac{sepd}, the simplified overlap intensity term $I_\text{smp}(x, y, \theta)$ is expanded in $\theta$. The \ac{qpd} is assumed to have no slits for simplicity.

As a first step we calculate the power detected by a \ac{qpd} half to be half of that in \cref{eq:SEPD-fin-power}. 
\begin{equation}
    P_{\text{QPD, }\circ, \, r/m} = \frac{1}{2} P_{0, \, r/m} \left(1 - \exp\left( -2\frac{r_\text{QPD}^2}{w_{r/m}^2}\right) \right) \label{eq:QPD-fin-power}
\end{equation}
As a second step, we integrate the expanded overlap intensity term onto a semicircle centered in $(x, y)=(0, 0)$. This calculation is performed in polar coordinates for convenience. Note that, unlike in the case for the finite \ac{sepd}, the odd terms in $y$ do not cancel out in the integration process, as the integration domain is not circularly symmetric. The resulting finite \ac{qpd} overlap power, obtained by integrating \cref{eq::itf_tilt_simp-2nd}, for the top segments is reported in \cref{eq:qpd-finite-power} to second-order in $\theta$.
\begin{small}
\begin{equation}
\begin{split}
P_{\text{QPD, top, }\circ}(\theta) &= \int_0^{r_\text{QPD}} r \dd{r} \int_{0}^{\pi} \dd{\phi} I_{\text{smp}-\theta2}(r, \phi, \theta) \\
&= \frac{w_\text{eff}^2}{16(1-i\rho)^2}\left( 8(1-i\rho)-k^2 w_\text{eff}^2 \theta^2 - e^{-2\frac{r_\text{QPD}^2}{w_\text{eff}^2}(1-i\rho)}\left( -i\frac{16k}{\pi} r_\text{QPD}(1-i\rho) + 8(1-i\rho) -2k^2\theta^2(r_\text{QPD}^2(1-i\rho) + w_\text{eff}^2) \right) \right) + \\
&\quad + \frac{k w_\text{eff}^3 \theta}{\left(2(1-i\rho)\right)^\frac{3}{2}\sqrt{\pi}} \erf \left( \frac{r_\text{QPD} \sqrt{2(1-i\rho)}}{w_\text{eff} } \right)+ \mathcal{O}(k\theta)^3\label{eq:qpd-finite-power}
\end{split}
\end{equation}
\end{small}

Also here, the expression for the bottom segments can be calculated by reversing the sign of $\theta$. Unlike in the case of the finite \ac{sepd}, the calculation of the absolute value of \cref{eq:qpd-finite-power} is analytically intractable, as the imaginary error function cannot be separated into real and imaginary parts in closed form if its argument is complex; a similar problem was encountered in \cite[subsection III.B]{alvise_2024}. To get around this issue, we further Maclaurin-expand in the parameter $\rho$. This choice is motivated by the fact that the values of $\rho$ are typically small \cite{Hechenblaikner:10, alvise_2024}. This procedure leads to a much more complicated series, which also contains odd terms in $\theta$. As this is a relevant case for \ac{lisa}, the resulting \ac{he} for a finite \ac{qpd} is reported to fourth-order in $\theta$ and second-order in $\rho$ in \cref{eq:heff-QPD-fin}.

Also here, new knowledge can be gained from \cref{eq:heff-QPD-fin} regarding the limit of the coefficients $t_0$-$t_4$ for small \ac{qpd} radii. This gives a similar result as in the case of the finite \ac{sepd}, with the \ac{he} tending to a constant value of one independently of $\rho$ and $\theta$. Note that the term $t_1$, which is related to the shift of the peak angle of the \ac{he} as argued in subsection~\ref{ssec:infinite-qpd}, also tends to zero. This implies that, by reducing the radius of the \ac{qpd}, the asymmetry created by a wavefront curvature mismatch is also reduced. This is reasonable, as the local phase offset between the beams, which is the ultimate cause of the \ac{he}'s degradation, is proportional to the distance from the \ac{qpd} center; limiting such distance also limits the maximum possible phase offset.
\begin{align}
    \lim_{\beta\rightarrow0} \she_\text{, 0} \, t_0 &= 1 \quad \forall \rho, \, \theta \label{eq:qpd-lim-t0}\\
    \lim_{\beta\rightarrow0} \she_\text{, 0} \, t_1 &= 0 \quad \forall \rho, \, \theta \label{eq:qpd-lim-t1} \\
    \lim_{\beta\rightarrow0} \she_\text{, 0} \, t_2 &= 0 \quad \forall \rho, \, \theta \label{eq:qpd-lim-t2}\\
    \lim_{\beta\rightarrow0} \she_\text{, 0} \, t_3 &= 0 \quad \forall \rho, \, \theta \label{eq:qpd-lim-t3}\\
    \lim_{\beta\rightarrow0} \she_\text{, 0} \, t_4 &= 0 \quad \forall \rho, \, \theta \label{eq:qpd-lim-t4}
\end{align}

\begin{figure}[!htpb]
    \centering
    \includegraphics{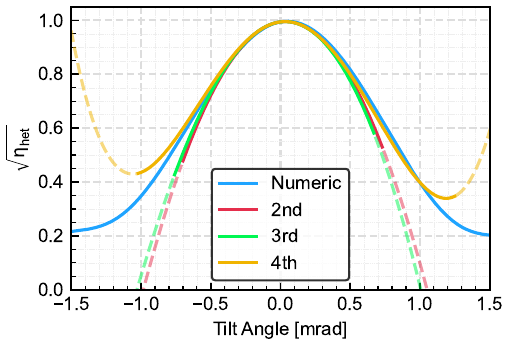}
    \caption{Plot of the \ac{he} as a function of the beam tilt calculated for the top segments of a finite \ac{qpd} for the parameters $\weff=1$\,mm, $r_\text{QPD}=1$\,mm and $\rho=0.2$. The \ac{he} is calculated both numerically and analytically at various orders. The analytic line is dashed when its value departs from the numerical one by more than 0.1. }
    \label{fig:heff-QPD-fin}
\end{figure}

\begin{figure*}[!htpb]
    \centering
    \begin{subfigure}{.32\textwidth}
        \centering
        \includegraphics[width=\columnwidth]{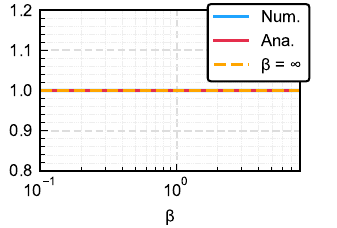}  
        \caption{$t_0$, $\rho=0$}
        \label{fig:15-00}
    \end{subfigure}
    \begin{subfigure}{.32\textwidth}
        \centering
        \includegraphics[width=\columnwidth]{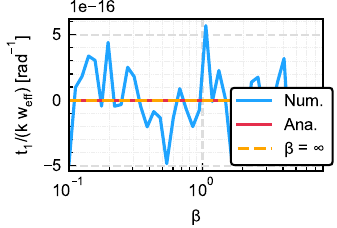}  
        \caption{$t_1$, $\rho=0$}
        \label{fig:15-01}
    \end{subfigure}
    \begin{subfigure}{.32\textwidth}
        \centering
        \includegraphics[width=\columnwidth]{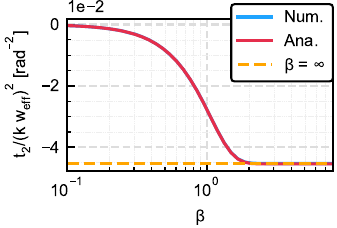}  
        \caption{$t_2$, $\rho=0$}
        \label{fig:15-02}
    \end{subfigure}
    \begin{subfigure}{.32\textwidth}
        \centering
        \includegraphics[width=\columnwidth]{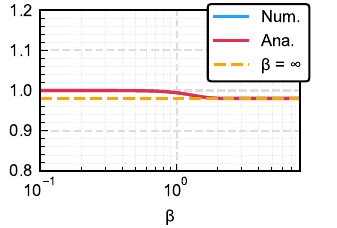}  
        \caption{$t_0$, $\rho=0.2$}
        \label{fig:het-eff-QPD-fin-d}
    \end{subfigure}
    \begin{subfigure}{.32\textwidth}
        \centering
        \includegraphics[width=\columnwidth]{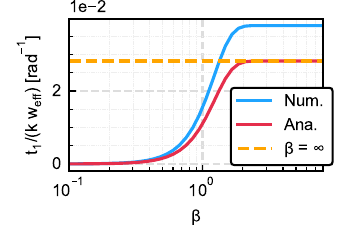}  
        \caption{$t_1$, $\rho=0.2$}
        \label{fig:het-eff-QPD-fin-e}
    \end{subfigure}
    \begin{subfigure}{.32\textwidth}
        \centering
        \includegraphics[width=\columnwidth]{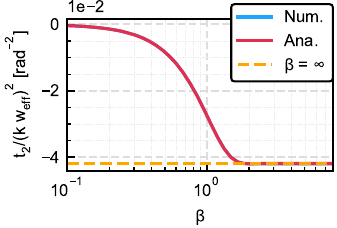}  
        \caption{$t_2$, $\rho=0.2$}
        \label{fig:het-eff-QPD-fin-f}
    \end{subfigure}
    \begin{subfigure}{.32\textwidth}
        \centering
        \includegraphics[width=\columnwidth]{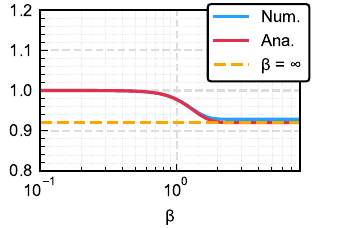}  
        \caption{$t_0$, $\rho=0.4$}
    \end{subfigure}
    \begin{subfigure}{.32\textwidth}
        \centering
        \includegraphics[width=\columnwidth]{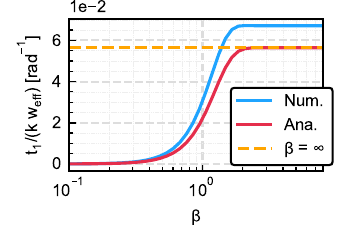}  
        \caption{$t_1$, $\rho=0.4$}
    \end{subfigure}
    \begin{subfigure}{.32\textwidth}
        \centering
        \includegraphics[width=\columnwidth]{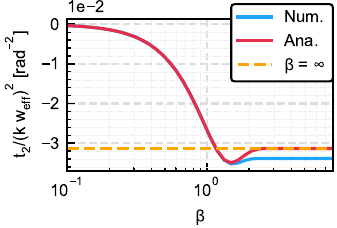}  
        \caption{$t_2$, $\rho=0.4$}
    \end{subfigure}
    \begin{subfigure}{.32\textwidth}
        \centering
        \includegraphics[width=\columnwidth]{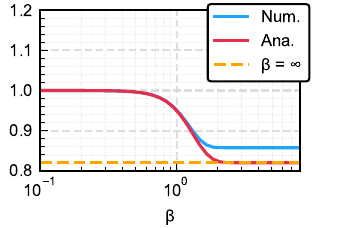}   
        \caption{$t_0$, $\rho=0.6$}
    \end{subfigure}
    \begin{subfigure}{.32\textwidth}
        \centering
        \includegraphics[width=\columnwidth]{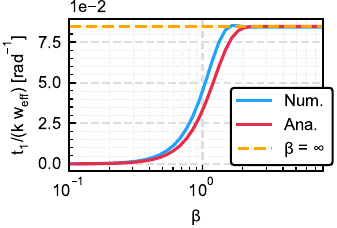}   
        \caption{$t_1$, $\rho=0.6$}
    \end{subfigure}
    \begin{subfigure}{.32\textwidth}
        \centering
        \includegraphics[width=\columnwidth]{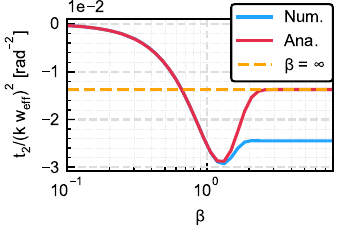}  
        \caption{$t_2$, $\rho=0.6$}
    \end{subfigure}
\caption{Comparison between the analytically obtained parameters $t_0$, $t_1$ and $t_2$ from \cref{eq:heff-QPD-fin} and the numerically calculated ones for four specific values of $\rho$ and as a function of the parameter $\beta = r_\text{QPD}/w_\text{eff}$. $t_1$ and $t_2$ are normalized respectively by the first and second power of $k w_\text{eff}$. The asymptotic analytical value for an infinite \ac{qpd} $\beta \rightarrow \infty$ is added as reference. In \cref{fig:15-00,fig:15-01}, for $\rho=0$, the numerical solution matches the analytically predicted value of zero up to its numerical precision. In all other subfigures, the finite analytical model is in good accordance with the numerical model up to $\beta \sim 1$, where the numerical model systematically underestimates the value of the coefficients. Note that the numerical estimation of $t_1$ improves at higher $\rho$ values (excluding $\rho=0$), while that of $t_2$ behaves oppositely, due to the expansion in $\rho$ itself. These departures of the numerical model are understood as systematic fitting error, as the \ac{he} develops an asymmetric and non-quadratic shape. The plot shows the same large change in \ac{he} around $\beta=1$ found in \Cref{fig:she_pe_SEPD_t0}.}
\label{fig:het-eff-QPD-fin}
\end{figure*}

Also here, we use numerical simulations (see subsection~\ref{subsection:heff-numeric-model} for more details) to assess the validity of the approximations for finite \ac{qpd} sizes. \Cref{fig:heff-QPD-fin} compares the \ac{he} on a finite \ac{qpd} of two beams calculated either numerically using the method described in subsection~\ref{subsection:heff-numeric-model}, or analytically to second-order using \cref{eq:heff-QPD-fin}. \Cref{fig:het-eff-QPD-fin} shows a comparison between the analytically calculated $t_0$, $t_1$ and $t_2$ coefficients using \cref{eq:heff-QPD-fin}, and the numerically calculated ones following the procedure in Section~\ref{subsection:heff-numeric-model}. The plots show that, for small values of $\beta$, both $t_1$ and $t_2$ are suppressed, as indicated by the limits in equations \cref{eq:qpd-lim-t0,eq:qpd-lim-t1,eq:qpd-lim-t2}, while $t_0$ tends to one.

\subsection{Top-Hat - Gaussian Beam on a Finite PD}\label{ssec:th-gb-pd-fin}
This subsection addresses the calculation of the heterodyne interference of a \ac{gb} with a tilting \ac{th} beam on a finite \ac{pd}; this case is what is needed to describe the \ac{isi} in \ac{lisa}, if the \ac{pd} is a \ac{qpd}. This result is built on subsection~\ref{ssec:th-gb-pd-inf}, with minor adaptations. The calculation of the \ac{he} requires, first, the beam powers impinging on the \ac{pd}. This is given by \cref{eq:SEPD-fin-power} for a \ac{gb} - the reference beam - on a finite size \ac{sepd}. The same calculation for the \ac{th} beam needs to be addressed more carefully: the radius of the \ac{th} beam $r_\text{TH}$ in \cref{eq:th-ef-boundary} needs to be compared to the radius of the \ac{pd} $r_\text{PD}$; the smallest of the two determines the detected \ac{th} beam's power. We formalize this as
\begin{equation}
    P_{\text{PD, }\circ \text{, TH}} = I_\text{TH} \pi r_\text{min}, \quad r_\text{min} = \text{min}(r_\text{PD}, \, r_\text{TH}) \label{eq:rmin-def}
\end{equation}

Second, we have to integrate the overlap intensity term on the \ac{pd}. As described in subsection~\ref{ssec:th-gb-pd-inf}, the overlap intensity term of the two beams does not vary with respect to the \ac{gb}-to-\ac{gb} case, and can be described with \cref{eq::itf_tilt_simp}. However, the integration domain is limited to $r_\text{min}$.
\begin{align}
    P_{\text{SEPD, }\circ}(\theta) &= \int_0^{r_\text{min}} r \dd{r} \int_{0}^{2\pi} \dd{\phi} I_\text{smp}(r, \phi, \theta), \\
    P_{\text{QPD, top, }\circ}(\theta) &= \int_0^{r_\text{min}} r \dd{r} \int_{0}^{\pi} \dd{\phi} I_\text{smp}(r, \phi, \theta)
\end{align}

The overlap power is hence described with the same equations \cref{eq:sepd-finite-power,eq:qpd-finite-power} in the case of a \ac{sepd} or a \ac{qpd} respectively, with the \textit{caveat} that the $\beta$ parameter must be adjusted to using the minimum of the two radii as in \cref{eq:rmin-def}:
\begin{equation}
    \beta   \rightarrow \frac{\sqrt{2}r_\text{min}}{\weff} \label{eq:beta-replacement}.
\end{equation}
The parameter $\beta_r$ is unchanged as defined in \cref{eq:beta-definitions}, as the reference \ac{gb} must be integrated on the whole \ac{pd} area, while $\beta_m$ is not needed anymore. The final expressions for the heterodyne efficiency \cref{eq:heff-fin-sepd,eq:heff-QPD-fin} also hold, with the only adjustment
\begin{equation}
    \frac{2 w_r w_m}{w_r^2 + w_m^2} \frac{1}{\sqrt{1-\exp(- \beta_r^2)}\sqrt{1-\exp(-\beta_m^2)}} \rightarrow \frac{\sqrt{2}w_r}{r_\text{min}}\frac{1}{\sqrt{1-\exp(- \beta_r^2)}}. \label{eq:thgb-fin-replacement}
\end{equation}

\begin{figure}
    \centering
    \includegraphics{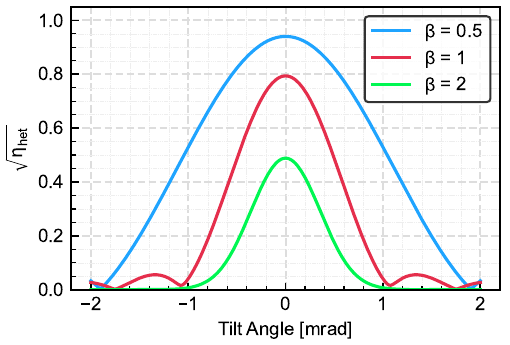}
    \caption{Plot of the \ac{he} as a function of the beam tilt numerically calculated for a \ac{gb}-\ac{th} beam pair impinging on a finite \ac{sepd}. The beam parameters are $\weff=1$\,mm, $r_\text{TH} = r_\text{SEPD}=[0.5,\,1,\,2]/\sqrt{2}$\,mm and $\rho=0.2$. Note that to a larger \ac{pd} size corresponds a lower \ac{he}. Note that in the $\beta=0.5$ and $\beta=1$ plots diffraction fringes are visible outside the main peak. These are the causes of the \ac{dws} sign reversal discussed in \cite[Figure 11]{alvise_2024}.}
    \label{fig:heff_TH_3betas}
\end{figure}

\begin{figure*}
    \centering
    \begin{subfigure}{0.45\textwidth}
        \centering
        \includegraphics[width=\columnwidth]{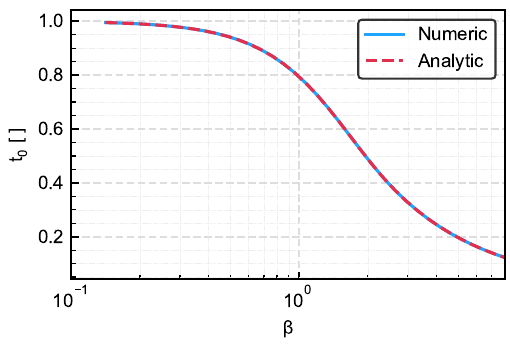}  
        \caption{$t_0$}
    \end{subfigure}
    \begin{subfigure}{0.45\textwidth}
        \centering
        \includegraphics[width=\columnwidth]{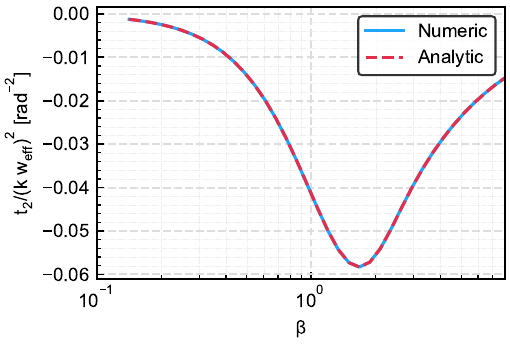}  
        \caption{$t_2$}
    \end{subfigure}
    \begin{subfigure}{0.45\textwidth}
        \centering
        \includegraphics[width=\columnwidth]{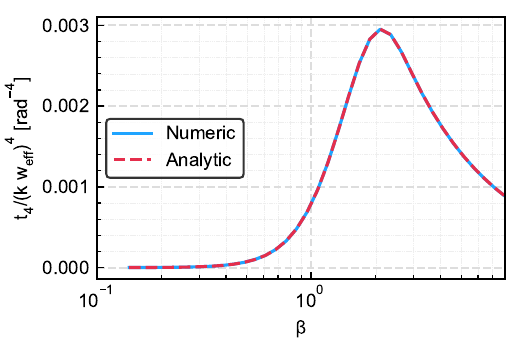}  
        \caption{$t_4$.}
    \end{subfigure}
    \begin{subfigure}{0.45\textwidth}
        \centering
        \includegraphics[width=\columnwidth]{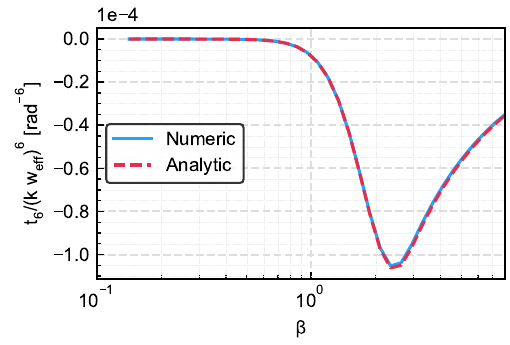}  
        \caption{$t_6$.}
    \end{subfigure}
    \caption{Plots of the $t_0$, $t_2$, $t_4$ and $t_6$ coefficients as a function of the ratio between \ac{sepd} radius and effective beam spot radius $\beta = \sqrt{2} r_\text{SEPD}/\weff$ calculated for a \ac{gb}-\ac{th} beam interference both analytically using \cref{eq:heff-fin-sepd}, and numerically (see subsection~\ref{subsection:heff-numeric-model}). The used beam parameters are $w_r = \weff=1$\,mm, $\rho=0.2$ and $r_\text{TH}=r_\text{SEPD}$. Note that all coefficients tend to zero for $\beta \rightarrow \infty$ due to the normalization to an infinitely growing power.}
    \label{fig:heff-TH-SEPD-fin-t0t2}
\end{figure*}

In the expression for a \ac{sepd}, \cref{eq:thgb-fin-replacement} leads to the same simplification which occurred in \cref{eq:het_eff_SEPD-inf-THGB}, where $w_r(1+\rho^2)^{-\frac{1}{2}} \rightarrow w_{0, \, r}$. As an example, the \ac{he} resulting from the interference of a \ac{gb}-\ac{th} beam pair on a finite \ac{sepd} is plotted in \Cref{fig:heff_TH_3betas} for three values of the parameter $\beta$. For simplicity, we set $r_\text{TH}=r_\text{SEPD}$. \Cref{fig:heff-TH-SEPD-fin-t0t2} compares the analytically calculated $t_0$, $t_2$, $t_4$ and $t_6$ coefficients to the numerically calculated ones for a specific value of $\rho=0.2$ and $r_\text{TH}=r_\text{SEPD}$. The numerical model confirms the analytically estimated coefficients. Note that, with respect to \Cref{fig:heff-SEPD-fin-t0t2}, all coefficients tend to zero for $\beta \rightarrow \infty$.

\subsection{Heterodyne efficiency numeric model for a generic PD} \label{subsection:heff-numeric-model}
\begin{figure*}[!htpb]
    \begin{subfigure}{.45\textwidth}
        \centering
        \includegraphics[width=\columnwidth]{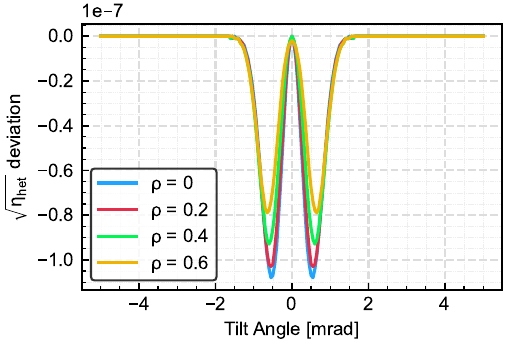}   
        \caption{\ac{sepd}}
        \label{fig:heff-full-sim-a}
    \end{subfigure}
    \begin{subfigure}{.45\textwidth}
        \centering
        \includegraphics[width=\columnwidth]{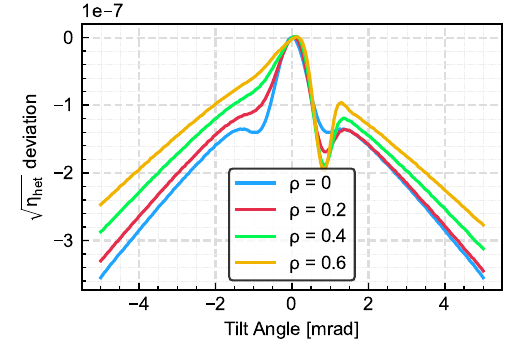}  
        \caption{\ac{qpd}}
        \label{fig:heff-full-sim-b}
    \end{subfigure}
    \begin{subfigure}{.45\textwidth}
        \centering
        \includegraphics[width=\columnwidth]{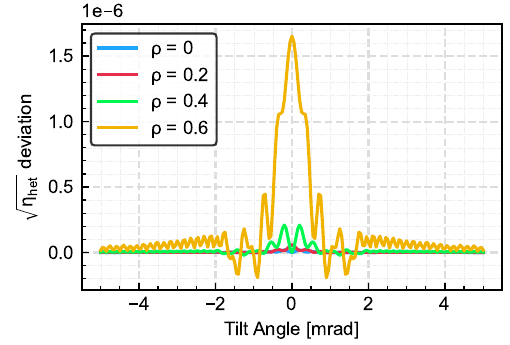}   
        \caption{\ac{sepd}}
        \label{fig:heff-full-sim-c}
    \end{subfigure}
    \begin{subfigure}{0.45\textwidth}
        \centering
        \includegraphics[width=\columnwidth]{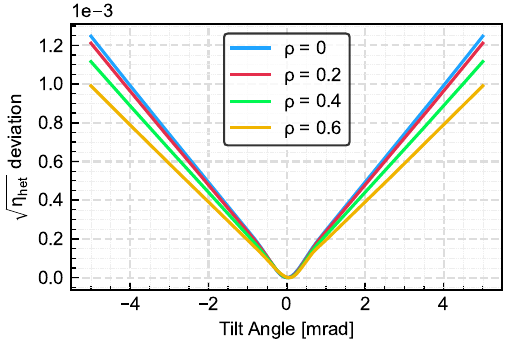}   
        \caption{\ac{qpd}}
        \label{fig:heff-full-sim-d}
    \end{subfigure}
    \caption{\textit{Above:} Plot of the difference between numerically calculated \ac{he} and the analytically calculated \ac{he} for a $w_\text{eff} = 1$\,mm beam impinging on an infinite \ac{sepd} (\Cref{fig:heff-full-sim-a}) and an infinite \ac{qpd} (\Cref{fig:heff-full-sim-b}). In both cases, the approximated expression for the intensity in equation \cref{eq::GB-tilt} is used. This results in the numerical integration's artifacts; note that the magnitude of such artifact depends on the spatial discretization $\Delta s^2$ used in the numerical integration; the result plotted in this figure uses $\Delta s=5\unit{\micro \meter}$. Note that \cref{fig:heff-full-sim-b} is symmetric under the $\theta \rightarrow -\theta$ transformation only for $\rho=0$, as wavefront mismatches break the tilt symmetry (see \Cref{fig:wavefront-curvature-interpret}). \textit{Below:} Plot of the difference between numerically calculated full \ac{he} and approximated \ac{he} for a $w_\text{eff} = 1$\,mm beam impinging on an infinite \ac{sepd} (\Cref{fig:heff-full-sim-c}) and an infinite \ac{qpd} (\Cref{fig:heff-full-sim-d}). The full \ac{he} uses \cref{eq::GB-full-tilt} for the tilted measurement beam, and the resulting intensity can be integrated only numerically. The approximatex \ac{he} uses \cref{eq::GB-tilt} instead. This discrepancy between the two outputs of the numerical integration is the effect of the approximations introduced between \cref{eq::GB-full-tilt} and \cref{eq::GB-tilt}. Such effect is very small in the \ac{sepd}-case, where a strong dependency on $\rho$ is also visible, while relatively large in the \ac{qpd}-case, where it diverges for large tilts. Still, the most relevant part, between -0.5\,mrad and 0.5\,mrad is addressed within $10^{-4}$ precision.}
    \label{fig:heff-full-sim}
\end{figure*}

A numerical model can be developed to address all \ac{he} calculations that cannot be carried out analytically, following the same approach as in \cite[Subsection III.C]{alvise_2024}. These include arbitrary \ac{pd} geometries and values of $\rho$ that extend beyond the validity range of the previous Maclaurin expansions. The model is also used to verify the validity of the previous analytical approximations, as those performed between \cref{eq::GB-full-tilt,eq::GB-tilt}. The power incident on each \ac{pd}/\ac{qpd}-segment is computed by numerically integrating the beam intensity (\cref{eq::itf_tilt_simp} in the approximated case), employing a discretization of both the \ac{qpd} surface and the tilt angles. A similar approach is followed by the AEI-developed simulation software IfoCAD \cite{WANNER20124831}, which is, furthermore, capable of simulating large-scale optics experiments. Due to the simplicity of our purpose, we opted for our own numerical integrator. The corresponding phase is then extracted as the argument of the resulting complex power $P_\text{num}(\theta)$.
\begin{equation}
    \begin{split}
    P_\text{num}(\theta) &= \int \int_{S_{\text{PD}}} \dd{x} \dd{y} \, I_{\text{smp}}(x, y, \theta)  \\
    &\approx \sum_i^{x_i \in S_{\text{PD}}} \sum_j^{y_j \in S_{\text{PD}}} I_{\text{smp}}(x_i, y_j, \theta) \Delta s^2 + \mathcal{O}(\Delta s^4)\label{eq:numeric-discretization}
    \end{split}
\end{equation}
As the amount of sums to be taken is proportional to $\Delta s^2$, the discretization introduces an overall numerical error that scales as $\Delta s^2$. We found that a spatial resolution of $\Delta s = 5~\unit{\micro\meter}$, which limits the integration's runtime to a few seconds, induces \ac{he} numerical artifacts of approximately $10^{-7}$ at most for a beam of effective spot radius 1\,mm when integrating on both a \ac{sepd} (see \Cref{fig:heff-full-sim-a}) or a \ac{qpd}-segment (see \Cref{fig:heff-full-sim-b}), which is well within the target precision of this model. From the numerically computed power $P_\text{num}(\theta)$, the \ac{he} is extracted by evaluating its absolute value and normalizing it by the power of the two beams according to \cref{eq:het-eff-definition}. Finally, the parameters $t_0$, $t_1$ ... $t_n$ are obtained by performing a linear fit of the resulting \ac{he}. We stress that this process is sensitive to the used angular fitting range, which must be adjusted accordingly to the order of the linear fit.

Numerical integrations are used to assess the validity of the approximations introduced between \cref{eq::GB-full-tilt} and \cref{eq::GB-tilt}. The discrepancy between the fully numerically integrated \ac{he} --- using \cref{eq::GB-full-tilt} for the measurement beam --- and the approximated numerically integrated \ac{he} --- using \cref{eq::GB-tilt} for the measurement beam --- is of the order of $10^{-6}$ in the case of a \ac{sepd} (see \Cref{fig:heff-full-sim-c}) and at most of $10^{-3}$ in the case of a \ac{qpd} (see \Cref{fig:heff-full-sim-d}). We conclude that this analytical model can estimate the \ac{he} with deviations smaller than $10^{-6}$ in the case of a \ac{sepd}. In the case of the \ac{qpd}, the introduced approximations cause a discrepancy of the order of $10^{-4}$ within $\pm0.5$\,mrad of tilt; such discrepancy grows linearly for larger tilts, in an area of lower interest for the \ac{lisa} mission.

\section{Heterodyne Efficiency and Imaging Systems} \label{section:heff-imaging-systems}
Consider a pair of beams, i.e., a measurement beam and a reference beam, impinging on a large \ac{pd}. Both beams are nominally incident perpendicular to the detector surface; however, the measurement beam is allowed to rotate by a small angle $\theta$ about the center of the \ac{pd}. The effective beam spot radius and the relative \ac{roc} at the \ac{pd} are denoted by $w_\text{eff}$ and $R_\text{rel}$, respectively. In this configuration, the \ac{he} of the beams on the \ac{pd} is given by \cref{eq:het_eff_SEPD-inf} in the case of a \ac{sepd}, or by \cref{eq:heff-QPD-inf-top} \cref{eq:heff-QPD-inf-bottom} in the case of a \ac{qpd}.

\begin{figure}[!htpb]
    \centering
    \includegraphics[width=0.9\columnwidth]{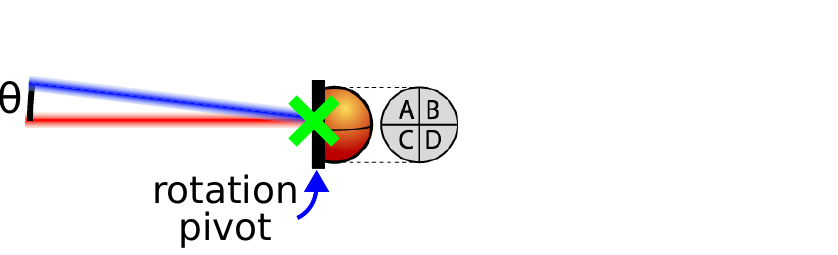}
    \includegraphics[width=0.9\columnwidth]{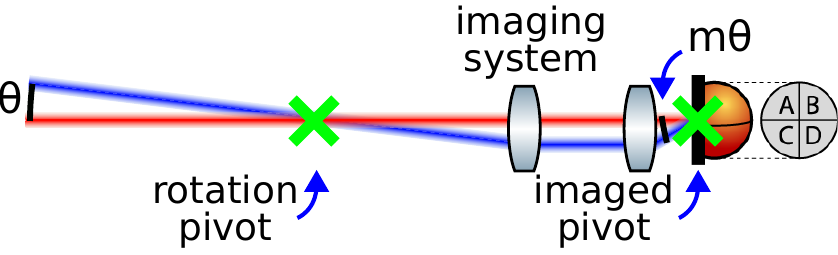}
    \caption{Drawing of the two systems compared in this section. \textit{Above}: two beams impinging on a \ac{qpd}, with the measurement beam (in blue) rotated about the center of the \ac{qpd} by an angle $\theta$. \textit{Below:} same system as above, with an ensemble consisting of an imaging system plus a \ac{qpd} being placed instead of the \ac{qpd}. Note that in this case the measurement beam's angle at the \ac{qpd} is magnified to $m \theta$.}
    \label{fig:he-with-wo-IS}
\end{figure}

Now, consider introducing a collimating imaging system as depicted in \Cref{fig:he-with-wo-IS}: the entrance pupil of the imaging system is placed at the original position of the \ac{pd}, while the \ac{pd} itself is relocated to the exit pupil. What is the resulting \ac{he} as a function of the tilt angle for this combined system of imaging optics and detector? The propagation of a \ac{gb} through a collimating imaging system with angular magnification $m$ modifies the beam parameters according to equations \cref{eq:IS-beam-waist,eq:IS-beam-roc}. These transformations apply not only to the individual beam parameters, but also to the effective beam spot radius $w_\text{eff}$, the relative radius of curvature $R_\text{rel}$, and the parameter $\rho$ \cite{alvise_2024}.
\begin{align}
    w_\text{eff, out} &= m^{-1}w_\text{eff, in}\\
    R_\text{rel, out} &= m^{-2}R_\text{rel, in}\\
    \rho_\text{out}   &= \rho_\text{in}
\end{align}
An important consequence is that the parameter $\rho$ (the effective-spot-radius-normalized wavefront curvature mismatch) is invariant under propagation through a collimating imaging system. In contrast, the tilt angle $\theta$ between the beams is magnified by a factor $m$, as described by \cref{eq::generic-is}. When computing the \ac{he} for such a beam configuration using an infinitely large \ac{pd}, the $m$ factors cancel in \cref{eq:het_eff_SEPD-inf,eq:heff-QPD-inf-top,eq:heff-QPD-inf-bottom}. In other words, the previously mentioned quantity $\Theta = k w_\text{eff} \theta$ is invariant under the presence of a collimating imaging system. Therefore, in the limit of small tilts and infinite detector size, the \ac{he} as a function of the tilt angle remains unchanged in the presence of the imaging system. For a finite-sized \ac{pd}, the \ac{he} remains invariant as long as the ratio $\beta = \sqrt{2} r_\text{PD}/w_\text{eff}$ is conserved. However, this conclusion breaks down when the magnified tilt becomes large enough to invalidate the linear approximation in $\theta$ used in \cref{eq::GB-tilt} \cite[Section III]{alvise_2024}. A similar argumentation is derived in \cite[Section IV]{alvise_2024} for the \ac{dws} signal.

\section{Experimental Results} \label{section:results}

This section presents the \ac{he} measurements performed using \ac{tdobs}, aimed at experimentally validating the \ac{he} model. Before discussing this in detail, it is important to stress that an experimental measurement of the \ac{he} in heterodyne interferometry is not trivial at all, or, at least, is way more complex than a \ac{dws} calibration. While to obtain the \ac{dws} signal, it is sufficient to measure the phase of the MHz heterodyne beat note, to measure the \ac{he} one has to compare the amplitude of the MHz heterodyne beat note to the DC powers of the two interfering beams. This is especially tricky, as the involved electronics usually present different \ac{tf} values at DC and at MHz. Furthermore, the \ac{tf} at MHz is frequency dependent. Moreover, for technical convenience, the AC and DC components of the photocurrents in \ac{tdobs} are often processed separately; this happens also in \ac{tdobs}, where the output signal of the \ac{tia} is fed to a second-stage amplifier, which splits it into a DC ($f<10$\,kHz) and AC ($f>10$\,kHz) components. These are then acquired by different devices: the DC component is digitized by the I/Q demodulation phasemeter, while the AC component is digitized by the \ac{dpll} phasemeter. 

It is, hence, fairly difficult to perform an absolute measurement of the \ac{he}. The \ac{he} is, in principle, $f_\text{het}$-independent; however, some $f_\text{het}$-dependency will unavoidably leak in due to the electronics' \ac{tf}, adding a frequency-dependent scaling factor $G(f)$. Relative variations of the \ac{he} measured at one specific frequency are, however, unaffected by this re-scaling issue. 

\subsection{Absolute Heterodyne Efficiency Measurement} \label{subsection:he-absolute-meas}
A solution that we found to compare the beat note and DC power amplitude is to process both through the DC readout chain. We have set the beat note frequency to a few tens of Hz\footnote{This is possible, as we generate the three different frequencies from the same laser source using three independent \acp{aom}.}; this enables measuring both DC powers and beat note using the DC chain~\cite[subsection 3.2.2]{me_thesis}. The \ac{he} can then be extracted by using \cref{eq:het-to-visibility}, adapted into \cref{eq:heff-absolute-meas}, by measuring the power of the two beams and the maximum and minimum power output independently. We remark that all DC offsets have been calibrated and removed in this procedure.
\begin{equation}
    \she = \frac{P_\text{PD, max} - P_\text{PD, min}}{4 \sqrt{P_\text{PD, r} P_\text{PD, m}}}
    \label{eq:heff-absolute-meas}
\end{equation}
This led to the \ac{he} measurements reported in \Cref{tab:heff_tdobs_absolute}, which were performed with parallel beams impinging on the center of each \ac{qpd}. The results in \Cref{tab:heff_tdobs_absolute} show good agreement between the model and the measurements.

\begin{table*}[!htpb]
    \centering
    \begin{tabular}{c|c|c|c|c}
    \hline
     Beams \& \acp{qpd}                  &$\beta$             & Measured        & Analytic Model & Numerical Model \\
     \hline
     \ac{rxgb} \& \ac{lo} @ \acp{refqpd} &  0.74  $\pm$ 0.02  & 1.007 $\pm$ 0.01 & 0.99939 $\pm$ \sn{1}{-5} & 0.9994 $\pm$ \sn{1}{-4} \\
     \ac{rxft} \& \ac{lo} @ \acp{refqpd} &  0.570 $\pm$ 0.005 & 1.038 $\pm$ 0.05 & 0.99196 $\pm$ \sn{1}{-5} & 0.9951 $\pm$ \sn{2}{-3}\\
     \ac{rxgb} \& \ac{lo} @ \acp{sciqpd} &  1.87  $\pm$ 0.01  & 0.98 $\pm$ 0.01 & 0.9853 $\pm$ \sn{2}{-4}  & 0.9853 $\pm$ \sn{2}{-3}\\ 
     \hline
    \end{tabular}
    \caption{Absolute \ac{he} measurements in \ac{tdobs}, obtained by a DC beatnote measurement using \cref{eq:heff-absolute-meas} using single channels of a \ac{qpd}, compared to a numerical model of the interference. The table also reports the value of $\beta = \sqrt{2}\rqpd/\weff$ of the analyzed case.}
    \label{tab:heff_tdobs_absolute}
\end{table*}

\subsection{Normalized Heterodyne Efficiency Measurement} \label{subsection:he-normal-meas}
We further proceed characterizing the dependency of the \ac{he} on the measurement beam's angle. For this purpose, we opt to use MHz \acp{hf} to be representative of \ac{lisa}. We must therefore neglect the exact value of the \ac{he}'s scaling factor $G(f)$ and characterize only the variations of the \ac{he}; we do this by normalizing the measured \ac{he} by its maximum. The main goal of this section is to characterize the asymmetry in the \ac{he} measured by the two \ac{qpd} halves as a function of the beam tilt; this is possible because the beam pairs present a wavefront curvature mismatch \cite{alvise_2024}. This procedure was followed using the beam couple \ac{rxgb}\&\ac{lo} on both the \acp{refqpd} and \ac{sciqpd}.

\begin{figure}[!htpb]
    \centering
    \includegraphics[width=0.9\columnwidth]{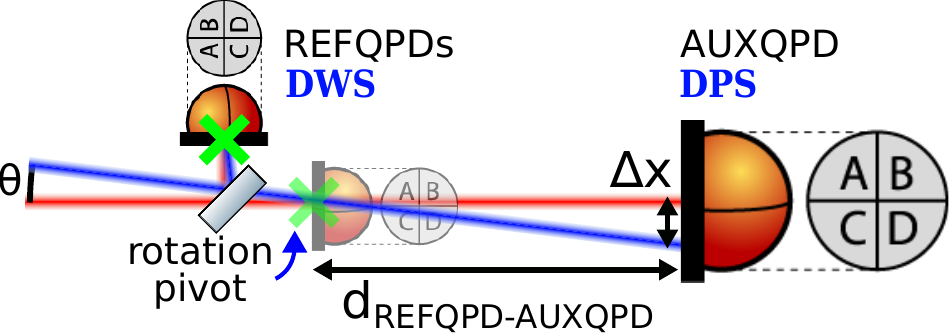}
    \caption{Drawing of the beam rotation process. The \ac{refqpd} is located on the left, while the \ac{auxqpd} is located on the right. The virtual position of the \ac{refqpd} along the beam path is also shown in transparency. The Rx beam's rotation pivot is indicated by the green cross. The lateral displacement $\Delta x$ is measured using the \ac{dps} signal (see \cite[Section V]{alvise_2024}) of the \ac{auxqpd}. The tilt angle of the Rx beam $\theta$ can be recovered trigonometrically using the \cref{eq::beam_tilt_dps}. The tilt angle in the figure is strongly exaggerated; this is in the range of $\pm 1$\,\unit{\milli \radian}. Figure from \cite{alvise_2024}. }
    \label{fig::beam_rotation}
\end{figure}

We performed this analysis analogously to the \ac{dws} calibrations in \cite[Section V]{alvise_2024}: the Rx beam is rotated about the center of the \ac{qpd} under test, i.e., either the \acp{refqpd} or the \acp{sciqpd}, using the steering mirrors on the \ac{ts}. The procedure is depicted in \Cref{fig::beam_rotation}; for a full description of this procedure, see \cite[Section V]{alvise_2024}. To establish the calibration, the measured beam angle is compared with the corresponding \ac{he}. \Cref{eq::beam_tilt_dps} is the relation used to recover the beam tilt angle from the \ac{dps} measurement, where $\Delta x_\text{AUXQPD}$ is the lateral displacement of the beam measured on the \ac{auxqpd}, and $d_\text{QPD-AUXQPD}$ is the distance between the two \acp{qpd} along the beam path. This procedure is the main (systematic) uncertainty source of the measurement, as $d_\text{REFQPD-AUXQPD}$ is known to within $\sim7$\% precision, as previously shown in \cite{alvise_2024}.
\begin{equation}
    \theta = \frac{\Delta x_\text{AUXQPD}}{d_\text{REFQPD-AUXQPD}}
    \label{eq::beam_tilt_dps}
\end{equation}

This measurement was performed at $f_\text{het}$=5\,MHz, with the difficulties in measuring the absolute \ac{he} mentioned above due to the little-known gain factor $G(f)$. To make the \ac{he} measurement independent of $G(f)$, we normalize the measured \ac{he} by its highest value. The measurements are then compared to numerical models built for the specific case. To quantitatively compare the data and the model, we fit both with a candidate function and compare the resulting fit parameters. For the \ac{rxgb} \& \ac{lo} @ \acp{refqpd} case, we fit a simple Gaussian curve in \cref{eq:candidate-gaussian}, as the tilt range is too small to grasp the full asymmetry of the \ac{he}. For the \ac{rxgb} \& \ac{lo} @ \acp{sciqpd} case, we use the full expression in \cref{eq:heff-QPD-inf-top,eq:heff-QPD-inf-bottom}, adapted into the fit-function form \cref{eq:candidate-gaussian-asym}. Note that a fit constant $c$ was added to this equation. Despite this term not being present in the analytical model, it proves necessary to fit the \ac{he} of a finite \ac{qpd} when using the curve for an infinite \ac{qpd}. We stress that this is needed for fitting both experimental data and the numerically simulated \ac{he}, leading to similar estimated values of $c$. The reason why this constant must be added is related to the finite size of the \ac{qpd}, which causes diffraction fringes to appear. If just a portion of these is included in the measurement's angular range, it manifests as an effective offset of the measured \ac{he}; a good example of this is \Cref{fig:heff-QPD-fin}, where the numerically calculated \ac{he} indeed looks like a Gaussian curve with an added positive constant taking a value of roughly 0.2 in the plotted range. The parameter $\rho$, which is known from previous measurements to be equal to $\rho=0.14\pm0.1$ in both analyzed cases \cite[Figure 3.9]{me_thesis}, is fed as a fixed parameter of \cref{eq:candidate-gaussian-asym} to the fit routine, as it would not converge otherwise. The parameter $s$, which has a theoretical value of $s_\infty=\frac{m}{k w_\text{eff}}=0.137\pm0.002$\,mrad for an infinite \ac{qpd}, is left free, as its effective value is affected by the finite size of the \ac{qpd}.
\begin{align}
    f(x) &= a \exp\left( - \left( \frac{x-x_0}{s}\right)^2\right) \label{eq:candidate-gaussian}\\
    g_\pm(x) &= a \exp\left( - \left( \frac{x-x_0}{s}\right)^2\right)\sqrt{1 \pm 2\Im\left[ \frac{x-x_0}{2^{3/2}s \sqrt{1-i\rho}} \right] + \left| \frac{x-x_0}{2^{3/2}s \sqrt{1-i\rho}} \right|^2} + c \label{eq:candidate-gaussian-asym}
\end{align}

\begin{figure*}[!htpb]
    \centering
    \begin{subfigure}{.45\textwidth}
        \centering
        \includegraphics[width=\columnwidth]{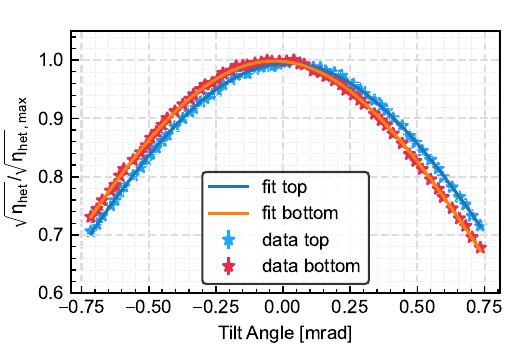}  
        \caption{}
        \label{fig:heff-exp-refqpds-a}
    \end{subfigure}
    \begin{subfigure}{.45\textwidth}
        \centering
        \includegraphics[width=\columnwidth]{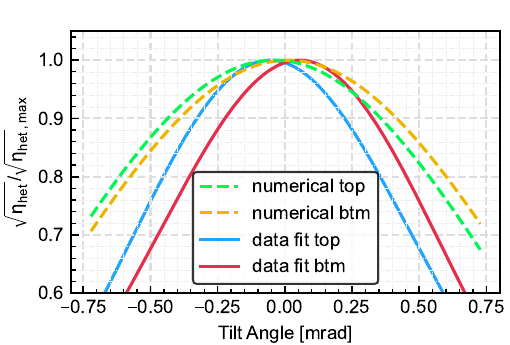}  
        \caption{}
        \label{fig:heff-exp-refqpds-b}
    \end{subfigure}
    \caption{\textit{Left:} Measurement of the \ac{he} as a function of the beam tilt angle using the \ac{rxgb}\&\ac{lo} beams on \ac{refqpd}1. We remind that the \ac{he} has been normalized to its maximum value. The \ac{rxgb} beam was tilted vertically, hence we show the \ac{he} measured on the top and bottom segments of the \ac{qpd}. The parameters of the two interfering beams are $w_r = 0.874\pm0.006$\,mm, $w_m = 1.038\pm0.008$\,mm, $\weff = 0.95\pm0.02$\, mm, $\rho=0.14\pm0.01$, $\beta=0.74\pm0.02$. The fit is performed using \cref{eq:candidate-gaussian}. The resulting fit parameters are reported in Table~\ref{tab:heff-SD-fitparameters-refqpds}. \textit{Right:} Comparison between the Gaussian fit in \Cref{fig:heff-exp-refqpds-a} and the numerical model.}
    \label{fig:heff-exp-refqpds}
\end{figure*}

\begin{figure*}[!htpb]
    \centering
    \begin{subfigure}{.45\textwidth}
        \centering
        \includegraphics[width=\columnwidth]{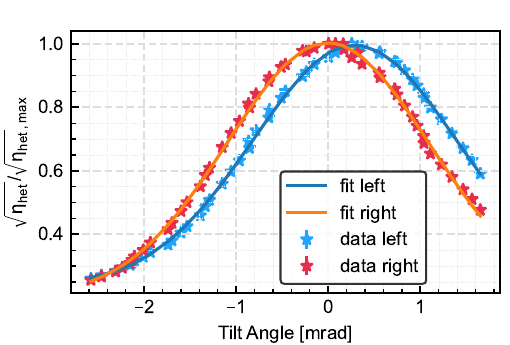}  
        \caption{}
        \label{fig:heff-exp-sciqpds-a}
    \end{subfigure}
    \begin{subfigure}{.45\textwidth}
        \centering
        \includegraphics[width=\columnwidth]{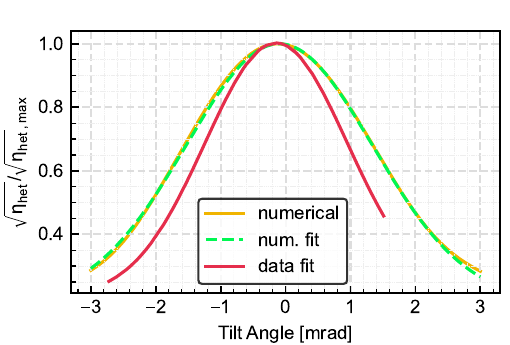}  
        \caption{}
        \label{fig:heff-exp-sciqpds-b}
    \end{subfigure}
    \caption{\textit{Left:}  Measurement of the \ac{he} as a function of the beam tilt angle in \ac{tdobs} using the \ac{rxgb}\&\ac{lo} beams on \ac{sciqpd}1. We remind that the \ac{he} has been normalized to its maximum value. The \ac{rxgb} beam was tilted horizontally, hence we show the \ac{he} measured on the left and right segments of the \ac{qpd}. The parameters of the two interfering beams are $w_r = 0.350\pm0.002$\,mm, $w_m = 0.415\pm0.003$\,mm, $\weff = 0.377\pm0.002$\, mm, $\rho=0.14\pm0.01$, $\beta=1.87\pm0.01$. The Fit is performed using \cref{eq:candidate-gaussian-asym}. The resulting fit parameters are reported in Table~\ref{tab:heff-SD-fitparameters-sciqpds}. \textit{Right:} Comparison between the Gaussian fit in \Cref{fig:heff-exp-sciqpds-a} and the numerical model.}
    \label{fig:heff-exp-sciqpds}
\end{figure*}

The measured and simulated \acp{he} are plotted in \Cref{fig:heff-exp-refqpds,fig:heff-exp-sciqpds}. The respective fit parameters are reported in \Cref{tab:heff-SD-fitparameters-refqpds,tab:heff-SD-fitparameters-sciqpds}. We note that this small value of $\rho$ is large enough to introduce a detectable difference in the peak angles of the \acp{he} of the two \ac{qpd} halves during a tilt. This effect is still visible, despite being suppressed by the small value of $\beta$, in the \acp{refqpd}. Note that, due to small beam misalignment, the tilt angle at which the two \acp{he} share the same value is not exactly zero. This must be taken into account when comparing the data with the simulation in \Cref{fig:heff-exp-sciqpds-b,fig:heff-exp-refqpds-b}; here we shift the $x$-axis by $\bar x_0 = \frac{1}{2}(x_{0 \text{, 1st half}} + x_{0 \text{, 2nd half}})$, where the parameters $x_{0 \text{, 1st half}}$ and $x_{0 \text{, 2nd half}}$ are estimated by fitting \cref{eq:candidate-gaussian,eq:candidate-gaussian-asym} to the \acp{he} of the two \ac{qpd} halves. 

\Cref{fig:heff-exp-refqpds-a} shows the normalized measured \ac{he} as a function of the beam tilt angle using the \ac{rxgb} \& \ac{lo} on \ac{refqpd}1. In this system, $\rho=0.14\pm0.1$ and $\beta=0.74 \pm 0.02$, meaning that the effects of the wavefront curvature mismatch on the two \ac{refqpd}1's halves are expected to be suppressed (see \Cref{fig:het-eff-QPD-fin-d,fig:het-eff-QPD-fin-e,fig:het-eff-QPD-fin-f}). Due to the limited angular range, the expected \ac{he} is expected to have a near-Gaussian shape; we therefore fit it using \cref{eq:candidate-gaussian}, which indeed proves a very valid fit of the data points. This measurement is able to capture a tiny difference in the peak angles of the \ac{he} of the two \ac{refqpd}1's top and bottom halves. \Cref{fig:heff-exp-refqpds-b} compares the resulting fit from \Cref{fig:heff-exp-refqpds-a} with the numerical model developed in subsection~\ref{subsection:heff-numeric-model}. \Cref{tab:heff-SD-fitparameters-refqpds} compares the parameters obtained by fitting the data in \Cref{fig:heff-exp-refqpds-a} and the model in \Cref{fig:heff-exp-refqpds-b} with \cref{eq:candidate-gaussian}. Measurement and simulation look quite different at first glance. However, neglecting the small uncertainties in the parameter furnished by the fit, note that the resulting $s$ parameters present a fairly small relative discrepancy. Such discrepancy is larger, but still very small in absolute value, for the $x_0$ parameter. 

\begin{table*}[!htpb]
    \centering
    \begin{tabular}{cc|cc|c|c}
    \hline
    Parameter & Units & \ac{qpd}-Top & \ac{qpd}-Bottom & Model & \makecell{Relative\\Discrepancy}\\
    \hline
    $a$   &  -   & 0.9945 $\pm$ 0.004  & 0.9986  $\pm$ 0.0005  & 1     $\pm$ 0.0001 & -\\
    $s$   & mrad & 1.245  $\pm$ 0.002  & 1.222  $\pm$ 0.002   & 1.293 $\pm$ 0.0007 & 4.6\%\\
    $x_0 - \bar x_0$ & mrad & 0.0258 $\pm$ 0.0007 & -0.0258 $\pm$ 0.0006  & $\pm$(0.017 $\pm$ 0.0003) & 51\%\\
    \hline
    \end{tabular}
    \caption{Table with the fit parameters from Figure~\ref{fig:heff-exp-refqpds}. The centroids of the Gaussian curves are already corrected by the average centroid shift $\bar x_0 = -6.7\pm0.5$\,\unit{\micro \radian}. This small deviation is attributable to a lateral misalignment of the beams with respect to the \ac{qpd}. The last column reports the relative discrepancy, defined as $\frac{theo - \overline{meas}}{theo}$, where $\overline{meas}$ is the average of the measured parameter for the two \ac{qpd} halves. Note that the discrepancy between the measured and modelled value of $s$ is large if compared to the errors, but relatively very small.}
    \label{tab:heff-SD-fitparameters-refqpds}
\end{table*}

\Cref{fig:heff-exp-sciqpds-a} shows the normalized measured \ac{he} as a function of the beam tilt angle using the \ac{rxgb} \& \ac{lo} on \ac{sciqpd}1. Note that this system differs with respect to the \ac{rxgb} \& \ac{lo} on \ac{refqpd}1 only due to the presence of a collimating imaging system, and therefore the two are related by the argument presented in Section~\ref{section:heff-imaging-systems}. Note that the angular range is magnified by the presence of the imaging system. In this system, $\rho=0.14\pm0.1$ and $\beta=1.87 \pm 0.01$, meaning that the effects of the wavefront curvature mismatch on the two \ac{sciqpd}1's halves are expected to be roughly half of those on an infinite \ac{qpd} (see \Cref{fig:het-eff-QPD-fin-d,fig:het-eff-QPD-fin-e,fig:het-eff-QPD-fin-f}), and way more noticeable than on the \acp{refqpd}. This measurement evidences a noticeable asymmetry between the two \ac{sciqpd} halves, and a clear difference in peak angle of the two \acp{he}; we hence use \cref{eq:candidate-gaussian-asym} to fit the experimental data. We observe that the model closely follows the experimental data, although not exactly. \Cref{fig:heff-exp-sciqpds-b} compares the resulting fit from \Cref{fig:heff-exp-sciqpds-a} with the numerical model developed in subsection~\ref{subsection:heff-numeric-model}. \Cref{tab:heff-SD-fitparameters-sciqpds} compares the parameters obtained by fitting the data in \Cref{fig:heff-exp-sciqpds-a} and by fitting the numerical model in \Cref{fig:heff-exp-sciqpds-b}; despite the discrepancies between the estimated parameters being much larger than the associated uncertainties, the relative discrepancies are fairly small. In particular, the position of the \ac{he}'s maximum, $x_0$, is predicted very accurately by the numerical model, while the $s$ parameter has a relative uncertainty as high as 23\%.

\begin{table*}[!htpb]
    \centering
    \begin{tabular}{cc|cc|c|c}
    \hline
    Parameter & Units & \ac{qpd}-Left & \ac{qpd}-Right & Model & \makecell{Relative\\Discrepancy}\\
    \hline
    $a$  &             -     & 0.759 $\pm$ 0.003 & 0.801   $\pm$ 0.003  & 0.817 $\pm$ 0.002 & 4.5\% \\
    $s$  &             mrad  & 1.50  $\pm$ 0.01  & 1.49    $\pm$ 0.01   & 1.950 $\pm$ 0.007 & 23\%  \\
    $x_0 - \bar x_0$ & mrad  & 0.123 $\pm$ 0.005 & -0.123  $\pm$ 0.005  & $\pm$(0.126 $\pm$ 0.001) & 2.4\%\\
    $c$  &             -     & 0.237 $\pm$ 0.005 & 0.201   $\pm$ 0.005  & 0.187 $\pm$ 0.003 & -17\%\\
    \hline
    \end{tabular}
    \caption{Table with the fit parameters from Figure~\ref{fig:heff-exp-sciqpds}. The centroids of the Gaussian curves are already corrected by the average centroid shift $\bar x_0 = -142\pm5$\,\unit{\micro \radian}. This small deviation is attributable to a lateral misalignment of the beams with respect to the \ac{qpd}. The last column reports the relative discrepancy, defined as $\frac{theo - \overline{meas}}{theo}$, where $\overline{meas}$ is the average of the measured parameter for the two \ac{qpd} halves. Note that the discrepancy between the measured and modelled values of all parameters is large if compared to the uncertainties, but relatively very small.}
    \label{tab:heff-SD-fitparameters-sciqpds}
\end{table*}

A comparison between these two measurements can be used to obtain a result which is independent of the systematic uncertainties due to \cref{eq::beam_tilt_dps}. This is possible, as mentioned, because these two setups measure the interference of the \ac{rxgb}\&\ac{lo} beams at the \acp{refqpd} and \ac{sciqpd}, and only differ due to the presence of a collimating imaging system. A similar method, using data from the same measurement, gave a very good result in \cite[eqs (68) and (69)]{alvise_2024} while comparing the \ac{dws} of the same optical system. Similarly, we can use this comparison to cross-check the resulting value of the $s$ parameter obtained in the two configurations. We measure and model
\begin{align}
    \frac{s_\text{RxGB-LO @ REFQPDs, meas}}{s_\text{RxGB-LO @ SCIQPDs, meas}}   &= 0.826 \pm 0.006 \\
    \frac{s_\text{RxGB-LO @ REFQPDs, model}}{s_\text{RxGB-LO @ SCIQPDs, model}} &= 0.663 \pm 0.004
\end{align}
These values are relatively close, although not as absolutely close as expected. Potential issues in the used analysis lie in the choice of the fit function in \cref{eq:candidate-gaussian-asym}, where the constant $c$ was necessary in order to correctly fit the data, or in the impact of beam lateral misalignments.

\section{QPD signal readout noise} \label{section:qpd-noise}
This section uses the knowledge derived in the previous Sections to calculate how beam tilts affect the phase noise in \ac{lisa}'s interferometers. As previously mentioned, \ac{lisa} employs \acp{qpd} to measure the output power of its interferometers, such that it can measure not only longitudinal signals but also angular signals via \ac{dws} and beam position via \ac{dps}. The four phases derived by the outputs of a \ac{qpd} are combined, after digitization, to obtain either the \ac{lps} or the \ac{dws} signals. During \ac{lisa}'s science operation, the measurement beams in the \ac{isi} and \ac{tmi} can be affected by tilts up to 1\,mrad, causing a significant reduction of the \ac{he}, and therefore of the beat note's amplitude. This causes the interferometric phase noise to increase; to predict how much it increases, a model of the \ac{he} as a function of the beam tilt angle is needed, which is what was carried out throughout this article.

\subsection{Interferometric Noise} \label{ssection:qpd-phase-noise}
The interferometric readout noise affecting the output phase can be distinguished into direct phase noise, additive noise, and multiplicative noise. The first and third categories are \ac{he} independent. Laser frequency noise is also common-mode in all segments, and will hence be neglected in this model. To the last category belongs $2f$-\ac{rin}\cite{RIN_lennart}. The additive noise category includes laser shot noise, electronic noise from the \ac{tia}, and $1f$-\acf{rin}\cite{RIN_lennart}, which are the main noise sources in \ac{lisa}, and is \ac{he}-dependent. $1f$-\ac{rin} is the coupling of relative intensity variations of the laser at the frequency of the beat note, while $2f$-\ac{rin} couples relative intensity variations of the laser at twice the $f_\text{het}$. 

We hence model how these noises affect the \ac{qpd}-signals' readout, depending on the incidence angle of the measurement beam, while assuming that both beams are always centered on the \ac{qpd} and that the reference beam is also always orthogonal to the detector's surface, as assumed in the framework introduced in \Cref{section:heff-model}. The reasoning presented in this subsection should also be extended to other additive noise, as digitization noise and \ac{pd} dark current noise, but will be limited to the previously mentioned noise sources for simplicity, but could also easily be included.

The amount of phase noise caused by a noise contributor can be quantified using the inverse of \cref{eq:SNR-def}
\begin{equation}
    \tilde \phi_\text{noise} = \text{SNR}^{-1} = \frac{\text{Noise ASD}}{\text{Signal RMS}}, \label{eq:phase-noise-def}
\end{equation}
where $\tilde \cdot$ indicates noise densities, "Noise \acs{asd}" indicates the \ac{asd} near and around the \ac{hf} $f_\text{het}$, and "Signal \acs{rms}" indicates the \ac{rms} of the signal $s(t)$ over a period $T$, defined as \cite{PhysRevApplied.14.054013}
\begin{equation}
\mathcal{RMS} \{ s(t) \} = \sqrt{\frac{1}{T} \int_0^T [s(t)]^2 dt}.
\label{equation::rms-signal-definition}
\end{equation}
The signal $s(t)$ is the last term of \cref{eq:power2} and its \ac{rms} is 
\begin{equation}
\begin{split}
 \mathcal{RMS} &\left\{ 2\sqrt{\he(\theta) \bar P_{\text{seg, }r} \bar P_{\text{seg, }m} } \cos (\omega_\text{het} t + \psi_\text{het}) \right\} \\
 &= \sqrt{ 2\eta_\text{het}(\theta) \bar P_{\text{seg, }r} \bar P_{\text{seg, }m}},  
 \end{split}
\end{equation}
where $\bar P_{\text{seg, }r}$, $\bar P_{\text{seg, }m}$ are the average powers of the reference and measurement beams impinging on each segment. Note that $\mathcal{RMS} \{ s(t) \}$ depends on the tilt angle $\theta$ via the \ac{he} $\eta_\text{het}(\theta)$. We further assume $\theta$ to vary at frequencies much lower than the \ac{lisa}-observation band. On the other hand, the \ac{asd}s of the various noises are specific to the particular noise cause and are discussed, for instance, in \cite[subsection 2.1.5]{me_thesis}, and are independent of the tilt angle. The ratio of these two quantities, which gives the phase noise, is hence tilt-dependent, and such dependency is driven only by the \ac{he}. The resulting phase noises caused by the mentioned noise sources at \ac{qpd}-segment level are given in \cref{eq:phase-noise-shot,eq::phase-noise-en,eq:1f-rin-phase-noise,eq::2f-rin-phase-noise} in units of \radsqrthz. 

\begin{align}
\tilde \phi_{\text{shot}}(\theta) &=\sqrt{ \frac{(\bar P_{\text{seg, }r} + \bar P_{\text{seg, }m}) h \nu_l }{\eqpd \he(\theta) \bar P_{\text{seg, }r}\bar P_{\text{seg, }m} }} \label{eq:phase-noise-shot}\\
\tilde{\phi}_{\text{en}}(f_\text{het}, \theta)     &= \frac{\tilde i_\text{en}(f_\text{het})}{A \sqrt{2 \he(\theta) \bar P_{\text{seg, }r} \bar P_{\text{seg, }m}}}  \label{eq::phase-noise-en}\\
\tilde \phi_{\text{RIN-}1f}(f_\text{het}, \theta) &= \frac{\bar P_{\text{seg, }r} \tilde r_r(f_\text{het}) \boxplus \bar P_{\text{seg, }m} \tilde r_m(f_\text{het})}{\sqrt{2 \he(\theta) \bar P_{\text{seg, }r} \bar P_{\text{seg, }m}}} \label{eq:1f-rin-phase-noise}\\
\tilde \phi_{\text{RIN-}2f}(f_\text{het}) &= \frac{\tilde r_r(2f_\text{het}) \boxplus \tilde r_m(2f_\text{het})}{2 \sqrt{2}}
\label{eq::2f-rin-phase-noise}
\end{align}
where $h$ is Planck's constant, $\nu_l$ is the frequency of the used laser beam, where we neglect the frequency difference as argued in \cite{alvise_2024}, $\eta_\text{PD}$ is the quantum efficiency of the \ac{qpd}, $A=\frac{q \eta_\text{QPD}}{h \nu_l}$ is the responsivity of the \ac{qpd}, $\tilde i_\text{en}(f_\text{het})$ is the equivalent input current noise density of the \ac{qpr} at the \ac{hf} given in \cref{eq::tia-eq-in-current-noise}, and $\tilde r_r$, $\tilde r_m$ are the relative power variations of the reference and measurement beam at $f_\text{het}$ for $1f-$\ac{rin} and at twice $f_\text{het}$ for $2f-$\ac{rin}. All such parameters are assumed to be the same for all \ac{qpd} segments in this model. The symbol $\boxplus$ indicates the root-squared sum, which is used as the power fluctuations of the two beams are uncorrelated \cite{RIN_lennart}. Except for $2f$-\ac{rin}, all these noise contributions depend on the \ac{he} and are therefore dependent on the tilt angle $\theta$. The total phase noise of the \ac{qpd} segment $i$, with  $i \in (A, \, B, \, C, \, D)$, is
\begin{equation}
\tilde \phi_i(f_\text{het}, \theta) = \sqrt{\tilde \phi_{\text{shot}}^2(\theta) + \tilde\phi_{\text{en}}^2(\theta) + \tilde \phi_{\text{RIN-}1f}^2(f_\text{het}, \theta) + \tilde \phi_{\text{RIN-}2f}^2(2f_\text{het})} \label{eq:seg-phase-noise}. 
\end{equation}
The $\theta$ dependency in the three $\eta_{\text{het}}(\theta)$ dependent noise sources can be collected by re-writing \cref{eq:seg-phase-noise} in the following form. 
\begin{equation}
\begin{split}
\tilde \phi_{\text{shot/en/ RIN-}1f, \, i} &= \Upsilon_{\text{shot/en/ RIN-}1f}/\she_{\,i}(\theta), \\
\Upsilon_\text{shot}       &= \sqrt{ \frac{(\bar P_{\text{seg, }r} + \bar P_{\text{seg, }m}) h \nu_l }{\eqpd \bar P_{\text{seg, }r}\bar P_{\text{seg, }m} }},\\
\Upsilon_\text{en}(f_\text{het})       &= \frac{\tilde i_\text{en}(f_\text{het})}{A \sqrt{2 \bar P_{\text{seg, }r}\bar P_{\text{seg, }m}}},\\
\Upsilon_{\text{RIN-}1f}(f_\text{het})   &=\frac{\bar P_{\text{seg, }r} \tilde r_r(f_\text{het}) \boxplus \bar P_{\text{seg, }m} \tilde r_m(f_\text{het})}{\sqrt{2 \bar P_{\text{seg, }r} \bar P_{\text{seg, }m}}}.
\end{split} \label{eq:noise-heff-norm}
\end{equation}
where the quantities $\Upsilon_{\text{shot/en/ RIN-}1f}$ are simply $\theta$- and segment-independent experimental parameters, which we can treat as constants related to the noise levels on a single \ac{qpd}-segment. Note that the segment-independency holds as long as both beams are centered on the \ac{qpd}, which is what we assume. This step allows to rewrite \cref{eq:seg-phase-noise} as
\begin{equation}
\tilde \phi_i(f_\text{het}, \theta) = \sqrt{  \frac{\Upsilon_\text{shot}^2 + \Upsilon_{\text{en}}^2(f_\text{het}) + \Upsilon_{\text{RIN-}1f}^2(f_\text{het})}{\eta_{\text{het, }i}(\theta)} + \tilde \phi_{\text{RIN-}2f}^2(f_\text{het})} \label{eq:seg-phase-noise-coll}\, .
\end{equation}

In the following, we drop the explicit dependency on $f_\text{het}$ for clarity. We hence calculate the phase noise of the \ac{qpd}-signals, i.e. combinations of the individual \ac{qpd}-segments' outputs as the \ac{dws} signals and the \ac{lps}. In particular, this section focuses on the dependence of phase noise at the \ac{qpd} level on the tilt angle $\theta$ and the beam parameters, with particular emphasis on the effective-beam-spot-radius wavefront curvature mismatch $\rho$. We apply for this purpose the \ac{he} models derived in \Cref{section:heff-model}. For simplicity, this analysis is carried out assuming only vertical tilts. Note that the proposed reasoning is independent of the \ac{qpd} size, whereas the resulting coupling factor is. However, as we want to explicitly calculate the resulting coupling factors as a function of the beam parameters, this is possible in closed form only for infinite \acp{qpd}. Considerations for a finite \ac{qpd} can be done either in approximated form using the series expansion of the finite \ac{qpd} \ac{he} in \cref{eq:heff-QPD-fin}, or numerically. 

We acknowledge that, especially, the assumption of vertical-only tilts does not give a full picture of what happens in reality; in other words, the effect of a beam tilt $\theta_0$ surely depends on its \textit{azimuthal} angle on the \ac{qpd}; we expect, however, the noise increase be well within one order of magnitude. A full derivation of this, however, would require a much longer analysis and is left for future work. This example focuses mainly on highlighting the new \ac{qpd} noise feature arising from asymmetry in \ac{he} due to a wavefront curvature mismatch.

Such assumptions simplify the problem greatly, as it implies that the two top segments of a \ac{qpd} (A and B) experience the same \ac{he} as discussed in subsection~\ref{ssec:infinite-qpd}. Hence, we simplify to $i \in (\text{top, btm})$ instead of the individual quadrants, and we name these $\sqrt{\eta_\text{top}}(\theta)$ for the top segments, given by \cref{eq:heff-QPD-inf-top}, and $\sqrt{\eta_\text{btm}}(\theta)$ for the bottom segments, given by \cref{eq:heff-QPD-inf-bottom}. Therefore, the \ac{qpd} can also be simplified as having only two segments; therefore, the two \ac{qpd}-signals, \ac{ap}-\ac{lps} and \ac{dws}, can be expressed as
\begin{align}
    \phi_\text{AP} &= \frac{1}{2}\left( \phi_\text{top} + \phi_\text{btm} \right) \label{eq:2seg-lps},\\
    \text{DWS} &=  \, \, \quad \phi_\text{top} - \phi_\text{btm} \label{eq:2seg-dws}.
\end{align}

The noise analysis must be carried out separately for noises that are correlated and uncorrelated across the \ac{qpd} segments. This is carried out in the following two subsections.

\subsection{Uncorrelated noises} \label{ssection:qpd-uncorr-noise}
Shot noise and electronic noise originate at the \ac{qpr} segment, and are therefore uncorrelated between the segments. Their noise contributions to \ac{qpd} signals can hence simply be added using the root square sum regardless of the signs in \cref{eq:2seg-lps,eq:2seg-dws}. This results in $\tilde \phi_\text{AP} = \frac{1}{2}\widetilde{\text{DWS}}$; we hence focus our analysis only on the \ac{dws} signal's noise, as that of the \ac{lps} can be recovered from it by applying the relation above. We can hence write
\begin{equation}
\begin{split}
\widetilde{\text{DWS}}_\text{unc}(\theta) &= \sqrt{\tilde \phi_\text{top}^2(\theta)+\tilde \phi_\text{btm}^2(\theta)}\\
&= \sqrt{\Upsilon_\text{shot}^2 + \Upsilon_\text{en}^2} \sqrt{\frac{1}{\eta_\text{top}(\theta)}+\frac{1}{\eta_\text{btm}(\theta)}}\\
&= \sqrt{\Upsilon_\text{shot}^2 + \Upsilon_\text{en}^2} \sqrt{2} \, \sheunc(\theta),
\end{split}
\end{equation}
where in the second line the \ac{dws} noise was rewritten using \cref{eq:noise-heff-norm}, which allows to isolate the $\theta$ dependent factors, and in the third line the \textit{\ac{qpd}-signal tilt-dependent coupling coefficient for uncorrelated noises} of a two-segment \ac{qpd} $\sheunc(\theta)$ was introduced.
\begin{equation}
    \sheunc(\theta) = \frac{1}{\sqrt{2}}\sqrt{\frac{1}{\eta_\text{top}(\theta)}+\frac{1}{\eta_\text{btm}(\theta)}} \label{eq:eta-eff-unc}
\end{equation}
By substituting the relations in \cref{eq:heff-QPD-inf-top,eq:heff-QPD-inf-bottom} in \cref{eq:eta-eff-unc}, one can get an expression for $\sheunc(\theta)$ for an infinite \ac{qpd} depending explicitly on the beam parameters:
\begin{equation}
   \sheunc^{-1}(\theta) = \frac{1}{\sqrt{1+\rho^2}} \frac{2 w_r w_m}{w_r^2 + w_m^2} \exp\left( -\frac{k^2 w_\text{eff}^2 \theta^2}{8(1+\rho^2)} \right) \sqrt{\frac{ \left(1 + \left|\alpha \right|^2 \right)^2 - 4 \left(\Im \left[ \alpha \right] \right)^2}{1 + \left| \alpha \right|^2}, } \label{eq:eta-eff-unc-simp}
\end{equation}
where, for convenience, the quantity
\begin{equation}    
   \alpha = \erfi \left( \frac{k \weff \theta}{2^\frac{3}{2}\sqrt{1 -i\rho}} \right) \label{eq:alpha-def}
\end{equation}
was defined. As \cref{eq:eta-eff-unc-simp} shows, even though we define the quantity $\sheunc(\theta)$ as the coupling coefficient, its reciprocal is much easier to handle. Furthermore, as it resembles the \ac{he}, $\sheunc^{-1}(\theta)$ can easily be compared to the \ac{he} of a \ac{qpd}-segment on a plot.

The \textit{\ac{qpd}-signal tilt-dependent coupling coefficient for uncorrelated noises} $\sheunc(\theta)$ captures the whole angular dependence of the uncorrelated noise sources affecting the \ac{qpd}; it depends, of course, on the angle $\theta$, as well as on all the geometrical parameters used to calculate the \ac{he}. The dependency of $\sheunc^{-1}(\theta)$ on the interferometric configuration's geometric parameters is similar to that of the \ac{he}; however the term in the square root gives additional relevance to the effective-spot-radius-normalized wavefront curvature mismatch $\rho$. On top of the already argued decrease of the \ac{he} due to wavefront curvature mismatch, $\sheunc(\theta)$ also takes into account that, if $\rho \neq 0$ and in the presence of a tilt, the heterodyne efficiencies of the individual \ac{qpd} segments take different values. This implies, looking at \Cref{fig:heff_QPD_inf}, that the phase noise of the \ac{qpd} segments with the lowest \ac{he} will contribute more to the overall phase noise than those with the highest \ac{he}. Such an effect becomes more noticeable when $\rho$ takes particularly large values, such as $\rho=0.6$.

As a function, $\sheunc^{-1}(\theta) \in [0, 1]$ as the \ac{he}. It must be symmetric about $\theta=0$ even in the presence of a wavefront curvature mismatch, and must therefore have a maximum at $\theta=0$. Note that $\theta=0$ implies $\alpha=0$, hence the argument of the square root in \cref{eq:heff-QPD-inf-top,eq:heff-QPD-inf-bottom} is zero and the expression simplifies to the \ac{he} of an infinite \ac{sepd} at $\theta=0$, or $\sheunc^{-1}(\theta=0) = \she_\text{SEPD}(\theta=0)$. Also, in the case of $\rho=0$, this implies $\Im[\alpha]=0$, and hence \cref{eq:eta-eff-unc} simplifies to the \ac{he} of an infinite \ac{qpd} reported in \cref{eq:hef-QPD-inf-rho0}. 

\begin{figure*}[!htpb]
    \begin{subfigure}{.45\textwidth}
        \centering
        \includegraphics[width=\columnwidth]{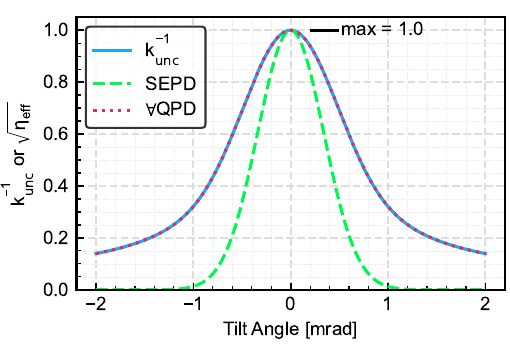}  
        \caption{$\rho=0$}
        \label{fig:het-eff-eff-a}
    \end{subfigure}
    \begin{subfigure}{.45\textwidth}
        \centering
        \includegraphics[width=\columnwidth]{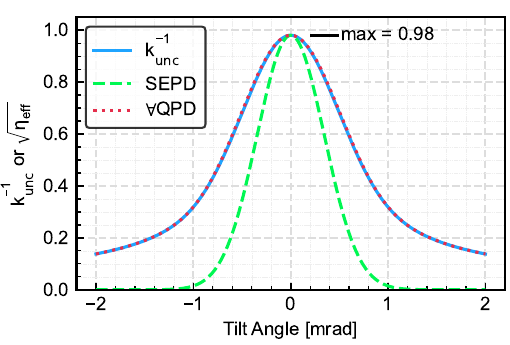}  
        \caption{$\rho=0.2$}
        \label{fig:het-eff-eff-b}
    \end{subfigure}
        \begin{subfigure}{.45\textwidth}
        \centering
        \includegraphics[width=\columnwidth]{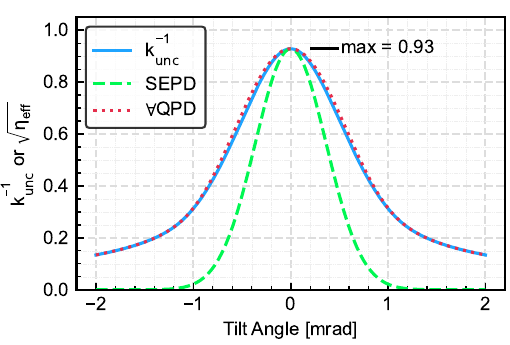}  
        \caption{$\rho=0.4$}
        \label{fig:het-eff-eff-c}
    \end{subfigure}
    \begin{subfigure}{.45\textwidth}
        \centering
        \includegraphics[width=\columnwidth]{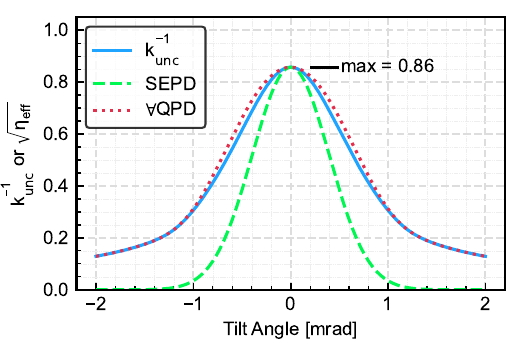}  
        \caption{$\rho=0.6$}
        \label{fig:het-eff-eff-d}
    \end{subfigure}
    \caption{Plot of the reciprocal of \ac{qpd}-signal tilt-dependent coupling coefficient for uncorrelated noises in \cref{eq:eta-eff-unc} for an infinite \ac{qpd} as a function of the beam tilt, compared with the \acp{he} of an infinite \ac{sepd} in \cref{eq:het_eff_SEPD-inf} and the whole-\ac{qpd} \ac{he} of an infinite $\she_{\forall\text{QPD, }\infty}$ \ac{qpd} in \cref{eq:heff-QPD-inf-tot}. The used beam parameters are $w_r = w_m = \weff=1$\,mm, and an infinite \ac{qpd} radius, and four different values of the effective-spot-radius-normalized wavefront curvature mismatch parameter $\rho$. This function is meant to capture the entire angular dependence of the uncorrelated noise sources affecting the \ac{qpd} signals. Note that a smaller value of $\sheunc^{-1}(\theta)$ corresponds to a higher noise coupling.}
    \label{fig:het-eff-eff}
\end{figure*}

A plot of $\sheunc^{-1}(\theta)$ for an infinite \ac{qpd} is shown in \Cref{fig:het-eff-eff} for four values of the effective-spot-radius-normalized wavefront curvature mismatch parameter $\rho$. This plot shows that $\sheunc^{-1}(\theta)$ is very close to the whole-\ac{qpd} \ac{he} $\she_{\forall \text{QPD, }\infty}(\theta)$, and the two are identical for $\rho=0$. On the other hand, $\sheunc^{-1}(\theta)$ is very different from the \ac{he} of a \ac{sepd}; therefore, we stress that the modelling of phase readout noise in \ac{lisa} must treat \ac{qpd} segments separately. An increase of $\rho$ causes, together with a higher coupling due to a decrease of the maximum achievable \ac{he}, a small discrepancy to arise between the whole-\ac{qpd} \ac{he} and $\sheunc^{-1}(\theta)$. \Cref{fig:het-eff-eff} evidences how, in the presence of an effective-spot-radius-normalized wavefront curvature mismatch of $\rho=0.6$, even in the absence of tilts, $\sheunc^{-1}(\theta=0)$ is at most 0.857. This corresponds to an increase in the uncorrelated noise coupling of $\sim 16.7\%$. To the \ac{lisa} collaboration's knowledge, the highest expected value of $\rho$ in \ac{lisa}'s interferometers is $\rho=0.469$ \cite[Table III]{alvise_2024}; if \ac{lisa} was to be built in such configuration, this would cause an increase of interferometric readout phase noise of $\sim10\%$. This could be easily avoided by better mode matching the beams in both the \ac{tmi} and \ac{isi}. Furthermore, in \Cref{fig:het-eff-eff-d}, one can see how, at a tilt angle of $\pm$ 0.5\,mrad, $\sheunc^{-1}(\theta)$ is relatively $\sim8\%$ lower than the whole-\ac{qpd} \ac{he}, implying increased noise coupling in a tilt range likely to happen in \ac{lisa}.

\subsection{Correlated noises} \label{ssection:qpd-corr-noise}
The relative intensity variations in the laser beams influence all \acp{qpd} and \ac{qpd}-segments in the same way; therefore, \ac{rin} are correlated among the segments of a \ac{qpd} \cite{RIN_lennart}. A similar argument can be made for straylight, as long as this impinges identically on all segments of the \ac{qpd}; as this is not a general case, we leave this out and focus solely on \ac{rin}. 

The phase noise for summation and subtraction of perfectly correlated phase noises is derived in \Cref{appendix:corr-phase-noise} in \cref{eq:asd-final}, and is given in \cref{eq:lps-corr,eq:dws-corr}
\begin{align}
    \tilde \phi_\text{AP, cor}(\theta)  &= \frac{1}{2} \sqrt{\tilde \phi_\text{top}^2(\theta) + \tilde \phi_\text{btm}^2(\theta) + 2 \tilde \phi_\text{top}(\theta) \tilde \phi_\text{btm}(\theta) \cos \Big(\bar \phi_\text{top}(\theta) - \bar \phi_\text{btm}(\theta) \Big)}, \label{eq:lps-corr}\\
    \widetilde{ \text{DWS} }_\text{cor}(\theta)    &= \sqrt{\tilde \phi_\text{top}^2(\theta) + \tilde \phi_\text{btm}^2(\theta) - 2\tilde \phi_\text{top}(\theta) \tilde \phi_\text{btm}(\theta) \cos \Big(\bar \phi_\text{top}(\theta) - \bar \phi_\text{btm}(\theta) \Big)}, \label{eq:dws-corr}
\end{align}
where $\bar \phi$ is the average phase, defined assuming that $\tilde \phi \ll 2\pi$. Note that it depends also on the difference of the average phases: $\tilde \phi_\text{AP, cor}$ is minimum if $\bar \phi_\text{top} - \bar \phi_\text{btm} = \pi + 2n\pi$, while $\widetilde{ \text{DWS} }_\text{cor}$ is minimum for $\bar \phi_\text{top} - \bar \phi_\text{btm} = 0 + 2n\pi$, $n \in \mathbb{Z}$. 

As a general experimental remark, note that the average phases are the resulting phases at the output of the measurement chain. They, therefore, include also eventual phase offsets originating along the analog- and phasemeter-signal processing of each \ac{qpd} segment. As these chains generally present slightly different phase \acp{tf}, this results in different phase outputs for the same optical input. Such issue is related to the \ac{dws}-bias mentioned in \cite[Figure 3.18]{me_thesis}. Such biases are, for simplicity, neglected in this model, and only average phase variations due to beam tilt and \ac{qpd} geometry are considered. With this assumption, the average phase difference $\bar \phi_\text{top}(\theta) - \bar \phi_\text{btm}(\theta)$ that appears as the argument of the cosine is simply the vertical \ac{dws} signal introduced in \cref{eq:2seg-dws}. This can therefore be expressed as a function of the beam's tilt as demonstrated in \cite[eqs. (29, 30)]{alvise_2024}, provided that the tilt is sufficiently small. We name this quantity $\text{DWS}(\theta)$.

By expressing \cref{eq:lps-corr,eq:dws-corr} using the $\Upsilon_{\text{RIN-}1f}$ parameter defined in \cref{eq:noise-heff-norm}, the average phase noise and \ac{dws} noise can be rewritten as
\begin{align}
\tilde \phi_\text{AP, cor}(\theta) &= \frac{1}{2} \, \Upsilon_{\text{RIN-}1f} \, \sqrt{2} \, \shecorS(\theta), \label{eq:noise-AP}\\
\shecorS(\theta) &= \frac{1}{\sqrt{2}} \sqrt{\frac{1}{\eta_\text{top}(\theta)}+\frac{1}{\eta_\text{btm}(\theta)} + \frac{2}{ \sqrt{\eta_\text{top}(\theta) \eta_\text{btm}(\theta)} } \cos \Big(\text{DWS}(\theta) \Big)}
\label{eq:eta-eff-corS},\\
\widetilde{ \text{DWS} }_\text{cor}(\theta) &= \Upsilon_{\text{RIN-}1f} \, \sqrt{2} \, \shecorD(\theta), \label{eq:noise-DWS}\\
\shecorD(\theta) &=  \frac{1}{\sqrt{2}} \sqrt{\frac{1}{\eta_\text{top}(\theta)}+\frac{1}{\eta_\text{btm}(\theta)} - \frac{2}{ \sqrt{\eta_\text{top}(\theta) \eta_\text{btm}(\theta)} } \cos \Big(\text{DWS}(\theta)\Big)},\label{eq:eta-eff-corD}
\end{align}
where the \textit{\ac{qpd}-signal tilt-dependent coupling coefficient for correlated noise for sum} $\shecorS(\theta)$ \textit{and difference} $\shecorD(\theta)$ were introduced in \cref{eq:eta-eff-corS,eq:eta-eff-corD}, respectively. Note that \cref{eq:noise-DWS} matches the equation reported for $1f$-\ac{rin} in \cite[Table I]{RIN_lennart} in the case of  $\eta_\text{top}=\eta_\text{btm}$.

By substituting the relations in \cref{eq:heff-QPD-inf-top,eq:heff-QPD-inf-bottom} in \cref{eq:eta-eff-corS,eq:eta-eff-corD}, one can get an expression for $k_{\text{cor, }\Sigma/\Delta}^{-1}(\theta)$ for an infinite \ac{qpd} depending explicitly on the beam parameters:
    \begin{equation}
    k_{\text{cor, }\Sigma/\Delta}^{-1}(\theta) = \frac{1}{\sqrt{1+\rho^2}} \frac{2 w_r w_m}{w_r^2 + w_m^2} \exp\left( -\frac{k^2 w_\text{eff}^2 \theta^2}{8(1+\rho^2)} \right) \sqrt{\frac{ \left(1 + \left|\alpha \right|^2 \right)^2 - 4 \left(\Im \left[ \alpha \right] \right)^2}{1 + \left| \alpha \right|^2 \pm \sqrt{ \left(1 + \left|\alpha \right|^2 \right)^2 - 4 \left(\Im \left[ \alpha \right] \right)^2} \cos(\text{DWS}(\theta))} }, \label{eq:eta-eff-cor-simp}
    \end{equation}
where the + and the - are meant to be used for sum and difference, respectively.

As the \ac{qpd}-signal tilt-dependent coupling coefficient for uncorrelated noises, these two functions capture the whole angular dependence of the correlated noise sources affecting the \ac{qpd} readout, which in this case comprises both the angular variation of the \ac{he} and noise correlation between the \ac{qpd}-segments. \Cref{eq:eta-eff-corS,eq:eta-eff-corD} are rather complicated, and further complex than \cref{eq:eta-eff-unc}. 

We now take a closer look at two particular situations of the derived coupling coefficient. The two coupling factors greatly simplify if either there is perfect beam mode-matching $\rho=0$, or in the case of no tilt $\theta=0$. In the first case, the \acp{he} of both the top and bottom segments coincident; we hence define $\she_\text{QPD, top}(\theta, \rho=0) = \she_\text{QPD, btm}(\theta, \rho=0) = \she_\text{QPD}(\theta)$. Furtheremore, for an infinite \ac{qpd}, $\Im[\alpha]=0$, and \cref{eq:eta-eff-cor-simp} simplifies to 
\begin{equation}
k^{-1}_{\text{cor, }\Sigma/\Delta}(\theta) = \frac{2 w_r w_m}{w_r^2 + w_m^2} \exp\left( -\frac{k^2 w_\text{eff}^2 \theta^2}{8} \right) \sqrt{\frac{1 + \alpha^2}{1 \pm \cos(\text{DWS}(\theta))}}, \label{eq:eta-eff-cor-rho0}
\end{equation}
while $k^{-1}_{\text{cor, }\Sigma/\Delta}(\theta)$, for a generic size \ac{qpd}, simplify to 
\begin{align}
\shecorS(\theta) &= \frac{\sqrt{2}}{\she_\text{QPD}(\theta)}\cos\left( \frac{1}{2} \text{DWS}(\theta)\right),\\
\shecorD(\theta) &= \frac{\sqrt{2}}{\she_\text{QPD}(\theta)}\sin\left( \frac{1}{2} \text{DWS}(\theta)\right),
\end{align}
in accordance with the result in \cite[eq. (51)]{RIN_lennart} and \cite[eqs. (4-7)]{PhysRevApplied.22.044048}. In the second case, for $\theta=0 \rightarrow \alpha =0$ this results, independently of $\rho$, in \begin{align}
    \shecorS(\theta=0) &= \frac{\sqrt{2}}{\she_\text{QPD}(\theta=0)},\\
    \shecorD(\theta=0) &= 0,
\end{align}
where we remind that $\text{DWS}(\theta=0) = 0$ \cite{alvise_2024}.

\begin{figure*}[!htpb]
    \centering
    \begin{subfigure}{.45\textwidth}
        \centering
        \includegraphics[width=\columnwidth]{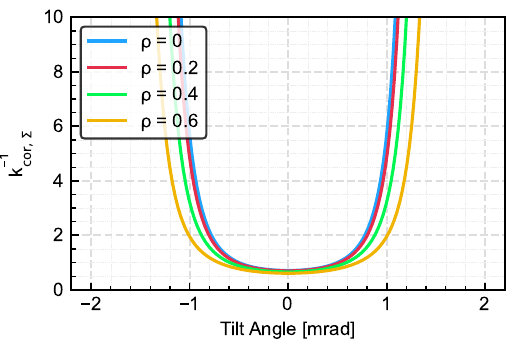}  
        \caption{$\shecorS^{-1}(\theta)$}
        \label{fig:heff-SD-a}
    \end{subfigure}
    \begin{subfigure}{.45\textwidth}
        \centering
        \includegraphics[width=\columnwidth]{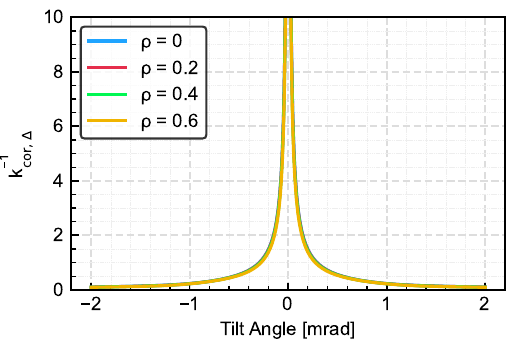}  
        \caption{$\shecorD^{-1}(\theta)$}
        \label{fig:heff-SD-b}
    \end{subfigure}
    \caption{Plot of the reciprocal of the \ac{qpd}-signal tilt-dependent coupling coefficient for correlated noises for sum $\shecorS^{-1}(\theta)$ (a) and difference $\shecorD^{-1}(\theta)$ (b) for an infinite \ac{qpd} from \cref{eq:eta-eff-corS,eq:eta-eff-corD}. The used beam parameters are $\weff=1$\,mm, and an infinite \ac{qpd} radius, and four different values of the effective-spot-radius-normalized wavefront curvature mismatch parameter $\rho$. Note that a smaller value of $k^{-1}_{\text{cor, }\Sigma/\Delta}(\theta)$ corresponds to a higher noise coupling; therefore, $k^{-1}_{\text{cor, }\Sigma/\Delta}(\theta)$ diverge when the correlated noise are expected to perfectly cancel. This happens at $\theta=0$ for $\shecorD^{-1}(\theta)$ and roughly at $\pm 1.2$\,mrad for $\shecorS^{-1}(\theta)$ for the parameters used in this plot.}
    \label{fig:heff-SD}
\end{figure*}

A plot of $\shecorS^{-1}(\theta)$ and $\shecorD^{-1}(\theta)$ for an infinite \ac{qpd} is shown in \Cref{fig:heff-SD}. As one can see, these functions present a fundamental difference with respect to the \ac{qpd}-signal tilt-dependent coupling coefficient for uncorrelated noises: they diverge at some specific tilt values. This is due to the fact that at those specific angles, these noises are anticorrelated and therefore cancel each other out. At such angles, the coupling of the correlated noise sources tends to zero, and therefore $k^{-1}_{\text{cor, }\Sigma/\Delta}(\theta)$ tend to infinity.

 A plot of the two functions $\shecorD^{-1}(\theta)$ and $\shecorD^{-1}(\theta)$ is shown in \Cref{fig:heff-SD-a,fig:heff-SD-b}, respectively, for four values of the effective-spot-radius-normalized wavefront curvature mismatch parameter $\rho$. These two manifest opposite angular behaviour, as $\shecorD^{-1}(\theta)$ diverges for $\theta=0$, whereas $\shecorD^{-1}(\theta)$ diverges for $\theta = \pm \pi$. \ac{lisa} operates close to $\theta=0$, meaning that \ac{rin} has a very small coupling to the \ac{dws} signal, while there is little to no mitigation for the \ac{lps}. In particular, a value of $\rho=0.6$ causes an increase of $16.6\%$ in the correlated noise in comparison to the case $\rho=0$, a situation close to that of the \ac{qpd}-signal tilt-dependent coupling coefficient for uncorrelated noises $\sheunc^{-1}(\theta)$. Furthermore, a higher value of $\rho$ broadens the angular range where $\shecorS^{-1}(\theta)$ does not diverge, effectively increasing the noise coupling in the whole operational angular range. In \Cref{fig:heff-SD-b}, $\shecorD^{-1}(\theta)$ is less affected by the presence of an effective-spot-radius-normalized wavefront curvature mismatch. Still, this causes a higher coupling of correlated noise sources. In general, scenarios with lower $\rho$ manifest lower optical noise.

\subsection{Total QPD readout noise} \label{ssection:qpd-tot-noise}
Combining together the results of subsections~\ref{ssection:qpd-uncorr-noise} and ~\ref{ssection:qpd-corr-noise}, the total readout noise of the \ac{qpd}-signals is recovered in \cref{eq:lps-eff,eq:dws-eff}, where we report this with both $\theta$ and $f_\text{het}$ dependency. The angular dependence of $2f$-\ac{rin}, which does not depend on the \ac{he}, is taken from \cite{RIN_lennart,PhysRevApplied.22.044048}. This final expression is a constructed using the coupling factors mentioned above and the individual phase noises for a single \ac{qpd}-segment.
\begin{align}
\tilde \phi_\text{AP}(f_\text{het}, \theta) &= \frac{1}{2}\sqrt{ 2\left(\Upsilon_\text{shot}^2 + \Upsilon_\text{en}^2(f_\text{het})\right) \, \sheunc^2(\theta) + 2 \Upsilon_{\text{RIN-}1f}^2(f_\text{het}) \, \shecorS^2(\theta) + 4 \tilde \phi_{\text{RIN-}2f}^2(f_\text{het}) \, \cos ^2 \big( \text{DWS} (\theta) \big)} \label{eq:lps-eff}\\
\widetilde{ \text{DWS}}(f_\text{het}, \theta) &= \sqrt{ 2\left(\Upsilon_\text{shot}^2 + \Upsilon_\text{en}^2(f_\text{het})\right)\, \sheunc^2(\theta) + 2 \Upsilon_{\text{RIN-}1f}^2(f_\text{het}) \, \shecorD^2(\theta) + 4 \tilde \phi_{\text{RIN-}2f}^2(f_\text{het}) \, \sin ^2 \big( \text{DWS} (\theta) \big)  } \label{eq:dws-eff}
\end{align}

\Cref{eq:eta-eff-corS,eq:eta-eff-corD} state that, in the calculation of the overall readout phase noise of \ac{qpd} signals as a function of the beam tilt, one must consider the \ac{qpd} as a whole, as the individual segments manifest different behaviours as a function of the tilt angle. The only exception to this is when $\rho=0$; then, all segments will present the same \ac{he}, as the symmetry presented in \Cref{fig:wavefront-curvature-interpret} is not broken. The functions $\sheunc(\theta)$, $\shecorS(\theta)$, $\shecorD(\theta)$ collect the whole angular dependence of the \ac{qpd} signals' noise, which is related both to the geometry of the interferometric topology and to the correlation of the noise sources, while the $\Upsilon_{\text{shot/en/ RIN-}1f}$ parameters collect the information about beam power and noise magnitude. 

In other words, in the case of perfectly matched beams, all segments manifest the same \ac{he} for all tilts; one can hence calculate the phase readout noise based solely on the \ac{he} of one segment. In the presence of wavefront curvature mismatch, this does not hold anymore, and the phase readout noise cannot be calculated exactly based only on the \ac{he} of one single segment. In this scenario, the segments have to be analyzed separately and their individual phase noises combined. We also stress that there is a substantial difference between the phase noise of a \ac{qpd}'s output and that of a \ac{sepd}. This is due to their different \acp{he}. 

The magnitude of the impact of this novel approach in the phase noise modelling of \ac{qpd}-signals depends on the value of the wavefront curvature mismatch; for small values of $\rho$, modelling all segments as having the same \ac{he} is indeed a good approximation; otherwise, individual dedicated modelling is to be preferred. A benchmark value of $\rho>0.2$ can be set to discriminate between these two cases. We further remark that this is a simplified model accounting only for vertical tilts. The predicted noise increase due to wavefront curvature mismatches becomes more significant in the presence of horizontal tilts, as this further increases the asymmetry between the measured power.

\section{LISA noise budget operating with tilted TM} \label{section:tm-tilt}
The performance of \ac{lisa} is affected, among other noises, by the alignment precision of the components on the \ac{lob}. Static misalignments that occur during the build of the \ac{ob} due to the imperfect positioning of optical components lead to departures from the nominal interferometric topology, thereby enhancing additive interferometric noise couplings. Such effects cannot be fully investigated experimentally on ground before the mission, and have to be mitigated in flight.

Let's suppose, for instance, that a static misalignment affects the \ac{lo} beam in the \ac{tmi}; this beam impinges at the center of the \acp{qpd}, but with a small tilt, of the order of a few tenths of \unit{\micro \radian}. At the same time, the Tx beam sits in its nominal position, impinging orthogonally on the \acp{qpd}. The result is a lower \ac{he}, and an enhancement of all \ac{he} dependent noise sources.

As the \ac{lob} lacks beam-angle correction mechanisms for the \ac{lo} beam, such an issue cannot be corrected. One possible solution is to make the Tx beam "follow" the misaligned \ac{lo} beam, thereby recovering an optimal \ac{he}. This can be achieved by intentionally tilting the \ac{tm}, which is the only actuatable mirror on the Tx beam's path. This means operating the \ac{tm} at a small tilt angle, or the order of a few\,\unit{\micro \radian}, such that the Tx beam, reflecting on the \ac{tm}, is made parallel to the \ac{lo} beam. Such an operation point is named \textit{\ac{dws}-zero}. 
In contrast, the operation without correction for residual beam tilt is called \textit{\ac{dws}-offset}. Such a method improves the optical noise balance, as the \ac{he} of the two beams increases; however, it also increases the \ac{ttl}-coupling, as tilting the \ac{tm}, which would otherwise sit in its optimal position, couples its lateral jitter into the interferometrically measured \ac{lps}. Such an approach was successfully used in \ac{lisa}-Pathfinder \cite{Wanner_2017, gudrun_thesis}, where the maximum tilt of the reference \ac{tm} tilt was of the order of 60\,\unit{\micro \radian} \cite[Table VII]{PhysRevD.108.102003}, translating to a beam tilt of 120\,\unit{\micro \radian} at \ac{qpd}-level.

This section aims to check whether operating at \ac{dws}-zero is also reasonable for \ac{lisa}. To answer this question, a detailed comparison of the two noise couplings -- additive interferometric noise and \ac{ttl} coupling -- is needed, in order to find the optimal operating point. As we lack the expertise to derive additional \ac{ttl} coupling due to \ac{tm} tilt, this section addresses only the optical noise considerations, which are linked to the \ac{he} as a function of beam tilt, which was extensively modelled in this article. We hope that the results of this section can help answer the full matter.

We approach this question not by examining how the tilts of the measurement beam affect the interferometric noise in the \ac{tmi}, which, neglecting \ac{qpd} tilts, yields a similar answer: the performance degradation that a measurement beam causes due to a tilt is hence assumed to be the improvement that "following" the \ac{lo} beam's tilt can achieve. We argue that \ac{qpd} tilts can be neglected as long as the \ac{qpd}'s radius is sufficiently larger than the beams' spot size; we show that this is the case in the \ac{tmi}.

\begin{figure}[!htpb]
    \centering
    \includegraphics{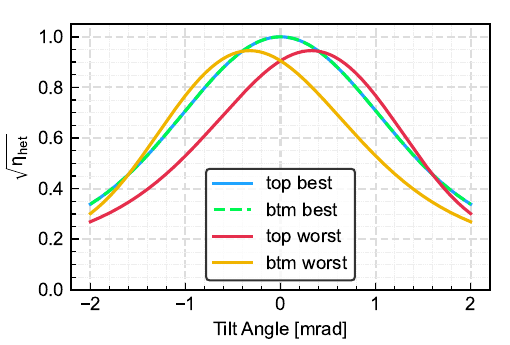}
    \caption{Plot of the best- and worst-case scenario \acp{he} of the \ac{tmi} in \ac{lisa}. Note that the $x$ axis is the Tx beam's tilt angle; the \ac{tm}'s angle can be obtained by rescaling the $x$ axis by a factor of $m_\text{IS}/2$, where $m_\text{IS}$ is the magnification of the imaging system in \ac{lisa}. As previously argued, the top and bottom \acp{he} are coincident in the best-case scenario, as $\rho=0$.}
    \label{fig:tmi-heff}
\end{figure}

We estimate here the predicted increase in optical noise at small angles of the Tx beam. Give the tilt range in \ac{lisa}-Pathfinder, we can hence reasonably suppose a \ac{lo} beam tilt of the order of 100\,\unit{\micro \radian} in \ac{lisa}, assuming once more only vertical tilts and consequently two \ac{qpd} halves. First, we characterize the \ac{he} of the \ac{tmi}: this depends on the beam and \ac{qpd} sizes. These values are reported in \cite[Section VI]{alvise_2024}, and are:
\begin{itemize}
    \item The Gaussian beam's (hence the Tx and \acs{lo}) parameters lie in the range $0.34 \text{ mm} < w_0 < 0.46$\,\unit{\milli \meter} and $|z_0| < 160$\,\unit{\milli \meter}, with $z_0 = 0$ \unit{\milli \meter} indicating the position of the \ac{qpd}. The two beams can be considered independent.
    \item The \ac{qpd}'s diameter is 1.5 mm. There are 20 \unit{\micro \meter} gaps between the segments \cite{eyes_of_lisa}. These gaps are approximately as sensitive as the \ac{qpd}'s active area, hence the \ac{qpd} can be considered as having no gaps.
\end{itemize}
From the latest \ac{lisa} design, a range of possible \acp{he} can be estimated. These can be split into best- and worst-case scenarios using \cref{eq:het_eff_SEPD-inf}. The best-case scenario corresponds to a perfect mode match, hence a minimum value of $|\rho|=0$ and $w_\text{Tx}=w_\text{LO}$, and a minimum value of \weff, to minimize the degradation of the \ac{he} as a function of the tilt. The worst-case scenario is more difficult to determine, as both the beam-size mismatch and the wavefront-curvature mismatch contribute to degrading the \ac{he}. Numerical comparisons show that the wavefront-curvature mismatch is the leading contributor; hence, the worst-case scenario is given by the maximum value of $|\rho|=0.469$ \cite{alvise_2024}; such value of $\rho$ is also reached when the waist size $w_0$ of the two \acp{gb} is minimum. The resulting interferometric parameters are listed in \Cref{tab:bw-parameters}. Note that these best- and worst-case scenarios for the \ac{he} do not exactly correspond, but are close, to the best- and worst-case scenarios for the \ac{dws} signal reported in \cite{alvise_2024}. Note that the values of $\beta = \sqrt{2}r_\text{QPD}/\weff$ are relatively large, hence justifying an infinite-\ac{qpd} approximation. This leads to the \acp{he} plotted in \Cref{fig:tmi-heff} for the two \ac{qpd} halves.

\begin{table}[!htpb]
    \centering
    \begin{tabular}{cc|cc}
    \hline
    Parameter & Unit  &  \multicolumn{2}{c}{Value}\\
              &       &  Best-case & Worst-case\\
    \hline  
    \weff     & mm  & 0.34 & 0.376\\
    $\rho$    & -     & 0    & 0.469\\
    $\beta$   & -     & 3.12 & 2.82\\
    \hline
    \end{tabular}
    \caption{Relevant parameters of the \ac{tmi}'s geometry in the best- and worst-case scenarios.}
    \label{tab:bw-parameters}
\end{table}

\begin{figure*}[!htpb]
    \centering
    \begin{subfigure}{.45\textwidth}
        \centering
        \includegraphics[width=\columnwidth]{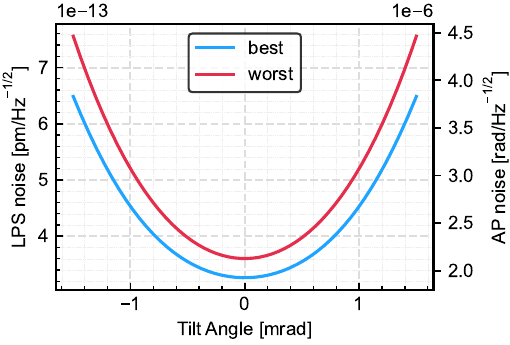}  
        \caption{\ac{ap} and \ac{lps} noise }
        \label{fig:LISA-noise-a}
    \end{subfigure}
    \begin{subfigure}{.45\textwidth}
        \centering
        \includegraphics[width=\columnwidth]{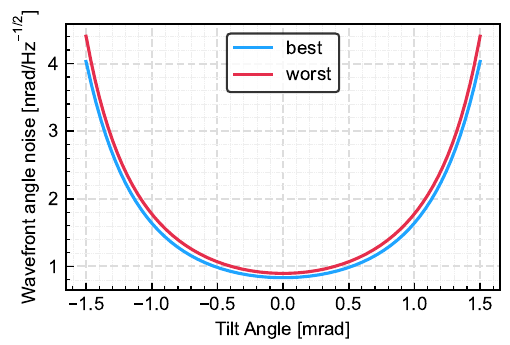}  
        \caption{\ac{wfa} noise}
        \label{fig:LISA-noise-b}
    \end{subfigure}
    \caption{Calculated \ac{lps} and \ac{wfa} noise as a function of the Tx beam's tilt angle for the \ac{tmi} in \ac{lisa}. These curves are based on equations \cref{eq:lps-eff,eq:dws-eff}, which assume an infinite \ac{qpd}. This approximation is justified by the values of $\beta$ in \Cref{tab:bw-parameters}, which are larger enough than one. The used noise input parameters are reported in \Cref{tab:LISA-TMI-parameters}. \Cref{fig:LISA-noise-a} also reports the \ac{ap} noise on a second y-axis; this is easily done, as the two y-axis are linearly related. The same does not hold for \ac{dws} and \ac{wfa}. In fact, note that the \ac{wfa} noise rapidly increases for tilts larger than 1\,mrad, as the \ac{dws} gain tends to zero.}
    \label{fig:LISA-noise}
\end{figure*}

From these estimations, a theoretical phase noise model can be derived. A comparison between the phase noise in these two cases is shown in \Cref{fig:LISA-noise}, which shows the \ac{ap} and \ac{ap}-\ac{lps} noise on the left, and \ac{wfa} noise on the right. The noise parameters for \ac{lisa}, which were used in this calculation, are reported in \Cref{tab:LISA-TMI-parameters}. 

First, the most noticeable result in \Cref{fig:LISA-noise} is that the \ac{wfa} noise sharply increases for tilts larger than 1\,mrad, due to the reduction of the \ac{dws} gain. Secondly, in the absence of tilts, the worst-scenario configuration results in a noise increase of $10.4\%$ in \ac{lps} and $7.7\%$ \ac{wfa} noise; we therefore recommend, once more, good mode matching between the interfering beams. Third, in the tilt range of $\pm 100$\,\unit{\micro \radian}, a range representative of the correction needed in \ac{lisa}-Pathfinder \cite[Table VII]{PhysRevD.108.102003}, the noise increase is modest. \Cref{tab:LISA-TMI-LPS-noise,tab:LISA-TMI-DWS-noise} quantitatively report the noise increase as a function of the tilt angle for \ac{lps} and \ac{wfa}, respectively. 

This result shows that beam tilts of the order of a few \unit{\micro \radian} cause very modest increases in the noise, with relative performance degradation of the order of one part per thousand. This noise increase corresponds to the improvement that operating at \ac{dws}-zero can hope to achieve. In comparison to such a small improvement, the \ac{ttl} of the \ac{tm} surely worsens; although this improvement has not been compared with the increase of \ac{ttl} noise, we think that it is small enough to argue that operation at a \ac{dws} offset is preferable. Further checks should still be performed comparing the interferometric and \ac{ttl} performances. 

Experimental measurements of the \ac{dws} performance as a function of beam tilt have already been performed in \ac{tdobs} and are in the process of publication to consolidate this model. 

\begin{table}[!htpb]
    \centering
    \begin{tabular}{c|cc}
    \hline
    Parameter             & Value & Units\\
    \hline
    $P_\text{Tx}$         & \sn{1.5}{-4} & \unit{\watt}\\
    $P_\text{LO}$         & \sn{8.3}{-7} & \unit{\watt}\\
    $\eta_\text{PD}$      &  0.93  & -\\
    $A$                   &  0.80  & \unit{\ampere \per \watt}\\
    $\tilde i_\text{en}(f_\text{het})$  &  2     & \unit{\pico \ampere \per \hertz^{1/2}}\\
    $\tilde r$            &  \sn{3}{-8}     & \unit{ 1 \per \hertz^{1/2}}\\
    \hline
    \end{tabular}
    \caption{Parameters of the \ac{lisa} mission used to calculate the interferometric noise. For simplicity, the electronic noise of the \ac{qpr} $\tilde i_\text{en}$ is set to be frequency independent and equal to the requirement. Source: \cite{me_thesis}.}
    \label{tab:LISA-TMI-parameters}
\end{table}

\begin{table*}
    \centering
    \begin{tabular}{c|cc|cc}
    \hline
    Tilt Angle [\unit{\micro \radian}]   & \multicolumn{2}{c|}{Best-case scenario} & \multicolumn{2}{c}{Worst-case scenario}\\
      & \ac{lps} noise [\unit{\pico \meter \per \hertz^{1/2}}]& Relative Increase & \ac{lps} noise [\unit{\pico \meter \per \hertz^{1/2}}] & Relative Increase\\
    \hline
	5  &  \sn{3.2633}{-1}  &  \sn{8.4304}{-6}  &  \sn{3.6043}{-1}  &  \sn{1.0033}{-5}\\
	10  &  \sn{3.2634}{-1}  &  \sn{3.3722}{-5}  &  \sn{3.6044}{-1}  &  \sn{4.0133}{-5}\\
	20  &  \sn{3.2637}{-1}  &  \sn{1.3489}{-4}  &  \sn{3.6048}{-1}  &  \sn{1.6054}{-4}\\
	50  &  \sn{3.2660}{-1}  &  \sn{8.4338}{-4}  &  \sn{3.6079}{-1}  &  \sn{1.0036}{-3}\\
	100  &  \sn{3.2743}{-1}  &  \sn{3.3777}{-3}  &  \sn{3.6187}{-1}  &  \sn{4.0180}{-3}\\
    \hline
    \end{tabular}
    \caption{\ac{ap}-\ac{lps} noise and relative noise increase as a function of the beam tilt angle for \ac{lisa}'s \ac{tmi}. This calculation is based on \cref{eq:lps-eff}, using the parameters in \Cref{tab:bw-parameters} for the best- and worst-case scenarios, and on \Cref{tab:LISA-TMI-parameters} for the \ac{tmi}'s parameters. These values in the table correspond to \Cref{fig:LISA-noise-a}. The relative noise increase is stated with respect to the no-tilt case.}
    \label{tab:LISA-TMI-LPS-noise}
\end{table*}

\begin{table*}
    \centering
    \begin{tabular}{c|cc|cc}
    \hline
    Tilt Angle [\unit{\micro \radian}]   & \multicolumn{2}{c|}{Best-case scenario} & \multicolumn{2}{c}{Worst-case scenario}\\
      & \ac{wfa} noise [\unit{\nano \radian \per \hertz^{1/2}}]& Relative & \ac{wfa} noise [\unit{\nano \radian \per \hertz^{1/2}}] & Relative \\
    \hline
	5  &  \sn{8.3153}{-1}  &  \sn{1.6796}{-5}  &  \sn{8.9543}{-1}  &  \sn{1.7825}{-5}\\
	10  &  \sn{8.3157}{-1}  &  \sn{6.7186}{-5}  &  \sn{8.9548}{-1}  &  \sn{7.1300}{-5}\\
	20  &  \sn{8.3174}{-1}  &  \sn{2.6877}{-4}  &  \sn{8.9567}{-1}  &  \sn{2.8522}{-4}\\
	50  &  \sn{8.3292}{-1}  &  \sn{1.6810}{-3}  &  \sn{8.9701}{-1}  &  \sn{1.7837}{-3}\\
	100  &  \sn{8.3712}{-1}  &  \sn{6.7353}{-3}  &  \sn{9.0181}{-1}  &  \sn{7.1451}{-3}\\
    \hline
    \end{tabular}
    \caption{\ac{wfa} noise and relative noise increase as a function of the beam tilt angle for \ac{lisa}'s \ac{tmi}. This calculation is based on \cref{eq:dws-eff}, using the parameters in \Cref{tab:bw-parameters} for the best- and worst-case scenarios, and on \Cref{tab:LISA-TMI-parameters} for the \ac{tmi}'s parameters. The values in the table correspond to \Cref{fig:LISA-noise-b}. The relative noise increase is stated with respect to the no-tilt case.}
    \label{tab:LISA-TMI-DWS-noise}
\end{table*}

\section{Summary and conclusion} \label{section:conclusion}
Gravitational spaceborne interferometers aim to measure the variation of distance between free-falling \acp{tm} with a precision of the order of $\pmsqrthz$ in the $10^{-4}-1$\,Hz band. Such performance is challenged by the presence of spacecraft and \ac{tm} jitter. The coupling of the consequent beam jitter into the interferometrically measured pathlength is an extremely relevant aspect in spaceborne interferometers such as \ac{lisa}. To mitigate this issue, while also ensuring constellation and \ac{tm} alignment, \ac{lisa} is implementing an angular readout using a technique named \ac{dws}. This technique, which requires the use of \acp{qpd}, enables \ac{lisa} to both actively control the tilts of the \ac{tm} and subtract the spurious pathlength contributions due to the tilt in post-processing. Therefore, the success of the mission depends on the achieved sensitivity not only of the \ac{lps}, but also of \ac{dws}, with a target of $\nradsqrthz$.

Such sensitivity is significantly affected by the dynamics we expect in \ac{lisa}. The \ac{tmi} and \ac{isi} measurement beams undergo tilts of up to 1\,mrad at the \ac{qpd}, causing a drop in the \ac{he} of up to a factor of five. This reduces the amplitude of the measured signal, consequently decreasing \ac{snr} and increasing readout noise. A correct modelling of how the \ac{he} is affected by tilts is hence crucial to understand the range of tilt motion that can be tolerated whilst maintaining performance.

In this paper, an approximated analytic description of the \ac{he} is proposed in \Cref{section:heff-model}. The resulting expression is capable of analytically deriving the \ac{he} of two beams impinging on infinite \acp{sepd} and \acp{qpd}, and to analytically derive the first few Maclaurin expansion coefficients of the \ac{he} of two beams impinging on a finite \acp{pd}. The references to the derived expressions are collected in \Cref{tab:summary}. The resulting \acp{he} derived by this model are compared to numerical integration, achieving excellent results for small wavefront curvature mismatches. The derivation is general, and can be used to estimate the beam overlap in any interferometric topology featuring beam tilts around the detector's center. These results are applicable to any interferometer, although the models are most relevant for spaceborne interferometers. In particular, the model developed in this paper predicts that, for two beams interfering on a \ac{qpd}, in the presence of a wavefront curvature mismatch, the \ac{he} measured by the individual segment differs; such a feature is reported for the first time in this work.
 
In \Cref{section:heff-imaging-systems}, we derived how the use of imaging systems affects the \ac{he}. In \Cref{section:results}, this model is finally compared to a \ac{he} measurement performed on an ultra-stable testbed described in \Cref{section:setup}. These experimental results confirm the prediction of the model.
 
In \Cref{section:qpd-noise}, the \ac{he} model is used to derive the behaviour of the interferometric readout noise of the \ac{qpd}-signals as a function of the beam tilt. In this context, the noise sources remain constant, whereas the signal amplitude degrades; due to the decrease in \ac{snr}, the phase noise increases solely as a result of the decrease in \ac{he}. The \ac{qpd} signals' readout noise is analyzed in a simplified model which uses only vertical tilts and only two segments. This model, albeit simplified, is able to grasp a new feature of the \ac{qpd} readout, which is induced by the prediction of the analytic \ac{he} model: in the presence of wavefront curvature mismatches, tilts affect the \ac{he} of the two segments differently, leading to different \acp{snr} and additive noise couplings between the two segments. This results in an increase in readout noise compared to a situation where such a difference is neglected. Therefore, in light of such considerations, the \ac{qpd} must be treated as a whole. A general conclusion of this article is to minimize the mode mismatch between the beams in order to improve \ac{snr}.

In \Cref{section:tm-tilt}, we calculate the possible benefits of intentionally tilting the \ac{tm} in the \ac{tmi} to correct for build imperfections of the \ac{lob}. A realignment of the \ac{tm} can reduce interferometric noise coupling, at the cost of increasing the coupling of the \ac{tm}'s lateral jitter to the measured \ac{lps}. We estimate the magnitude of the possible interferometric noise mitigation using the models developed in \Cref{section:heff-model}, which have to be compared to the \ac{ttl} noise increase to obtain an ultimate answer. However, this leads, in our opinion, to negligible gains in interferometric noise. Experimental results confirming this claim are in the process of publication.

We see several possibilities for further expansion and use of this work, which would benefit the \ac{lisa} mission, future space interferometers add potentially also any interferometer. For instance, a good model of the amplitudes of \ac{lisa}'s interferometric signals can also be used, either in postprocessing or in the phasemeter, to monitor and correct for cycle slips in the phasemeter \cite[chapter 4]{francis_thesis}. These occur in the \ac{dpll} when the magnitude of the phase error exceeds 0.5 cycles, driving the loop to find stability one or more cycles apart. In such a case, the phase output of the phasemeter is subject to a jump of $2\pi n$ with $n \in \mathbb{Z}$. This jump would carry over to the \ac{lps} and \ac{dws} signals, being interpreted as a sudden change in pathlength or beam angle, respectively. As we have demonstrated during this article, a real beam tilts also implies a change of \ac{he}\footnote{This is true, unless the tilt simply swaps the sign of the resulting beam angle}. Counter-checking variations of signal amplitude can hence be used to discriminate between real and cycle-slip-originated phase changes. A second use of a \ac{he} model would be to combine it with the \ac{dws} signal. As demonstrated in \cite{alvise_2024}, the \ac{dws} signal is a very powerful \ac{qpd}-signal for monitoring the beam tilt-angle. This technique, however, has a limited range due to the finite size of the used \acp{qpd} and the inevitable presence of wavefront curvature mismatches. These cause the \ac{dws} signal to \textit{wrap} and invert its sign, as soon as it crosses $\pm \pi$ (see \cite[Figure 11]{alvise_2024}). This can also occur for relatively small tilts. On the other hand, the \ac{he} is not subject to periodic jumps, as it is not a phase measurement. Combining the two quantities can, therefore, broaden the range of operation of the \ac{dws} signal beyond its first wrapping.

The model used in this article considers simply two beams, either a pair of \acp{gb} or a \ac{gb}-\ac{th} beam couple. Any interferometer using such beam pairs which is subject to tilts around the detector's center might find this work useful to evaluate the amplitude of the resulting signal. Future space interferometers can benefit from this work to improve their design.

We hope that this model will be beneficial not only to the space-borne interferometry community but also to any optics experiments that are subject to beam tilts.

\setlength{\tabcolsep}{6pt}
\begin{table*}[!htpb]
    \centering
    \begin{tabular}{c|c|cc|cc}
    \hline
    \hline
    Beams                            & Quantity  &  \multicolumn{2}{c|}{\ac{sepd}} & \multicolumn{2}{c}{\ac{qpd}} \\
                                     &                         & infinte                         & finite                 & infinite & finite \\
    \hline
    \multirow{3}{*}{\ac{gb}-\ac{gb}} & \multirow{2}{*}{$\she$} &  \cref{eq:het_eff_SEPD-inf}     & \cref{eq:heff-fin-sepd}& \cref{eq:heff-QPD-inf-top,eq:heff-QPD-inf-bottom}  & \cref{eq:heff-QPD-fin}\\
                                     &                         &  \Cref{fig:heff_SEPD_inf}       & \Cref{fig:het-eff-SEPD-fin,fig:heff_GB_3betas,fig:heff-SEPD-fin-t0t2} & \Cref{fig:heff_QPD_inf} & \Cref{fig:heff-QPD-fin,fig:het-eff-QPD-fin} \\
    \cline{2-6}
                                     & $\phe$                  &  \cref{eq:het_psi_SEPD-inf}     & not calculated         & \cref{eq:peff-QPD-inf-top,eq:peff-QPD-inf-bottom}  & not calculated        \\
    \hline
    \hline
    \multirow{3}{*}{\ac{gb}-\ac{th}} & \multirow{2}{*}{$\she$} &  \cref{eq:het_eff_SEPD-inf-THGB} & \makecell{\cref{eq:heff-fin-sepd} adjusted \\ with \cref{eq:beta-replacement,eq:thgb-fin-replacement}}& \cref{eq:het_eff_QPD-top-inf-THGB}  & \makecell{\cref{eq:heff-QPD-fin} adjusted \\ with \cref{eq:beta-replacement,eq:thgb-fin-replacement}}\\
                                     &                         &                                  & \Cref{fig:heff_TH_3betas,fig:heff-TH-SEPD-fin-t0t2} &   &   \\
    \cline{2-6}
                                     & $\phe$ &  \cref{eq:het_psi_SEPD-inf}      & not calculated         & \cref{eq:peff-QPD-inf-top,eq:peff-QPD-inf-bottom}  & not calculated        \\
    \hline
    \hline
    \end{tabular}
    \caption{Summary of the analytical results derived in this work for the \ac{he} and heterodyne phase shift. The main assumption of the analytical framework is that the beam tilts on the \ac{pd} are sufficiently small to allow linearization of the resulting electric field, performed between \cref{eq::GB-full-tilt} and \cref{eq::GB-tilt}. The resulting intensity is integrated over circular \acp{pd} of either infinite or finite radius, leading to the results in this table, which are calculated either for a \ac{gb}-\ac{gb} or \ac{gb}-\ac{th} beam interference. These cases are representative of the \ac{tmi} and \ac{isi} in \ac{lisa}. Both $\she$ and $\phe$ depend on $\theta, \weff, \rqpd$. An \ac{gb}-\ac{gb} interferometric topology can be considered to have an infinite  \ac{pd} if $\beta = \sqrt{2}\rqpd/\weff > 3$ (see \Cref{fig:heff-SEPD-fin-t0t2,fig:het-eff-QPD-fin}). The \ac{gb}-\ac{th} case can realistically be treated only in the finite \ac{pd} case. Stating such a rule, in general, for an \ac{gb}-\ac{th} interferometric topology is less trivial; cases with relatively high \ac{he} must be described with a finite \ac{pd} size.Z}
    \label{tab:summary}
\end{table*}


\section*{Acknowledgements}
We gratefully acknowledge support by the Deutsches Zentrum für Luft- und Raumfahrt (DLR) with funding of the Bundesministerium für Wirtschaft und Klimaschutz with a decision of the Deutsche Bundestag (DLR Project Reference No. FKZ 50 OQ 1801) and FKZ 50OQ2301. Additionally, we acknowledge funding by the European Space Agency within the project “Optical Bench Development for \ac{lisa}” (22331/09/NL/HB), support from UK Space Agency, University of Glasgow, Scottish Universities Physics Alliance (SUPA).

\appendix*
\section{Correlated Phase Noises} \label{appendix:corr-phase-noise}
We derive expressions for the \acp{asd} of the sum and the difference of two potentially correlated noise contributions with arbitrary  amplitudes and a specified relative phase or delay. The derivation follows \cite{BendatPiersol2010} and begins with the standard Fourier-domain definitions of auto- and cross-power spectral densities and coherence, ultimately leading to compact formulas that explicitly include the phase offset. 

\subsection{Setup and Definitions}

Let $x(t)$ and $y(t)$ be (wide-sense) stationary, zero-mean random processes. We hence define the Fourier transforms $X_T(f)$ and $Y_T(f)$ of these processes over a finite time period [0, T] as:

\begin{align}
    X_T(f) &= \int_{0}^{T} x(t) e^{-i 2 \pi f t} \dd{t}, \label{eq:finite-fourier-x}\\ 
    Y_T(f) &= \int_{0}^{T} y(t) e^{-i 2 \pi f t} \dd{t} \label{eq:finite-fourier-y}
\end{align}

We use \cref{eq:finite-fourier-x,eq:finite-fourier-y} to define the \acp{psd}, in the limit that $T \rightarrow \infty$, to approach the underlying processes' properties. The one-sided auto- and cross-power spectral densities $G_{\cdot \cdot}(f)$, for $x(t)$ and $y(t)$, are defined from the noise processes' ensemble averages in the aforementioned limit (following \cite[chapter 5]{BendatPiersol2010}), 
\begin{align}
    G_{xx}(f) &=  2 \lim_{T\to\infty}\frac{1}{T}\,\mathbb{E}\!\left\{X_T^{*}(f)X_T(f)\right\},\\
    G_{yy}(f) &= 2 \lim_{T\to\infty}\frac{1}{T}\,\mathbb{E}\!\left\{Y_T^{*}(f)Y_T(f)\right\},\\
    G_{xy}(f) &=  2 \lim_{T\to\infty}\frac{1}{T}\,\mathbb{E}\!\left\{X_T^{*}(f) 
    Y_T(f)\right\}.
\end{align}
Here, $\mathbb{E}\!\left\{\cdot\right\}$ is the expectation value operator. The one-sided cross-power spectral density $G_{xy}(f)$ can be expressed in polar form as 
    
\begin{equation}\label{eq:Gxy-polar}
    G_{xy}(f) = \abs{G_{xy}(f)} e^{-i \theta_{xy}(f)}, \quad 0 < f < \infty,
\end{equation}
with magnitude factor $ \abs{G_{xy}(f)}$ and $\theta_{xy}(f)$ as the relative phase between $x$ and $y$ at frequency $f$. Further, the coherence squared function is (as defined in  \cite[eq. (5.90)]{BendatPiersol2010})
\begin{equation}\label{eq:complex-coherence}
    \gamma_{xy}^2(f) = \frac{\abs{G_{xy}(f)}^2}{G_{xx}(f) G_{yy}(f)},
\end{equation}
where $0\le\gamma_{xy}(f)\le1$ is the magnitude of the coherence at frequency $f$.

\subsection{Sum and Difference Processes}
We define the sum and difference processes as
\begin{equation}
    z_{\pm}(t) = x(t)\pm y(t),
\end{equation}
and their corresponding (and existing) time-limited Fourier domain expressions (as given by \cref{eq:finite-fourier-x})
\begin{equation}
\qquad Z_{\pm,T}(f) = X_T(f) \pm Y_T(f).
\end{equation}
    
This gives the following expression by direct expansion and linearity of the expectation operator in the Fourier domain for the corresponding spectral density, $G_{z\pm  z\pm}(f)$,
\begin{align}
    G_{z\pm z\pm}(f)
    &= 2
    \lim_{T\to\infty}\frac{1}{T}\,\mathbb{E}\!\left\{Z_{\pm,T}^{*}(f)Z_{\pm,T}(f)\right\}
     \nonumber\\
    &= G_{xx}(f) + G_{yy}(f)\pm\big(G_{xy}(f)+G_{yx}(f)\big) \nonumber\\
    &= G_{xx}(f)+G_{yy}(f)\pm 2\,\mathrm{Re}\!\left[G_{xy}(f)\right].
    \label{eq:Gzz-basic}
\end{align}
This identity is a (slightly expanded) standard property of power spectra for linear combinations as given as an example in \cite[5.7]{BendatPiersol2010}. Using \cref{eq:Gxy-polar} and the definition of the coherence function, \cref{eq:complex-coherence},
\begin{equation}
\begin{split}
    \mathrm{Re}\!\left[G_{xy}(f)\right] &= \abs{G_{xy}(f)}  \cos\left(\theta_{xy}(f)\right) \\
    &= \gamma_{xy}(f)  \sqrt{G_{xx}(f)  G_{yy} (f)} \cos\left(\theta_{xy}(f)\right).
    \end{split}
\end{equation}
    
\Cref{eq:Gzz-basic} then leads to the following general form; the one-sided power spectral density for the sum and difference processes,
\begin{equation}
    G_{z_{\pm}z_{\pm}}(f) =G_{xx}(f)+G_{yy}(f) \pm 2\,\gamma_{xy}(f)\,\sqrt{G_{xx}(f)G_{yy}(f)}\,\cos\left(\theta_{xy}(f)\right).
    \label{eq:psd-final}
\end{equation}

\subsection{Amplitude spectral densities}
As we often work with \acp{asd}, we define $\tilde{n}_x(f)=\sqrt{G_{xx}(f)}$ 
and  $\tilde{n}_y(f)=\sqrt{G_{yy}(f)}$. The \acp{asd} of the sum and difference, $\tilde{n}_{\pm}(f) =\sqrt{G_{z_{\pm}z_{\pm}}(f)}$, follow from \cref{eq:psd-final} (dropping the indices for the coherence and phase factor):
\begin{equation}
    \tilde{n}_{\pm}(f)=\sqrt{\tilde{n}_x^2(f)+\tilde{n}_y^2(f) \pm 2 \gamma(f) \tilde{n}_x(f) \tilde{n}_y(f)\cos\left(\theta(f)\right)}.
    \label{eq:asd-final}
\end{equation}
    
\Cref{eq:asd-final} is the desired result used in this work. It reduces to common special cases (dropping the explicit frequency dependency):
\begin{description}
    \item[Fully coherent case ($\gamma=1$)] $\tilde{n}_{\pm}=\sqrt{\tilde{n}_x^2+\tilde{n}_y^2\pm 2 \tilde{n}_x \tilde{n}_y \cos(\theta)}$.
        \begin{description}
            \item[In phase $\theta=0$]
            $\tilde{n}_{+}=|\tilde{n}_x+\tilde{n}_y|$, 
            $\tilde{n}_{-}=|\tilde{n}_x-\tilde{n}_y|$.
            \item[Anti-phase $\theta=\pi$] 
            $\tilde{n}_{+}=|\tilde{n}_x-\tilde{n}_y|$, 
            $\tilde{n}_{-}=\tilde{n}_x+\tilde{n}_y$.
            \item[Quadrature $\theta=\pi/2$] 
            $\tilde{n}_{+}=\tilde{n}_{-}=\sqrt{\tilde{n}_1^2+\tilde{n}_2^2}$ 
            (same as uncorrelated).
        \end{description}
    
    \item[Uncorrelated noises ($\gamma=0$)]
    $\tilde{n}_{\pm}=\sqrt{\tilde{n}_x^2+\tilde{n}_y^2}$ 
    (the usual root squared sum).
\end{description}

\begin{small}
\begin{equation}
\begin{split}
    \she_{\text{, SEPD, }\circ} &= \she_\text{, 0}\left(\sum_{i=0}^6 t_i \theta^i\right) + \mathcal{O}(\theta^8)\\
    \she_\text{, 0} &= \frac{1}{\sqrt{1 + \rho^2}} \frac{2 w_r w_m}{w_r^2 + w_m^2} \frac{1}{\sqrt{1-\exp(- \beta_r^2)}\sqrt{1-\exp(-\beta_m^2)}}\\
    t_0 &= \sqrt{1 - 2 e^{-\beta^2}\cos \left( \beta^2 \rho \right) + e^{-2\beta^2}}\\
    t_2 &= \frac{(k w_\text{eff})^2 \left( 1 - \left(2 + \beta^2(1 + \rho^2) \right)\cos \left( \beta^2 \rho \right)e^{-\beta^2} + \left(1 + \beta^2(1 + \rho^2) \right)e^{-2\beta^2}  \right) }{8(1 + \rho^2) \left( 1 - 2 e^{-\beta^2}\cos \left( \beta^2 \rho \right) + e^{-2\beta^2}\right)^\frac{1}{2} }\\
    t_4 &= \frac{(k w_\text{eff})^4}{128(1+\rho^2)^2\left( 1 - 2 e^{-\beta^2}\cos \left( \beta^2 \rho \right) + e^{-2\beta^2}\right)^\frac{3}{2}} \left(\sum_{n=0}^4 e^{-n\beta^2}\tau_n \right)\\
    &\quad \tau_0 = 1\\
    &\quad \tau_1 = -\frac{1}{2} \cos(\beta^2 \rho) \left( 8 + \beta^2(1 + \rho^2)(2 + \beta^2(1+\rho^2))\right) -  \sin(\beta^2 \rho) \beta^2 \rho (1 + \rho^2) \\
    &\quad \tau_2 = 4 + \frac{1}{2}\beta^2(1+\rho^2)(4 +3\beta^2(1+\rho^2)) +(2 +\beta^2(1+\rho^2))\cos(2\beta^2 \rho) +\\
    &\quad \quad \quad -\beta^2 \rho (1 + \rho^2)\sin(2 \beta^2 \rho) \\
    &\quad \tau_3 = -\frac{1}{2} \left(8 +3\beta^2(1+\rho^2)(2 +\beta^2(1+\rho^2))\right)\cos(\beta^2\rho) -\beta^2\rho \sin( \beta^2 \rho)\\
    &\quad \tau_4 = 1 + \beta^2(1+\rho^2) \frac{1}{2}\beta^4(1+\rho^2)^2\\
    t_6 &= \frac{(k w_\text{eff})^6}{9216(1+\rho^2)^3 \left( 1 - 2 e^{-\beta^2}\cos \left( \beta^2 \rho \right) + e^{-2\beta^2}\right)^\frac{1}{2} } \left(\sum_{n=0}^6 e^{-n\beta^2}\tau_n \right)\\
    &\quad \tau_0 = -12 +\frac{9}{\left( 1 -2e^{-\beta^2}\cos \left( \beta^2 \rho \right) + e^{-2\beta^2}\right)^2}\\
    &\quad \tau_1 = \frac{1}{2}\left( (48 + \beta^2(1+\rho^2)(6(7-\rho^2) +\beta^2(1+\rho^2)(12 + \beta^2(1+\rho^2))))\cos(\beta^2 \rho) + 6\beta^2\rho(1+\rho^2)(4+\beta^2(1+\rho^2)) \right) \times \\
    &\quad \quad \quad \times \sin(\beta^2 \rho) -\frac{9 \left( \left(12 + \beta^2(1+\rho^2)(4+\beta^2(1+\rho^2)) \right)\cos(\beta^2 \rho) + 2\beta^2\rho(1+\rho^2)\sin(\beta^2 \rho) \right)}{2\left( 1 - 2 e^{-\beta^2}\cos \left( \beta^2 \rho \right) + e^{-2\beta^2}\right)^2}\\
    &\quad \tau_2 = -12 -\beta^2(1+\rho^2)\left(3(7-\rho^2) +5\beta^2(1+\rho^2)\left(3 + \beta^2(1+\rho^2) \right) \right) + \frac{9}{4\left( 1 -2e^{-\beta^2}\cos \left( \beta^2 \rho \right) + e^{-2\beta^2}\right)^2} \left( 36 \right.+\\
    &\quad \quad \quad \left. + \beta^2(1+\rho^2)(4+\beta^2(1+\rho^2))(6+\beta^2(1+\rho^2)) + (2+\beta^2(1+\rho^2))(12 + \beta^2(1+\rho^2)(2+\beta^2(1+\rho^2)) ) \cos(2\beta^2 \rho) \right. + \\
    &\quad \quad \quad \left. 2\beta^2\rho(1+\rho^2)(4+\beta^2(1+\rho^2))\sin(2\beta^2\rho) \right)\\
    &\quad \tau_3 = -\frac{9}{2\left( 1 -2e^{-\beta^2}\cos \left( \beta^2 \rho \right) + e^{-2\beta^2}\right)^2}\left( 2+\beta^2(1+\rho^2)\right)\left( (18+\beta^2(1+\rho^2)(9+4\beta^2(1+\beta^2))\cos(\beta^2\rho) + \right.\\
    &\quad \quad \quad \left. +2\cos(3\beta^2\rho) + \beta^2(1+\rho^2)\left(\cos(3\beta^2 \rho) + \rho \left(3\sin(\beta^2 \rho) + \sin(3\beta^2 \rho) \right) \right) \right)\\
    &\quad \tau_4 = \frac{9}{4\left( 1 -2e^{-\beta^2}\cos \left( \beta^2 \rho \right) + e^{-2\beta^2}\right)^2} \left( 36 + \beta^2(1+\rho^2)( 48 + \beta^2(1+\rho^2)( 29 +9\beta^2(1+\rho^2))) +\right. \\
    &\quad \quad \quad \left.(2+\beta^2(1+\rho^2))(12 +\beta^2(1+\rho^2)(10 +3\beta^2(1+\rho^2)))\cos(2\beta^2 \rho) +2\beta^2\rho(4+3\beta^2(1+\rho^2))\sin(2\beta^2 \rho) \right) \\
    &\quad \tau_5 = -\frac{9}{2\left( 1 -2e^{-\beta^2}\cos \left( \beta^2 \rho \right) + e^{-2\beta^2}\right)^2}\left( \left(12 + \beta^2(1+\rho^2)(20 +\beta^2(1+\rho^2)(13 +4\beta^2(1+\rho^2))) \right)\cos(\beta^2 \rho) \right.\\
    &\quad \quad \quad \left.+2\beta^2 \rho (1+\rho^2)(1+\beta^2(1+\rho^2)) \sin(\beta^2 \rho)\right)\\
    &\quad \tau_6 = \frac{9}{2\left( 1 -2e^{-\beta^2}\cos \left( \beta^2 \rho \right) + e^{-2\beta^2}\right)^2}( 1  +\beta^2(1+\rho^2) )(2 + \beta^2(1+\rho^2)(2 + \beta^2(1+\rho^2)))
\end{split}
\label{eq:heff-fin-sepd}
\end{equation}
\end{small}

\begin{small}
\begin{equation*}
\begin{split}
    \she_{\text{, QPD, top, }\circ} &= \she_\text{, 0} \left(\sum_{i=0}^4 t_i \theta^i\right) + \mathcal{O}(\theta^5, \rho^3) \\
    \she_\text{, 0} &= \frac{2 w_r w_m}{w_r^2 + w_m^2} \frac{1}{\sqrt{1-\exp(-\beta_r^2)}\sqrt{1-\exp(-\beta_m^2)}}\\
    t_0 &= \frac{1}{2\left(1-e^{-\beta^2}\right)} \left( 2 - \rho^2 + \left( -4 + 2\rho^2 + \beta^4 \rho^2 \right)e^{-\beta^2} + (2-\rho^2)e^{-2\beta^2} \right)\\
    t_1 &= \frac{(k w_\text{eff})\rho}{4\pi\left(1-e^{-\beta^2}\right)}\left( \sqrt{\pi} \erf(\beta) -\left( 2\beta (1 + 2 \beta^2) + \sqrt{\pi}(1-2\beta^2) \erf(\beta) \right)e^{-\beta^2}  + 2 \beta e^{-2\beta^2} \right)\\
    t_2 &= \frac{(k w_\text{eff})^2}{16\pi^2\left(1-e^{-\beta^2}\right)^3} \sum_{n=0}^4 \tau_n e^{-n\beta^2} \\
    &\quad \tau_0 = \pi \left(\left(4-5 \rho^2\right) \erf^2(\beta)+\pi  \left(3 \rho^2-2\right)\right)\\
    &\quad \tau_1 = \pi ^2 \left(2 \beta ^2-\left(\beta ^6+\beta ^4+\beta ^2+12\right) \rho ^2+8\right)-2 \pi  \left(\left(\beta ^4+2 \beta ^2-5\right) \rho ^2+4\right) \erf^2(\beta)+ \\
    &\quad \quad \quad +4 \sqrt{\pi } \beta  \left(\left(2 \beta ^4+\beta ^2+5\right) \rho ^2-4\right) \erf(\beta)\\
    &\quad \tau_2 = \beta ^2 \left(4-\left(4 \beta ^4+2 \beta ^2+5\right) \rho ^2\right)+\pi^2 \left(\left(\beta ^6+2 \beta ^4+3 \beta ^2+18\right) \rho ^2-6 \left(\beta ^2+2\right)\right)+\\
    &\quad \quad \quad +8 \sqrt{\pi } \beta  \left(\left(\beta ^4+\beta ^2-5\right) \rho ^2+4\right) \erf(\beta)+\pi  \left(\left(-4 \beta ^4+4 \beta ^2-5\right) \rho ^2+4\right) \erf^2(\beta)\\
    &\quad \tau_3 = -8 \beta ^2 \left(\left(\beta ^4-5\right) \rho ^2+4\right)+\pi ^2 \left(6 \beta ^2-\left(\beta ^4+3 \beta ^2+12\right) \rho ^2+8\right)+4 \sqrt{\pi } \beta  \left(\left(2 \beta ^4-3 \beta ^2+5\right) \rho ^2-4\right) \erf(\beta)\\
    &\quad \tau_4 = 4 \beta ^2 \left(\left(2 \beta ^2-5\right) \rho ^2+4\right)+\pi ^2 \left(\beta ^2 \left(\rho ^2-2\right)+3 \rho ^2-2\right) \\
    t_3 &= - \frac{(k w_\text{eff})^3 \rho}{144 \sqrt{2} \pi^3\left(1-e^{-\beta^2}\right)^3} \sum_{n=0}^4 \tau_n e^{-n\beta^2} \\
    &\quad \tau_0 = 9 \pi^\frac{3}{2} \erf(\beta)^3\\
    &\quad \tau_1 = \pi \left(-4 \pi \beta^3 (3 + 2 \beta^2) + 3 \erf(\beta) (\pi^\frac{3}{2} \beta^2 (2 + 3 \beta^2) - 6 \beta (3 + 2 \beta^2) \erf(\beta) + 3 \sqrt{\pi} (-1 + 2 \beta^2) \erf^2(\beta)) \right) \\
    &\quad \tau_2 = -2\pi^2 \beta^2(\beta^2-12) + 3\sqrt{\pi} \beta^2 \left(34 + 48\beta^2 -\pi^2(4+3\beta^2) \right)\erf(\beta) + 18\pi\beta(3- 4\beta^2)\erf^2(\beta)\\
    &\quad \tau_3 = -2\beta^2\left( 6(6+\pi^2)\beta + (72 -5\pi^2)\beta^3 -3\sqrt{\pi}(\pi^2-18+12\beta^2)\erf(\beta)  \right)\\
    &\quad \tau_4 = 72\beta^3\\
    t_4 &= \frac{(k w_\text{eff})^4}{2304  \pi^4\left(1-e^{-\beta^2}\right)^3} \sum_{n=0}^4 \tau_n e^{-n\beta^2} -\frac{(k w_\text{eff})^4 \rho^2}{4608 \pi^4\left(1-e^{-\beta^2}\right)^5} \sum_{n=0}^6 \sigma_n e^{-n\beta^2}\\
    &\quad \tau_0 = \pi^2 \left(9\pi^2 -12\pi \erf^2(\beta) -36 \erf^4(\beta) \right)\\
    &\quad \tau_1 = -\frac{9}{2} \pi^4\left( 8 +2\beta^2 +\beta^4\right) +4\pi^\frac{3}{2} \erf(\beta) \left(4\pi\beta(3+4\beta^2) +3 \pi^\frac{3}{2}(2-3\beta^2)\erf(\beta) +72\beta\erf^2(\beta) \right)\\
    &\quad \tau_2 = -16\pi^2\beta^2(3+8\beta^2) +\frac{27}{2}\pi^4(4+2\beta^2+\beta^4) +4\pi\erf(\beta)\left(4\pi^\frac{3}{2}\beta (\beta^2-6) -3\pi^2 \erf(\beta) +9(\pi^2-24)\beta^2 \erf(\beta) \right)\\
    &\quad \tau_3 = 16\pi^2 \beta^2(6+7\beta^2) -\frac{9}{2}\pi^4(8+6\beta^2+3\beta^4) +16\sqrt{\pi}\beta(72\beta^2 +\pi^2(3-5\beta^2))\erf(\beta)\\
    &\quad \tau_4 = -576\beta^4 +12\pi^2\beta^2(-3+\beta^2) +\frac{9}{2}(2+2\beta^2 +\beta^4)\\
    &\quad \sigma_0 = 3\pi^2 \left( 15\pi^2 -10\pi \erf^2(\beta) -72 \erf^4(\beta) \right)\\
    &\quad \sigma_1 = -\frac{9}{2}\pi^4(60 +6\beta^2 -\beta^4 +2\beta^6 +\beta^8) + 2\pi^\frac{3}{2}\erf(\beta)\left(4\pi\beta(15+5\beta^2 +6\beta^4+8\beta^6) -3\pi^\frac{3}{2}(-20 +11\beta^2 +14\beta^4 +\right.\\
    &\quad \quad \quad \left. +6\beta^6)\erf(\beta) +72\beta(12 +5\beta^2 +2\beta^4)\erf^2(\beta) -54\sqrt{\pi}(-4+4\beta^2 +\beta^4)\erf^3(\beta) \right)
\end{split} 
\end{equation*}
\end{small}

\begin{small}
\begin{equation}
\begin{split}
    &\sigma_2 = -8 \pi^2 \beta^2 (15 + 10 \beta^2 + 20 \beta^4 + 32 \beta^6) -\frac{9}{2} \pi^4 (-150 - 30 \beta^2 + 3 \beta^4 -6 \beta^6 + \beta^8) + \\
    &\quad \quad \quad 2 \pi \erf(\beta) \left(4 \pi^\frac{3}{2} \beta (-60 + 13 \beta^2 + 47 \beta^4 +  64 \beta^6) +  9 \erf(\beta) \left(-10 \pi^2 + (-288 + 11 \pi^2) \beta^2 + 4 (-60 + \pi^2) \beta^4 + \right. \right.\\
    &\quad \quad \quad \left. \left.- 10 (16 + \pi^2) \beta^6 + 16 \sqrt{\pi} \beta (-12 + 7 (\beta^2 + \beta^4)) \erf(\beta) - 12 \pi (1 - 2 \beta^2 + 2 \beta^4) \erf^2(\beta) \right) \right)\\
    &\sigma_3 = -8 \pi^2 \beta^2 (-60 - 7 \beta^2 + 20 \beta^4 + 78 \beta^6) + \frac{9}{2} \pi^4 (-200 - 60 \beta^2 + 2 \beta^4 - 6 \beta^6 + 5 \beta^8) + \\
    &\quad \quad +6 \sqrt{\pi} \erf(\beta) \left(96 \beta^3 (12 + 15 \beta^2 + 14 \beta^4) -4 \pi^2 \beta (-30 + 23 \beta^2 + 29 \beta^4) + \sqrt{\pi} \left(-48 \beta^2 (-36 + 6 \beta^2 + 25 \beta^4) \right. \right. +\\
    &\quad \quad \left. \left.+ \pi^2 (20 - 33 \beta^2 + 18 \beta^4 + 24 \beta^6) \right) \erf(\beta) + 24 \pi \beta (12 - 19 \beta^2 + 14 \beta^4) \erf^2(\beta) \right)\\
    &\sigma_4 = -1152 \beta^4 (3 + 5 \beta^2 + 6 \beta^4) + \frac{9}{2} \pi^4 (150 + 60 \beta^2 + 2 \beta^4 + 2 \beta^6 - 3 \beta^8) + 24 \pi^2 \beta^2 \left(-30 + \beta^2 \left(13 +30 (\beta^2 + \beta^4) \right)\right) +\\
    &\quad \quad - 8 \sqrt{\pi} \beta \left(-144 \beta^2 (-12 - 3 \beta^2 + 7 \beta^4) + \pi^2 (60 - 79 \beta^2 - 9 \beta^4 + 62 \beta^6) \right) \erf(\beta) + 6 \pi \left(\pi^2 (-5 + 6 \beta^2) (1 - \beta^2 + 2 \beta^4) +\right.\\
    &\quad \quad \left. -48 \beta^2 (18 - 21 \beta^2 + 10 \beta^4) \right) \erf^2(\beta)\\
    &\sigma_5 = -\frac{27}{2} \pi^4 (20 + 10 \beta^2 + \beta^4) - 576 \beta^4 (-12 - 8 \beta^2 + 3 \beta^4) + 8 \pi^2 \beta^2 (60 - 59 \beta^2 - 40 \beta^4 + 20 \beta^6) + \\
    &\quad \quad - 8 \sqrt{\pi} \beta \left(-72 \beta^2 (12 - 9 \beta^2 + 2 \beta^4) + \pi^2 (-15 + 28 \beta^2 - 25 \beta^4 + 10 \beta^6) \right) \erf(\beta)\\
    &\sigma_6 = 1152\beta^4(-3+\beta^2) +\frac{9}{2} \pi^4(10+6\beta^2 +\beta^4) -8\pi^2 \beta^2(15-23\beta^2 +10\beta^4)
\end{split} \label{eq:heff-QPD-fin}
\end{equation}
\end{small}

\newpage
\bibliography{paper.bib}

\begin{acronym}
\acro{lisa}[LISA]{Laser Interferometer Space Antenna}
\acro{lpf}[LPF]{\acs{lisa} Pathfinder}
\acro{ids}[IDS]{Interferometric Detection System}
\acro{dws}[DWS]{differential wavefront sensing}
\acro{wfa}[WFA]{wavefront angle}
\acro{dps}[DPS]{differential power sensing}
\acro{lps}[LPS]{longitudinal pathlength signal}
\acro{ap}[AP]{average phase}
\acro{ttl}[TTL]{tilt-to-length}
\acro{ng-ttl}[NG-TTL]{non-geometric \ac{ttl}}
\acro{tdobs}[TDOBS]{\acs{ttl} coupling and \acs{dws} Optical Bench Simulator}
\acro{ob}[OB]{optical bench}
\acrodefplural{ob}[OBs]{optical benches}
\acro{lob}[LOB]{\acs{lisa}-\acl{ob}}
\acro{ts}[TS]{telescope simulator}
\acro{aom}[AOM]{acousto-optic modulator}
\acro{fios}[FIOS]{fiber injector optical sub-assembly}
\acrodefplural{fios}[FIOSs]{fiber injector optical sub-assemblies}
\acro{pbs}[PBS]{polarizing beam splitter}
\acro{tia}[TIA]{trans-impedance amplifier}
\acro{tm}[TM]{test mass}
\acrodefplural{tm}[TMs]{test masses}
\acro{tmi}[TMI]{\acl{tm} interferometer}
\acro{isi}[SCI]{science interferometer}
\acro{rfi}[RFI]{reference interferometer}
\acro{rxgb}[Rx-GB]{Rx Gaussian-beam}
\acro{rxft}[Rx-FT]{Rx flat-top}
\acro{pr}[PR]{photoreceiver}
\acro{qpr}[QPR]{quadrant \acl{pr}}
\acro{pd}[PD]{photodiode}
\acro{sepd}[SEPD]{single-element \acs{pd}}
\acro{qpd}[QPD]{quadrant \acl{pd}}
\acro{cqp}[CQP]{Calibrated \acs{qpd} Pair}
\acro{lo}[LO]{Local Oscillator}
\acro{ligo}[LIGO]{Laser Interferometer Gravitational-Wave Observatory}
\acro{esa}[ESA]{European Space Agency}
\acro{dfacs}[DFACS]{Drag-Free and Attitude Control System}
\acro{roc}[RoC]{radius of curvature}
\acro{rf}[RF]{reference frame}
\acro{snr}[SNR]{signal-to-noise ratio}
\acro{fft}[FFT]{fast-Fourier transform}
\acro{dpll}[DPLL]{digital phase-locked loop}
\acro{adc}[ADC]{analog-to-digital converter}
\acro{refqpd}[REFQPD]{reference \acs{qpd}}
\acro{sciqpd}[SCIQPD]{science \acs{qpd}}
\acro{refsepd}[REFSEPD]{reference \acs{sepd}}
\acro{auxqpd}[AUXQPD]{auxiliary \acs{qpd}}
\acro{rin}[RIN]{relative intensity noise}
\acro{asd}[ASD]{amplitude spectral density}
\acro{psd}[PSD]{power spectral density}
\acrodefplural{psd}[PSDs]{power spectral densities}
\acro{lasd}[LASD]{logarithmic \acs{asd}}
\acro{rms}[RMS]{root mean square}
\acro{snr}[SNR]{signal-to-noise ratio}
\acro{tdi}[TDI]{time delayed interferometry}
\acro{th}[TH]{top-hat}
\acro{he}[HE]{heterodyne efficiency}
\acro{hf}[HF]{heterodyne frequency}
\acro{gb}[GB]{Gaussian beam}
\acro{tf}[TF]{transfer function}
\acro{ingaas}[InGaAs]{indium gallium arsenide}
\end{acronym}

\end{document}